\documentclass[useAMS,referee,usenatbib]{biom}

\usepackage{rotating,tabularx,listings} 
\usepackage{url}
\usepackage{algorithm} 
\usepackage{algpseudocode} 
\usepackage{amsmath}
\usepackage{amssymb}
\usepackage[mathscr]{euscript}
\usepackage[colorlinks=true,linkcolor=red,citecolor=blue]{hyperref}
\usepackage{url}
\usepackage{natbib}
\usepackage{multirow}
\usepackage{booktabs}
\usepackage{hhline}
\usepackage{enumerate}
\usepackage{verbatim}
\usepackage{dsfont}
\usepackage[colorlinks=true,linkcolor=red,citecolor=blue]{hyperref}
\usepackage{threeparttable}
\usepackage{array}
\usepackage{float}
\usepackage{xcolor}
\usepackage{subcaption}

\usepackage{xcolor}
\usepackage{soul}
\usepackage{tikz}
\usetikzlibrary{calc}

\newcommand{\bW}{\boldsymbol{W}}

\newcommand{\bX}{\boldsymbol{X}}
\newcommand{\bbeta}{\boldsymbol{\beta}}
\newcommand{\bY}{\boldsymbol{Y}}

\newcommand{\bO}{\boldsymbol{O}}

\newcommand{\bfI}{\mathbf{I}}
\newcommand{\bone}{\boldsymbol{1}}
\newcommand{\bzero}{\boldsymbol{0}}

\newcommand{\bfU}{\mathbf{U}}
\newcommand{\bfZ}{\mathbf{Z}}

\newcommand{\bfV}{\mathbf{V}}
\newcommand{\bDelta}{\boldsymbol{\Delta}}
\newcommand{\bLambda}{\boldsymbol{\Lambda}}

\newcommand{\btheta}{\boldsymbol{\theta}}
\newcommand{\bPhi}{\boldsymbol{\Phi}}
\newcommand{\bpsi}{\boldsymbol{\psi}}
\newcommand{\bmu}{\boldsymbol{\mu}}

\def\bSig\mathbf{\Sigma}

\title[Model-robust inference in SW-CRTs]{How to achieve model-robust inference in stepped wedge trials with model-based methods?}

\author{Bingkai Wang$^{1,*}$\email{bingkai.w@gmail.com}, 
Xueqi Wang$^{2,3}$, and Fan Li$^{2,4,**}$\email{fan.f.li@yale.edu}\\
$^{1}$Department of Biostatistics, School of Public Health, University of Michigan, Ann Arbor, MI, U.S.A.\\
$^{2}$Department of Biostatistics, Yale School of Public Health, New Haven, CT, U.S.A. \\
$^{3}$Department of Internal Medicine, Yale School of Medicine, New Haven, CT, U.S.A. \\
$^{4}$Center for Methods in Implementation and Prevention Science, Yale School of Public Health,\\ 
New Haven, CT, U.S.A.}

\begin{document}


\date{{\it Received October} 2007. {\it Revised February} 2008.  {\it
Accepted March} 2008.}



\pagerange{\pageref{firstpage}--\pageref{lastpage}} 
\volume{64}
\pubyear{2008}
\artmonth{December}


\doi{10.1111/j.1541-0420.2005.00454.x}


\label{firstpage}


\begin{abstract}
A stepped wedge design is a unidirectional crossover design where clusters are randomized to distinct treatment sequences. While model-based analysis of stepped wedge designs is standard practice to evaluate treatment effects accounting for clustering and adjusting for covariates, their properties under misspecification have not been systematically explored. In this article, we focus on model-based methods, including linear mixed models and generalized estimating equations with an independence, simple exchangeable, or nested exchangeable working correlation structure. We study when a potentially misspecified working model can offer consistent estimation of the marginal treatment effect estimands, which are defined nonparametrically with potential outcomes and may be functions of calendar time and/or exposure time. We prove a central result that consistency for nonparametric estimands usually requires a correctly specified treatment effect structure, but generally not the remaining aspects of the working model (functional form of covariates, random effects, and error distribution), and valid inference is obtained via the sandwich variance estimator. Furthermore, an additional g-computation step is required to achieve model-robust inference under non-identity link functions or for ratio estimands. The theoretical results are illustrated via several simulation experiments and re-analysis of a completed stepped wedge cluster randomized trial. 
\end{abstract}



\maketitle


%

\section{Introduction}
\label{s:intro}
Stepped wedge designs represent a popular class of study designs that sequentially transition experimental units from the control to the intervention conditions; and many stepped wedge designs in health science research randomize entire clusters of individuals. Adoption of a stepped wedge cluster randomized trial (SW-CRT) often reflects the need to ensure full roll-out of an intervention with perceived benefits to all clusters during the study period, and sometimes the need to efficiently allocate finite resources to implement the intervention within a subset of clusters at a time \citep{li2022stepped}. 

An essential task in analyzing SW-CRTs is to estimate the treatment effects, for which purpose the linear mixed model and generalized estimating equations (GEE) represent two mainstream regression techniques \citep{li2021mixed,nevins2023adherence}. Despite their wide use, their statistical properties for \emph{estimand-aligned inference} have not been fully elucidated in the context of SW-CRTs, especially when the estimands are defined separately from the regression model parameters. To address this gap, we investigate a fundamental question. That is, if we define the treatment effect estimands nonparametrically under the potential outcomes framework, when will typical model-based analyses of SW-CRTs---those offered by linear mixed models and GEE---be robust to working model misspecification? Through this investigation, we aim to improve the transparency in model-based analyses of SW-CRTs. 

To start with, \textit{model robustness} refers to valid statistical inference in large samples when the working models are misspecified. For example, the true outcome model may be nonlinear and non-additive functions of baseline covariates, include non-normally distributed errors, or have complex within-cluster correlation structures. Since the true outcome distribution is unknown, model robustness is a desirable property for a treatment effect estimator in a randomized trial. For SW-CRTs, several previous studies \citep{kasza2019inference,bowden2021inference,voldal2022model} have pointed out that misspecification of the correlation structure under linear mixed models can lead to substantial bias in the model-based variance estimator and hence bias inference about the treatment effect. Under a constant treatment effect structure, \citet{ouyang2023maintaining} recently demonstrated via simulations that the sandwich variance estimator provides nominal coverage in linear mixed models under correlation structure misspecification in SW-CRTs. However, the general theory regarding the model robustness property for linear mixed models and GEE, beyond correlation misspecification, remains largely unknown in SW-CRTs. 

Because the SW-CRT is a longitudinal cluster randomized design, the investigation of the model robustness property is further complicated by the potential variation of treatment effects over time. While it has been a convention to assume a regression model with an immediate and constant treatment effect structure \citep{hussey2007design}, the true treatment effect structure may vary by the duration of the intervention or exposure time (e.g., delayed or learning effects), by the calendar time (e.g., seasonality or external shocks), or both. \citet{kenny2022analysis} and \citet{maleyeff2022assessing} have pointed out that a constant treatment effect specification may lead to severely biased estimates of the true treatment effect in the presence of exposure-time treatment effect heterogeneity. Beyond that, there has been little investigation on treatment effect heterogeneity as a function of calendar time or both the calendar and exposure time. In fact, the definition of all possible treatment effect estimands, through the lens of potential outcomes, has not been fully elaborated for SW-CRTs. In light of recent interest in estimand-aligned inference for parallel-arm designs \citep{su2021model,wang2024model}, it is of natural interest to query when typical regression models for SW-CRTs can provide robust inference about more complex treatment effect estimands that may depend on the calendar time and the exposure time.

{Our primary contributions to model-robust inference in SW-CRTs are two-fold. First, we articulate a super-population potential outcomes framework to define estimands that allow for treatment effect heterogeneity across calendar time and/or treatment duration. We provide a causal interpretation of each estimand, echoing the emphasis on clarity of estimands under the ICH-E9 (R1) Addendum for the analysis of clinical trials \citep{ICH_E9}. Second, for each estimand, we show how to adapt linear mixed models and GEE to construct consistent estimators that are robust to working model misspecification. Our central finding is that consistency for nonparametric estimands usually requires a correctly specified treatment effect structure, but generally not the remaining aspects of the working model (functional form of covariates, random effects, and residual distribution). We also provide a generalization of our methods to accommodate ratio estimands. 
Overall, this work simultaneously addresses estimands and baseline covariate adjustment---areas where current guidance for SW-CRTs is relatively sparse. The development of model-robust methods in SW-CRTs can help prevent potential bias from misspecified regression models and ensure that the causal inference rigor for SW-CRTs is comparable to that of individually randomized trials.}

\section{Notation and assumptions}
Consider an SW-CRT with $I$ clusters, and each cluster $i$ contains $N_i$ individuals in its source population. We assume $N_i$ can vary by clusters, and takes values in a bounded subset of positive integers, but remains constant across calendar time in the study. We assume data from each cluster are collected in $J+1$ discrete, equally-spaced periods indexed by $j=0,\dots, J$. 
{Here, we use time $0$ to indicate the pre-rollout or baseline period; in this period, all measured variables are baseline variables, whereas the outcomes of interest are measured in periods $j=1,\dots, J$.}
As time proceeds, each cluster will start receiving treatment in a period randomly chosen among $\{1,\dots, J\}$ such that all clusters will be treated in period $J$ (post-rollout). 
For each individual $k\in\{ 1,\dots, N_i\}$ in cluster $i$, we define $Y_{ijk}$ as their outcome in period $j$, $j\ge 1$, and  $\bX_{ik}$ as their vector of baseline covariates. 
In practice, not all individuals in the cluster source population will be included in the study, 
and we define $S_{ijk}$ as the enrollment indicator such that $S_{ijk}=1$ if individual $k$ from cluster $i$ is included in period $j$; and $N_{ij} = \sum_{k=1}^{N_i}S_{ijk}$ is the observed cluster-period size. Under this framework, we allow each individual to appear in one, multiple or even no periods, thereby accommodating cross-sectional, closed-cohort, and open-cohort designs \citep{li2021mixed}. Furthermore, we define $Z_{i}=j$ if cluster $i$ starts receiving treatment in the beginning of period $j \in \{1,\dots, J\}$.

We pursue the potential outcomes framework to define treatment effect estimands. Focusing on the post-baseline periods $j\in\{1,\ldots,J\}$, 
{we let $Y_{ijk}(z)$ denote the potential outcome of individual $k$ in cluster $i$ during period $j$ had the cluster been first treated in period $z$ for $1\le z\le j$. If $z > j$, we assume no anticipation and $Y_{ijk}(0)$ denotes the untreated potential outcome.}
We connect the observed outcome $Y_{ijk}$ and potential outcomes via the following:
\begin{equation}\label{eq:consistency}
{Y_{ijk} = \sum_{z=1}^j I\{Z_i = z\} Y_{ijk}(z) + I\{Z_i > j\} Y_{ijk}(0),~~~~~j \in \{1,\dots, J\}.}
\end{equation}

The observed data for each cluster are denoted as $\bO_i = \{Y_{ijk},\bX_{ik}, Z_i: S_{ijk} = 1, j = 1,\dots, J,k = 1,\dots, N_i\}$; the source population size $N_i$, however, needs not to be known. The complete, but not fully observed, data vector for each cluster $i$ is denoted as {$\bW_i = \{(Y_{ijk}(0), Y_{ijk}(z), S_{ijk},\bX_{ik}, Z_{i}, N_i): k = 1,\dots, N_i, 1\le z\le j \le J\}$}. To proceed, we make the following assumptions on $\bW_i$:

\vspace{5pt}

\noindent \textbf{A1}. (\emph{Super-population sampling})  $\{\bW_i,i = 1,\dots, I\}$ are independent and identically distributed draws from a population distribution $\mathcal{P}$ with finite second moments. Further, within each cluster $i$, the data vectors $(Y_{i1k}(0), \dots, Y_{iJk}(J), \bX_{ik})$ for $k=1,\dots, N_i$ are identically distributed given the source population size $N_i$.

\noindent \textbf{A2}. (\emph{Staggered randomization}) $Z_i$ is independent of all other random variables in $\bW_i$. In addition,  $P(Z_i = j) = \pi_j \in(0,1)$ for $j = 1,\dots, J$ and $\sum_{j=1}^{J} \pi_j = 1$.

\noindent \textbf{A3}. (\emph{Non-informative enrollment})  $\{S_{ijk}:j = 1,\dots, J,k = 1,\dots, N_i\}$ is independent of all other random variables in $\bW_i$ given $N_i$, and $(N_{i1},\dots, N_{iJ})$ is independent of $N_i$.
\vspace{5pt}

Assumption A1 is typical for causal inference under a super-population framework. Although the dimension of $\bW_i$ varies by $N_i$, the data can be generated via the mixture model $\mathcal{P}=\mathcal{P}^{\bW|N} \times \mathcal{P}^N$. In addition, this assumption requires that individual-level complete data vector has the same expectation conditional on $N_i$; this allows us to construct estimands based on marginal expectations of individual-level potential outcomes. Importantly, although A1 assumes between-cluster independence, it allows for arbitrary within-cluster correlation structures among potential outcomes and covariates. Assumption A2 holds by the stepped wedge design. 
Assumption A3 describes a random enrollment scheme and assumes away selection bias. Under this assumption, the enrollment indicator $S_{ijk}$ in a period is assumed to be irrelevant to the potential outcomes, treatment, or baseline variables, whereas arbitrary correlations among $S_{ijk}$'s are allowed. For example, a closed-cohort design can thus be accommodated by simply setting $S_{i1k}=S_{i2k}=\dots=S_{iJk}$ for all $i$ and $k$. Finally, A3 allows $N_{ij}$ to vary over calendar time either due to randomness or period effects (e.g., arising from exogenous factors not related to $\bW_i$). 

\section{Treatment effect estimands for stepped wedge designs}\label{sec:estimands}
{We define the marginal cluster-average treatment effect as a function of potential outcomes. Given the treatment adoption time $z$ and in period $j$, we denote $d=j-z+1$ as the duration of treatment or exposure time. Then, $Y_{ijk}(j-d+1)$ is the individual potential outcome in period $j$ had cluster $i$ been treated for $d$ periods (or equivalently, had the treatment been first adopted in period $z=j-d+1$).
The treatment effect estimand is then defined as}
\begin{equation}\label{eq: estimand}
    \Delta_{j}(d)=E\left\{\frac{1}{N_i}\sum_{k=1}^{N_i}Y_{ijk}({j-d+1})\right\}-E\left\{\frac{1}{N_i}\sum_{k=1}^{N_i}Y_{ijk}(0)\right\}=E\{Y_{ijk}({j-d+1})\} - E\{Y_{ijk}(0)\},
\end{equation}
for $1\le d \le j \le J$, where the second equality is due to Assumption A1. Here, the expectation is taken over the distribution of clusters, with $j$ and $d$ being fixed quantities; that is, $E[f(\boldsymbol{W}_i)] = \int f(\boldsymbol{w}) d \mathcal{P}(\boldsymbol{w})$ for any integrable function $f$. 
The interpretation of \eqref{eq: estimand} is similar to that in \citet{kahan2023estimands} for parallel-arm designs, except that each estimand now corresponds to the average causal effect had all clusters been treated for $d$ periods at calendar period $j$. Definition \eqref{eq: estimand} is model-free and fully accommodates the treatment effect heterogeneity due to calendar time and exposure time. {However, it is not design-free as the number and length of periods can be specific to each study.} 
One may make the following simplifications of estimands based on content knowledge. 

The \textit{constant treatment effect structure} assumes that 
$\Delta_{j}(d)$ is invariant across all $j$ and $d$, i.e., $\Delta_{j}(d)\equiv \Delta$, and leads to a univariate target estimand $\Delta$. This is a reasonable assumption when the treatment effect is expected to be immediate and sustained; for example, a study evaluating a new emergency operation to improve the survival rate. The constant treatment effect structure has been the default in the stepped wedge design literature \citep{hussey2007design,li2021mixed} and current statistical practice \citep{nevins2023adherence}. 

The \textit{duration-specific treatment effect structure} considers $\Delta_{j}(d)$ to be constant across $j$, but vary by $d$, i.e., $\Delta_{j}(d) = \Delta(d)$. That is, the treatment effect may vary by exposure time, but not the calendar time at which the treatment effect is evaluated. Examples include a new teaching strategy to improve reading scores, or other treatments that may exhibit a delayed or learning effect. \citet{hughes2015current} referred to this as the time-on-treatment effect; \citet{kenny2022analysis} and \citet{maleyeff2022assessing} pointed out that ignoring the duration-specific treatment effect structure can lead to severe estimation bias. The target estimand in this case can be any function of $\bDelta^D = (\Delta(1), \dots, \Delta(J))^\top$; the exposure-time average treatment effect is, for example, $\Delta^{D\text{-avg}}=J^{-1}\sum_{d=1}^J\Delta(d)$.

The \textit{period-specific treatment effect structure} allows the treatment effect to vary by the calendar period, but not the treatment duration, i.e.,$\Delta_{j}(d) = \Delta_j$. The estimand $\Delta_j$ addresses the treatment-by-period interaction effects, which may be present if the effect of treatment is seasonal, or is affected by external events (pandemic or other concurrent intervention) that are correlated with the calendar time. 
{Of note, although we can conceptualize $\Delta_J$ as the treatment effect in period $J$, it is not identifiable because $Y_{iJk}(0)$ is never observable by design \citep{chen2023model}. Therefore, we do not further address $\Delta_J$ in the period-specific treatment effect setting.} The target estimand in this case can be any function of $\bDelta^P = (\Delta_1, \dots, \Delta_{J-1})^\top$, and the calendar-time average treatment effect is defined as $\Delta^{P\text{-avg}}=(J-1)^{-1}\sum_{j=1}^{J-1} \Delta_j$.

\begin{figure}[htbp]
     \centering
     \begin{subfigure}[b]{0.45\textwidth}
         \centering
         \includegraphics[width=\textwidth]{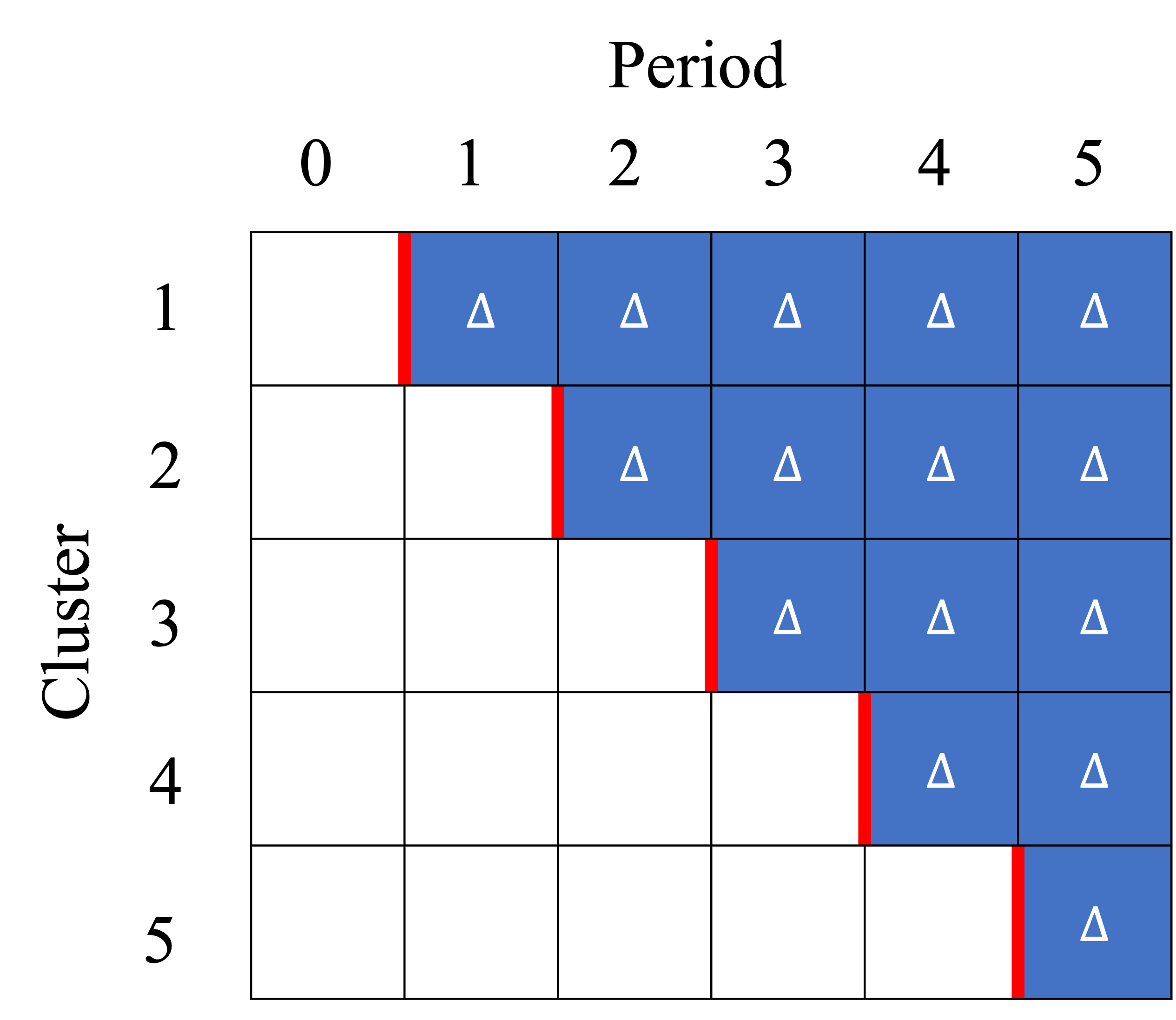}
         \caption{Constant treatment effect}
     \end{subfigure}
     \begin{subfigure}[b]{0.45\textwidth}
         \centering
         \includegraphics[width=\textwidth]{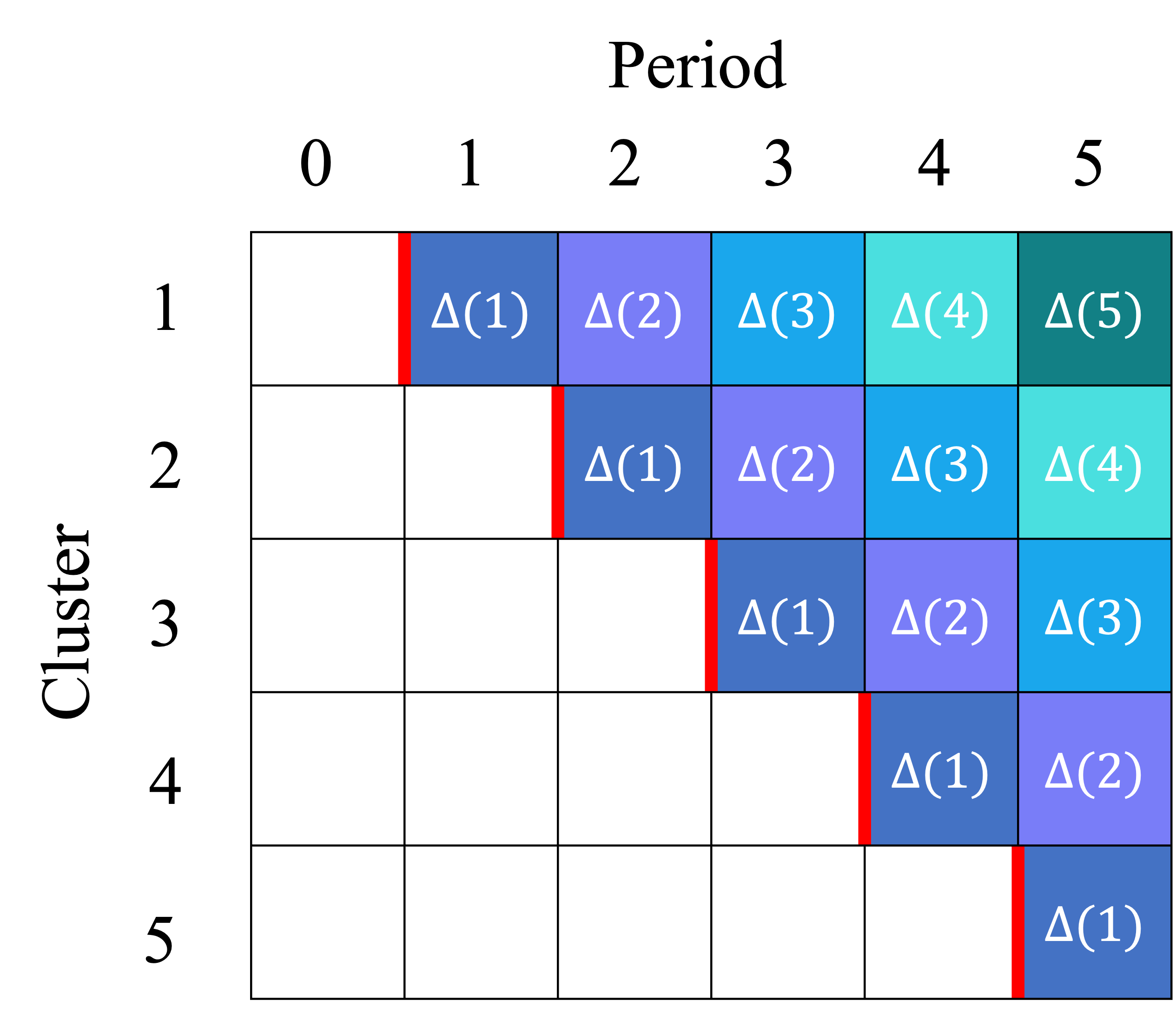}
         \caption{Duration-specific treatment effect}
         \label{fig:three sin x}
     \end{subfigure}
     \hfill \\
     \begin{subfigure}[b]{0.45\textwidth}
         \centering
         \includegraphics[width=\textwidth]{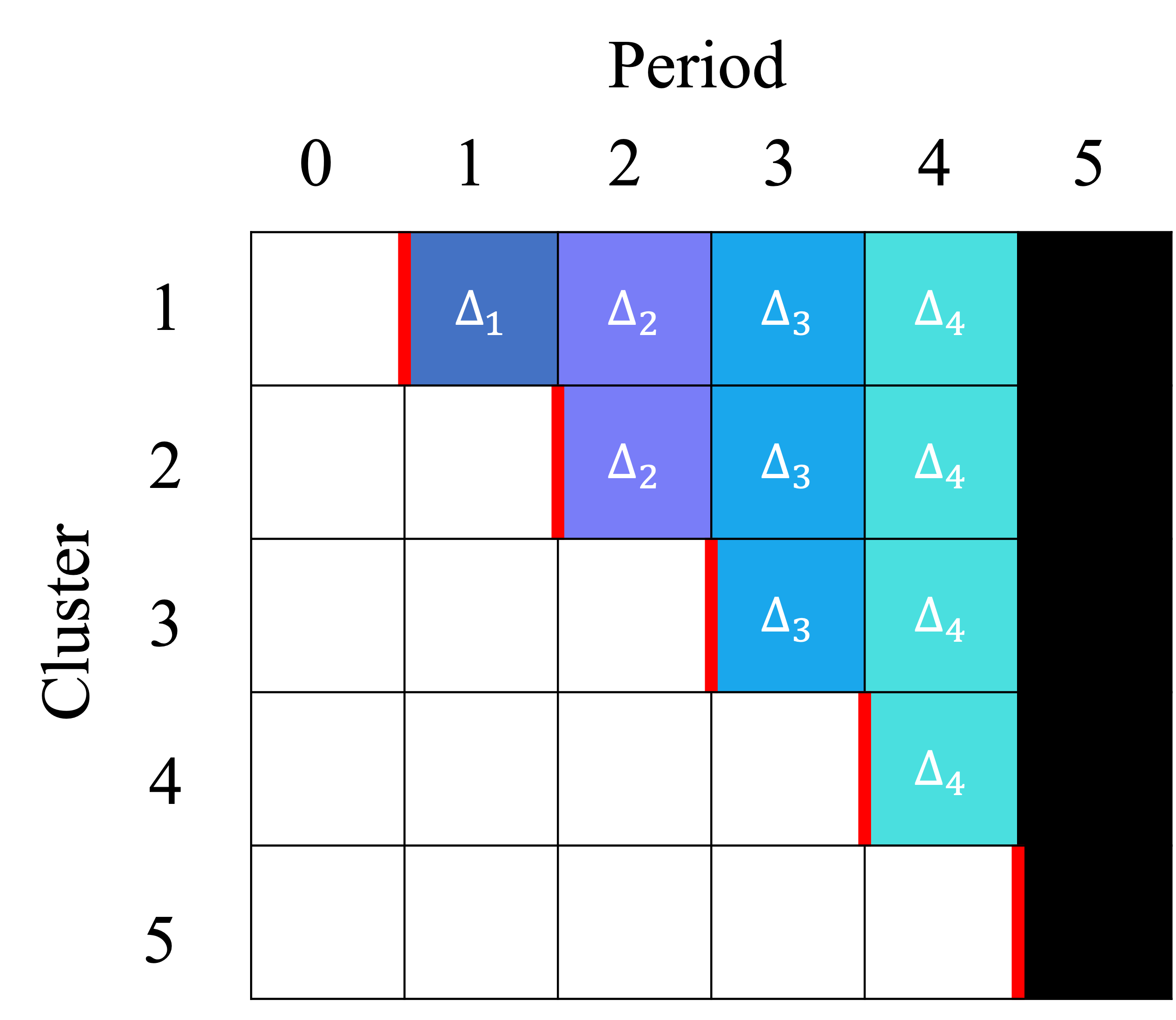}
         \caption{Period-specific treatment effect}
     \end{subfigure}
     \begin{subfigure}[b]{0.45\textwidth}
         \centering
         \includegraphics[width=\textwidth]{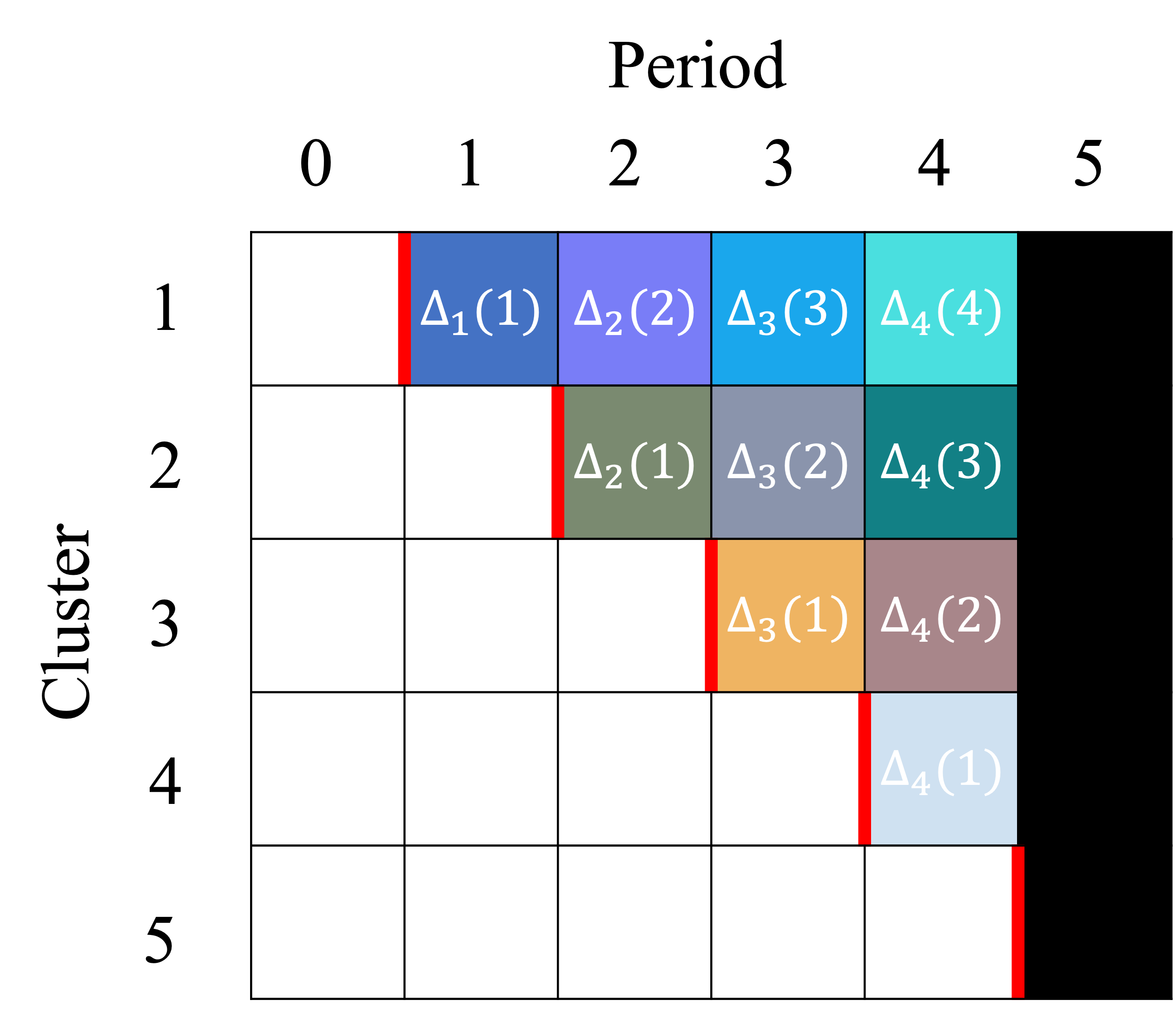}
         \caption{Saturated treatment effect}
     \end{subfigure}
     \vspace{0.2in}
        \caption{Visualization of four different treatment effect structures in SW-CRT with $J=5$ and $I=5$. The red bars represent the unique time point of treatment crossover. Cells with different colors represent potentially different treatment effects. For the period-specific treatment effect and saturated treatment effects (lower panels), causal estimands are not defined for the last period (black cells) due to non-identifiability under the stepped wedge design.}
        \label{fig:treatment-effects}
\end{figure}

Finally, the \textit{saturated treatment effect structure} can be used when the treatment effects are expected to vary by both the calendar time and exposure time. {In this assumption-lean setup, we write $\boldsymbol{\Delta}^{S} = (\Delta_{1}(1), \Delta_2(1), \Delta_2(2), \dots, \Delta_{J-1}(1),\dots,\Delta_{J-1}(J-1))^\top$, and an overall treatment effect estimand that may be of greater interest can be defined as a summary function of $\boldsymbol{\Delta}^{S}$. Similar to the period-specific treatment effect structure, the treatment effects in period $J$, $(\Delta_J(1),\dots, \Delta_J(J))$, are not identifiable and thus excluded from $\boldsymbol{\Delta}^{S}$.}
The saturated treatment effect structure has been considered in the dynamic causal difference-in-differences literature \citep{roth2021efficient}, and an example summary estimand is $\Delta^{S\text{-avg}}=\{(J-1)J\}^{-1}2\sum_{j=1}^{J-1}\sum_{d=1}^j\Delta_j(d)$. For illustration, Figure~\ref{fig:treatment-effects} provides a schematic of the four distinct treatment effect structures applicable to a standard stepped wedge design.

\section{Model-robust inference with linear mixed models}\label{sec:lmm}
We consider the linear mixed model commonly used in analyzing SW-CRTs \citep{li2021mixed}: 
\begin{equation}\label{eq: mixed model}
    Y_{ijk} = \beta_{0j} + TE_{ij}  + \bbeta_{\bX}^\top \bX_{ik} + RE_{ij} + \varepsilon_{ijk}.
\end{equation}
Here, $\beta_{0j}$ is the secular trend parameter (also referred to as the period effect), $TE_{ij}$ encodes the treatment effect structure, $ \bbeta_{\bX}$ is the coefficient for baseline variables, $RE_{ij}$ encodes the random-effects structure, and $\varepsilon_{ijk} \sim N(0, \sigma^2)$ is the residual random error. Model \eqref{eq: mixed model} includes the unadjusted analysis as a special case by setting $\bX_{ik}=\emptyset$. {Importantly, model \eqref{eq: mixed model} is only a working model and therefore accommodates continuous, binary, or count outcomes.}


We address the following specifications for $TE_{ij}$. For estimating the constant treatment effect $\Delta$, we specify $TE_{ij} = \beta_Z I\{Z_i \le j\}$ with coefficient $\beta_Z$ targeting $\Delta$. For the duration-specific treatment effect $\bDelta^D$, we set $TE_{ij} =  \sum_{d=1}^{j} \beta_{Zd} I\{Z_i =j-d+1\}$ such that $\beta_{Zd}$ targets $\Delta(d)$. For the period-specific treatment effect $\bDelta^P$, we set $TE_{ij} =  \beta_{jZ} I\{Z_i \le j\}$ such that $\beta_{jZ}$ targets $\Delta_j$. Finally, for estimating the saturated treatment effect $\bDelta^S$, we define $TE_{ij} =  \sum_{d=1}^{j} \beta_{jZd} I\{Z_i =j-d+1\}$ with $\beta_{jZd}$ targeting $\Delta_j(d)$. 
Furthermore, we consider the random-effects structure $RE_{ij}$ to be any of the following: (1) $ RE_{ij} = 0$ without random effects (hence estimation proceeds via ordinary least squares); (2) $ RE_{ij} = \alpha_i$ with $\alpha_i \sim N(0, \tau^2)$, an exchangeable working correlation structure adopted in \citet{hussey2007design}; (3) $ RE_{ij} = \alpha_i + \gamma_{ij}$  with $\alpha_i \sim N(0, \tau^2)$ and $\gamma_{ij} \sim N(0, \kappa^2)$, a nested exchangeable working correlation structure adopted in  \citet{hooper2016sample}. 
All the random effects are assumed to be mutually independent and independent of $\varepsilon_{ijk}$.

For model~(\ref{eq: mixed model}) with specified $TE_{ij}$ and $RE_{ij}$, we consider maximum likelihood estimators based on the observed data, $\{\bO_1,\dots, \bO_I\}$. {Of note, when estimating $\bDelta^P$ and $\bDelta^S$, data from period $J$ should be removed from model~\eqref{eq: mixed model} to avoid colinearity.} We generically denote all parameters in the linear mixed model as $\btheta$, which are estimated by solving estimating equations $\sum_{i=1}^I \bpsi(\bO_i;\btheta) = \bzero$, where $\bpsi$ is the Gaussian-likelihood score function based on model (\ref{eq: mixed model}). We further construct the sandwich variance estimator  for $\widehat{\btheta}$ as
\begin{equation}\label{Eq: sandwich}
   \widehat{\bfV}_{\btheta} = \left\{ \sum_{i=1}^I \frac{\partial \bpsi(\bO_i,\widehat{\btheta})}{\partial \btheta}  \right\}^{-1} 
    \left\{\sum_{i=1}^I \bpsi(\bO_i,\widehat{\btheta})\bpsi(\bO_i,\widehat{\btheta})^\top\right\}
    \left\{ \sum_{i=1}^I \frac{\partial \bpsi(\bO_i,\widehat{\btheta})}{\partial \btheta}  \right\}^{-1}.
\end{equation}
The point and variance estimators for the treatment effect parameters in $TE_{ij}$ are then the corresponding entries of $\widehat{\btheta}$ and $\widehat{\bfV}_{\btheta}$. To proceed, for each of the estimands $\Delta$, $\bDelta^D$, $\bDelta^P$, $\bDelta^S$, we denote the corresponding point estimators as $\widehat{\beta}_Z$, $\widehat{\bbeta}_Z^{D} =(\widehat\beta_{Z1}, \dots, \widehat\beta_{ZJ})^\top$,  $\widehat{\bbeta}_Z^P =(\widehat\beta_{1Z}, \dots, \widehat\beta_{J-1,Z})^\top$, and $\widehat{\bbeta}_Z^S =(
\widehat\beta_{1Z1}, \widehat\beta_{2Z1}, \widehat\beta_{2Z2}, \dots, \widehat\beta_{J-1,Z,1}, \dots, \widehat\beta_{J-1,Z,J-1})^\top$. Their respective sandwich variance estimators are given by $\widehat{V}$, $\mathbf{\widehat{V}}^{D}$, $\mathbf{\widehat{V}}^{P}$, $\mathbf{\widehat{V}}^{S}$. A concise summary of the estimands, model specifications, and estimators is given in Table~\ref{tab:summary_model_spec}.

\begin{table}[htbp]
    \caption{Summary of estimands, model specifications, and estimators when a working linear mixed model is considered.}
    \label{tab:summary_model_spec}
\renewcommand{\arraystretch}{2}
    \centering
    \begin{tabular}{lccc}
    \hline
        \begin{tabular}[c]{@{}c@{}}Treatment effect\\     structure\end{tabular}  & $TE_{ij}$ in model (\ref{eq: mixed model})  & Point estimator & \begin{tabular}[c]{@{}c@{}}Variance\\     estimator\end{tabular}\\
            \hline
    Constant  & $ \beta_Z I\{Z_i \le j\}$ & $\widehat{\beta}_Z$ & $\widehat{V}$ \\

   Duration-specific & $ \sum_{d=1}^{j-1} \beta_{Zd} I\{Z_i =j-d+1\} $ & $\widehat{\bbeta}_Z^{D} =(\widehat\beta_{Z1}, \dots, \widehat\beta_{ZJ})^\top$& $\mathbf{\widehat{V}}^{D}$ \\   
    Period-specific & $ \beta_{jZ} I\{Z_i \le j\} $ & $\widehat{\bbeta}_Z^{P} =(\widehat\beta_{1Z}, \dots, \widehat\beta_{J-1,Z})^\top$& $\mathbf{\widehat{V}}^{P}$ \\   
   Saturated & $\sum_{d=1}^{j-1} \beta_{jZd} I\{Z_i =j-d+1\}$ &  $\widehat{\bbeta}_Z^{S}=(
\widehat\beta_{1Z1}, \dots, \widehat\beta_{J-1,Z,J-1})^\top $   & $\mathbf{\widehat{V}}^{S}$ \\   \hline
    \end{tabular}
\end{table}

\begin{theorem}\label{thm:1}
   Under Assumptions A1-A3 and standard regularity conditions in the Supplementary Material, the following Central Limit Theorems hold. That is, (a) under a true constant treatment effect structure, $\widehat{V}^{-1/2}(\widehat{\beta}_Z-\Delta) \xrightarrow{d} N(0,1)$; (b) under a true duration-specific treatment effect structure, $\left(\mathbf{\widehat{V}}^{D}\right)^{-1/2} \left(\widehat{\bbeta}_Z^{D} - \boldsymbol{\Delta}^{D}\right)\xrightarrow{d} N(\bzero, \bfI_J)$; (c) under a true period-specific treatment effect structure, $\left(\mathbf{\widehat{V}}^{P}\right)^{-1/2} \left(\widehat{\bbeta}_Z^{P} - \boldsymbol{\Delta}^{P}\right)\xrightarrow{d} N(\bzero, \bfI_{J-1})$; (d) under a true saturated treatment effect structure, $\left(\mathbf{\widehat{V}}^{S}\right)^{-1/2} \left(\widehat{\bbeta}_Z^{S} - \boldsymbol{\Delta}^{S}\right)\xrightarrow{d} N(\bzero, \bfI_{(J-1)J/2})$; in (a)-(d), $\bfI_q$ is a $q\times q$ identity matrix.
\end{theorem}

Theorem \ref{thm:1} suggests that assuming a correct treatment effect structure, the linear mixed model provides consistent and asymptotically normal estimators for the respective {marginal} cluster-average treatment effect estimands, even if any other model components (covariates, random effects or residual distribution) are arbitrarily misspecified. It highlights a central message that robust inference for the treatment effect estimands typically requires a correct treatment effect structure in the working model, rather than any other remaining model aspects. This is a unique result for stepped wedge designs that differs from previous findings on model-robust inference for parallel-arm designs \citep{wang2021mixed}. 

From a theoretical point of view, the saturated treatment effect structure offers the most flexibility since inference can be based on weighted combinations of any component estimands, i.e., $\sum_{d\le j} w_{jd} \Delta_j(d)$ satisfying $\sum_{d\le j}w_{jd} =1$. For example, letting $w_{jd} = 2I\{d>1\}/{\{(J-1)(J-2)\}}$, we can quantify the average treatment effect when clusters have been treated for more than one period. Denoting the vector of weights as $\boldsymbol{w}$, the estimator for $\sum_{d\le j} w_{jd} \Delta_j(d)$ is $\boldsymbol{w}^\top \widehat{\bbeta}_Z^S$ with variance estimator $\boldsymbol{w}^\top \mathbf{\widehat{V}}^{S} \boldsymbol{w}$. By Theorem \ref{thm:1}(d), the quantity $(\boldsymbol{w}^\top \mathbf{\widehat{V}}^{S} \boldsymbol{w})^{-1}\{\boldsymbol{w}^\top \widehat{\bbeta}_Z^S-\sum_{d\le j} w_{jd} \Delta_j(d)\}$ follows a standard normal distribution asymptotically, providing a basis for robust inference about a weighted average treatment effect estimand.

\begin{remark}
\emph{When the true treatment effect structure is constant but estimated via the model assuming a saturated treatment effect structure, then $\sum_{d\le j} w_{jd} \widehat{\beta}_{jZd}$ is consistent to $\Delta$ under any specification of the weights. In this case, it is possible to specify the vector of weights as the eigenvector corresponding to the smallest eigenvector of $\mathbf{\widehat{V}}^{S}$ such that the variance is minimized. On the contrary, when the true treatment effect differs across periods and duration but is estimated under a constant treatment effect structure, the estimator $\widehat{\beta}_Z$ will converge to $\sum_{d\le j} \widetilde{w}_{jd} \Delta_j(d)$ for some weights $\widetilde{w}_{jd}$. By closer inspection, the weights have complex forms and can even be negative, thereby lacking a scientifically relevant interpretation. The same phenomenon also occurs when the true treatment effect structure is duration-specific or period-specific. Therefore, the result in \citet{kenny2022analysis} coincides with our results when the true correlation structure is exchangeable, when there are no covariates, and when the true treatment effect structure is duration-specific.}
\end{remark}

From a practical point of view, a likelihood ratio test can help aid in the decision on the choice of the treatment effect structure---the essential aspect for model-robust inference in SW-CRTs as revealed in Theorem \ref{thm:1}. For example, we can test $H_0: \beta_{Z1}=\dots =\beta_{ZJ}$ to compare duration-specific versus the constant treatment effect \citep{maleyeff2022assessing}. This test can be based on the likelihood ratio statistic $\Gamma_I = -2 \sum_{i=1}^I \log p(\bO_i;\btheta) - \log p^D(\bO_i;\btheta^D)$, where $p(\bO_i;\btheta)$ and $p^D(\bO_i;\btheta^D)$ are the likelihood function under the constant and duration-specific treatment effect model, respectively. Then $\Gamma_I$ converges in distribution to $\chi^2(J-1)$.
Similar $\chi^2$-tests can be considered to compare constant versus period-specific, constant versus saturated, duration-specific versus saturated, and period-specific versus saturated treatment effects, with appropriate degrees of freedom specifications. 


\section{Model-robust inference with generalized estimating equations}
Despite its wide applicability, the linear mixed model may not be a natural model choice when the outcome is binary or count. In this section, we study robust inference for our marginal causal estimands via marginal models fitted by GEE, which naturally handles non-continuous outcomes via non-identity link functions. In parallel to model \eqref{eq: mixed model}, we consider a marginal mean model for the individual-level data as
\begin{equation}\label{eq:gee-mean-model}
    E[Y_{ijk}|Z_i,\bX_{ik}] = g^{-1}(\beta_{0j} + TE_{ij} + \bbeta_{\bX}^\top \bX_{ik}),
\end{equation}
where $g$ is the link function and $TE_{ij}$ is the treatment effect structure specified in the same way as in model \eqref{eq: mixed model}. Following the most common practice, we assume that $g$ is the canonical link function, e.g., logit link for binary outcomes and log link for count outcomes. The vector of regression coefficients $\bbeta$ is estimated by $\widehat{\bbeta}$, a solution to the following GEE:
\begin{equation}\label{eq:GEE}
    \sum_{i=1}^I \bfU_i^\top \bfZ_i^{-1/2} \mathbf{R}_i^{-1} \bfZ_i^{-1/2} (\bY_i^o - \bmu_i^o) = \bzero,
\end{equation}
where $\bY_i^o = \{Y_{ijk}: j=1,\dots, J, k =1,\dots, N_i, S_{ijk} = 1\}$ is the observed outcome vector across periods for cluster $i$, $\bmu_i^o = \{g^{-1}(\beta_{0j} + TE_{ij} + \beta_{\bX}^\top \bX_{ik}): j=1,\dots, J, k =1,\dots, N_i, S_{ijk} = 1\} $ is the mean function for all observed individuals in cluster $i$, $\bfU_i = {\partial \bmu_i^o}/{\partial \bbeta}$ is the derivative matrix, $\bfZ_i = \text{diag}\{v(Y_{ijk}): j=1,\dots, J, k =1,\dots, N_i, S_{ijk} = 1\}$ is the diagonal matrix of natural variance functions $v(Y_{ijk})$, and $\mathbf{R}_i$ encodes the working correlation structure for the observed outcomes in cluster $i$. 
We consider the nested-exchangeable working correlation structure, $\mathbf{R}_i= (1-\rho_1-\rho_2)\bfI_{M_i} + \rho_1 \bone_{M_i} \bone_{M_i}^\top + \rho_2 \oplus_{j=1}^J \bone_{N_{ij}}\bone_{N_{ij}}^\top$, where $M_i = \sum_{j=1}^J N_{ij}$, $\bone_q$ is a $q$-dimensional vector of ones, ``$\oplus$'' is the block  diagonal operator, and $(\rho_1+\rho_2, \rho_1)$ are called the within-period and between-period correlation parameters \citep{li2018sample}. 
The correlation parameters $(\rho_1,\rho_2)$ can be estimated by moment estimators $(\widehat\rho_1, \widehat\rho_2)$ or simply set to 0 under the working independence assumption. 

To target each estimand defined in Section \ref{sec:estimands}, we first construct the average potential outcome estimator for period $j$ with a g-computation formula:
\begin{equation}\label{eq: gee-g-computation}
    \widehat{\mu}_j(b) = \frac{\sum_{i=1}^I\sum_{l=1}^J\sum_{k: S_{ilk}=1}  g^{-1}(\widehat\beta_{0j} + b  + \widehat\beta_{\bX}^\top \bX_{ik})}{\sum_{i=1}^I\sum_{l=1}^J N_{il}},
\end{equation}
where $b$ can be $\widehat\beta_Z, \widehat\beta_{Zd}, \widehat\beta_{jZ}, \widehat\beta_{jZd}$ depending on the model fit based on the assumed treatment effect structure, or $0$. For example, $\widehat{\mu}_j(\widehat\beta_Z)$ is used to estimate $E[Y_{ijk}(1)]$ under the constant treatment effect structure, and $\widehat{\mu}_j(0)$ targets $E[Y_{ijk}(0)]$ in the absence of treatment. This procedure is referred to as the g-computation, through which one can obtain estimates for {model-free} marginal estimands by standardizing across the covariate distribution in the target population. 
Ideally, as in a parallel-arm design, the final treatment effect estimator should be a direct comparison of $\widehat{\mu}_j(b)$ and $\widehat{\mu}_j(0)$. However, in SW-CRTs, the constant and duration-specific treatment effect estimands are defined across multiple periods, and thus the corresponding estimators will be a weighted average of $\widehat{\mu}_j(b)-\widehat{\mu}_j(0)$ across $j$. Specifically for estimating the constant treatment effect $\Delta$, we propose to use 
\begin{align*}
   \widehat{\Delta}_{\textrm{GEE-g}}=
\frac{\sum_{i=1}^I\sum_{j=1}^J\widehat{\lambda}_{ij}\{ \widehat{\mu}_j(\widehat{\beta}_Z)- \widehat{\mu}_j(0)\}}{\sum_{i=1}^I\sum_{j=1}^J \widehat{\lambda}_{ij}},
\end{align*}
where the weights are explicitly defined as functions of randomization probability, observed cluster-period sizes and correlation estimates,
\begin{align*}
    \widehat{\lambda}_{ij} &= \frac{\pi_j^s(1-\pi_j^s)}{1+N_{ij}\widehat\rho_2} - \widehat\rho_1\left(1 +\sum_{l=1}^J \frac{N_{il}\widehat\rho_1}{1+N_{il}\widehat\rho_2}\right)^{-1}\sum_{l=1}^J \frac{N_{il} (\pi_{\min\{j,l\}}^s-\pi_j^s\pi_{l}^s)}{1+N_{il}\widehat\rho_2}
\end{align*}
with $\pi_j^s = \sum_{l=1}^j \pi_{l}$ being the proportion of treated clusters in period $j$. The construction of $\widehat{\lambda}_{ij}$ is to intentionally balance the contribution of data from each period, thereby preventing the default precision weighting in GEE to target an ambiguous estimand under model misspecification. Finally, under working independence, the weights reduce to $\widehat{\lambda}_{ij} = \pi_j^s(1-\pi_j^s)$, which is the tilting function that generates the overlap propensity weights embedded within the independence GEE estimator under an identity link function \citep{tian2024information}. 


Next, for estimating the duration-specific treatment effects $\bDelta^D$, we define an estimator
\begin{align*}
    \widehat{\bDelta}_{\textrm{GEE-g}}^D = \left(\sum_{i=1}^I \sum_{d=1}^J \widehat\bLambda_i(d)\mathbf{H}_d\right)^{-1} \sum_{i=1}^I \sum_{d=1}^J \widehat\bLambda_i(d)\left(\begin{array}{c}
        \widehat{\mu}_1(\widehat{\beta}_{Z1}) - \widehat{\mu}_1(0)  \\
        \vdots \\
        \widehat{\mu}_J(\widehat{\beta}_{ZJ}) - \widehat{\mu}_J(0) 
    \end{array}\right)
\end{align*}
where, for each $d$, $\widehat\bLambda_i(d) \in \mathbb{R}^{J\times J}$ is a matrix function of $(N_{ij}, \pi_j, \widehat\rho_1, \widehat\rho_2)$ defined in Equation (11) of the Supplementary Material, and $\mathbf{H}_d \in \mathbb{R}^{J\times J}$ has the $d$-th column equal to $\bone_{J}$ and all other elements 0. {As an extension of $\widehat{\lambda}_{ij}$ under the constant treatment effect structure, $\widehat{\mathbf{\Lambda}}_i(d)$ is a weighting matrix designed to combat bias due to model misspecification, through balancing the contribution of data from each period and under different treatment duration.}
Interestingly, unlike the constant and duration-specific treatment effects, robust estimation of period-specific and saturated treatment effects is considerably simpler since no weighting across periods is required.
{However, a necessary modification here is to drop observed data from period $J$ from the estimating equations~\eqref{eq:GEE} such that the working model avoids colinearity.}
For $\bDelta^P$, we define an estimator $ \widehat{\bDelta}_{\textrm{GEE-g}}^P$, whose $j$-th entry is $\widehat{\mu}_j(\widehat\beta_{jZ}) - \widehat{\mu}_j(0)$. For $\bDelta^S$, we define the estimator $ \widehat{\bDelta}_{\textrm{GEE-g}}^S = (\widehat{\mu}_1(\widehat\beta_{1Z1}) - \widehat{\mu}_1(0), \dots, \widehat{\mu}_{J-1}(\widehat\beta_{J-1,Z,J-1}) - \widehat{\mu}_{J-1}(0))^\top$.

Our proposed estimators for the {marginal} estimands based on GEE can be cast as solutions to (\ref{eq:GEE}). Similar to the linear mixed models, we use the sandwich variance estimation~(\ref{Eq: sandwich}) to quantify the uncertainty of the point estimators, and we write the variance estimators for $\widehat{\Delta}_{\textrm{GEE-g}}, \widehat{\bDelta}_{\textrm{GEE-g}}^D, \widehat{\bDelta}_{\textrm{GEE-g}}^P, \widehat{\bDelta}_{\textrm{GEE-g}}^S$ as $\widehat{V}_{\textrm{GEE-g}}, \widehat{\bfV}^D_{\textrm{GEE-g}}, \widehat{\bfV}^P_{\textrm{GEE-g}}, \widehat{\bfV}^S_{\textrm{GEE-g}}$, respectively.

\begin{theorem}\label{thm:2}
Assume Assumptions A1-A3, standard regularity conditions in the Supplementary Material, and at least one of the following supplemental conditions: (I) $\rho_1 = \rho_2$ are set to be 0 such that working independence is assumed, (II) $g$ is the identity link with constant working variance $v(Y_{ijk}) = \sigma^2$, (III) $\rho_1$ is assumed to be 0 and $\bX_{ik}$ only contains cluster-level information, i.e.,  $\bX_{ik} = \bX_{ik'}$ for $k\ne k'$, or (IV) the mean model~(\ref{eq:gee-mean-model}) is correctly specified. Then, the following Central Limit Theorems hold. That is, (a) under a true constant treatment effect structure, $\widehat{V}_{\textrm{GEE-g}}^{-1/2}(\widehat{\Delta}_{\textrm{GEE-g}}-\Delta) \xrightarrow{d} N(0,1)$; (b) under a true duration-specific treatment effect structure, $\left(\widehat{\bfV}_{\textrm{GEE-g}}^{D}\right)^{-1/2} \left(\widehat{\bDelta}_{\textrm{GEE-g}}^{D} - \boldsymbol{\bDelta}^{D}\right)\xrightarrow{d} N(\bzero, \bfI_J)$; (c) under a true period-specific treatment effect structure, $\left(\widehat{\bfV}_{\textrm{GEE-g}}^{P}\right)^{-1/2} \left(\widehat{\bDelta}_{\textrm{GEE-g}}^{P} - \boldsymbol{\bDelta}^{P}\right)\xrightarrow{d} N(\bzero, \bfI_{J-1})$; (d) under a true saturated treatment effect structure, $\left(\widehat{\bfV}_{\textrm{GEE-g}}^{S}\right)^{-1/2} \left(\widehat{\bDelta}_{\textrm{GEE-g}}^{S} - \boldsymbol{\bDelta}^{S}\right)\xrightarrow{d} N(\bzero, \bfI_{(J-1)J/2})$. 
\end{theorem}

Theorem~\ref{thm:2} is the counterpart of Theorem~\ref{thm:1} and covers generalized linear models for continuous and categorical outcomes. It reveals the same message that correctly specifying the treatment effect structure remains essential for model-robust inference via GEE in SW-CRTs. However, model robustness for GEE additionally requires g-computation (under non-identity link) and conditions (I)-(IV) in Theorem \ref{thm:2}. Under condition (II), $\widehat{\Delta}_{\textrm{GEE-g}}, \widehat{\bDelta}_{\textrm{GEE-g}}^D, \widehat{\bDelta}_{\textrm{GEE-g}}^P, \widehat{\bDelta}_{\textrm{GEE-g}}^S$ reduce to $\widehat{\beta}_Z$, $\widehat{\bbeta}_Z^{D}$,  $\widehat{\bbeta}_Z^P$, $\widehat{\bbeta}_Z^S$ defined in Section~\ref{sec:lmm}, and the asymptotic results in Theorems~\ref{thm:1} and \ref{thm:2} coincide (in other words, the treatment effect coefficients in the linear mixed model \eqref{eq: mixed model} coincide with a g-computation routine). Beyond this special case, conditions (I) and (III) put restrictions on the working correlation structure to be fully independent or block-diagonal. In practice, condition (I) may be a preferred specification because working independence is straightforward to implement, and condition (III) does not allow for adjustment for individual-level covariates. Finally, results under condition (IV) are expected but a fully correct data-generating mean model with covariates is often impossible to specify. In summary, Theorem \ref{thm:2} provides a strong justification for pursuing the independence GEE as a convenient and feasible strategy for model-robust inference in SW-CRTs. This aligns with existing robustness properties of independence GEE under an alternative finite-population perspective; for example, in \citet{su2021model} for parallel-arm designs, and in \citet{chen2023model} for inferring period-specific treatment effects in SW-CRTs.

\section{Generalizations to treatment effect estimands with other summary measures}\label{sec: other-scales}
An important aspect of the treatment effect estimands is the summary measure \citep{ICH_E9}. While we have mainly addressed the difference estimands, ratio estimands are common for binary and count outcomes. We next extend our methods to accommodate alternative effect measures, focusing on the saturated treatment effect structure (see Remark 2). 
Generalizing \eqref{eq: estimand}, we define the saturated treatment effect estimands as
\begin{equation}\label{eq:gen-estimand}
    \Phi_j(d) = f\bigg(E[Y_{ijk}({j-d+1})], E[Y_{ijk}(0)]\bigg),~~~~~1\le d\le j\le J-1,
\end{equation}
for a user-defined function $f$. For example, $f(x,y) = x/y$ defines the causal risk ratio and $f(x,y) = \log\{x/(1-x)\} -\log\{y/(1-y)\}$ defines the log causal odds ratio. 
The vector of estimands is defined as $\bPhi^S = (\Phi_1(1), \dots, \Phi_{J-1}(J-1))$.
We can use the GEE~(\ref{eq:GEE}) with $TE_{ij} =  \sum_{d=1}^{j} \beta_{jZd} I\{Z_i =j-d+1\}$ to estimate $\bPhi^S$, and we define the set of estimators as $\widehat{\bPhi}^S =\left(f\left\{\widehat{\mu}_1(\widehat{\beta}_{1Z1}), \widehat{\mu}_1(0)\right\}, \dots, f\left\{\widehat{\mu}_{J-1}(\widehat{\beta}_{J-1,Z,J-1}), \widehat{\mu}_{J-1}(0)\right\}\right)$. {Again, observed data from period $J$ are dropped during model-fitting.} In other words, $\Phi_j(d)$ is estimated by $f\left\{\widehat{\mu}_j(\widehat{\beta}_{jZd}), \widehat{\mu}_j(0)\right\}$. Denoting the sandwich variance estimator for $\widehat{\bPhi}^S$ as $\widehat{\bfV}_{\Phi}^S$, we obtain the following result that $\widehat{\bPhi}^S$ is robust to working model misspecification. 

\begin{corollary}\label{corollary1}
    Under conditions in Theorem~\ref{thm:2}, $\left(\widehat{\bfV}_{\Phi}^S\right)^{-1/2} \left(\widehat{\bPhi}^S - \bPhi^S\right)\xrightarrow{d} N(\bzero, \bfI_{(J-1)J/2})$. 
\end{corollary}


Corollary~\ref{corollary1} also implies that linear mixed models can be used to formalize a model-robust estimator for $\Phi^S$ in a similar way. The asymptotic result can be considered as an application of Corollary~\ref{corollary1} under supplemental condition (II) of Theorem \ref{thm:2}, where we choose the identity link and a constant variance function. 

\begin{remark}
\emph{Period-specific treatment effects with a general summary measure $f$ can be defined similarly and estimated following a similar procedure, but there may be conceptual challenges associated with developing model-robust estimators for constant or duration-specific treatment effects defined with a ratio summary measure. 
This is because constant and duration-specific treatment effects are defined across multiple periods, which may be incompatible with a non-linear summary measure $f$. In other words, the treatment effect structure can be affected by the specification of summary measure in the estimands definition. 
Therefore, to address a general summary measure, we recommend using the saturated treatment effect structure and taking a weighted average of $\Phi_j(d)$ to target the constant or duration-specific treatment effect estimands, and model-robust inference can be based on Corollary~\ref{corollary1}.}
\end{remark}

\section{Simulations}
We conduct three simulation studies to illustrate our theoretical findings, with the first two focusing on estimating the marginal average treatment effect with continuous outcomes and the third focusing on marginal estimands defined on the odds ratio scale with binary outcomes. 
{A summary of goals and designs for each simulation is provided in Table~\ref{tab:simulation-overview}.}

\begin{table}[]
    \caption{An overview of simulation designs and settings. The target parameter(s) in each simulation is the marginal estimand(s) defined with potential outcomes.}
    \label{tab:simulation-overview}
    \centering
    \begin{tabular}{p{1cm}p{7cm}p{7cm}}
    \hline
        Design &  Key questions to address & Simulation settings\\
    \hline
    A     &         
    When the constant treatment effect model is correctly specified by a linear mixed model,
    \begin{itemize}
        \item Is the proposed estimator robust to model misspecification?
        \item Does the model-based or sandwich variance estimator provide valid inference? 
        \item Can covariate adjustment improve precision?
    \end{itemize} & \vspace{-0.2in}\begin{itemize}
        \item Continuous outcomes
        \item Constant treatment effect $\Delta = 2$
        \item  6 misspecified working linear mixed models differed by covariate adjustment and random-effects specification
        \item Misspecification of working linear mixed models:
        moderate (scenario A1) or severe (scenario A2). 
        Data-generating distribution in A1 includes Gaussian random effects and error, but in B1 includes Gamma random effects and Poisson error.
    \end{itemize}\\
            \hline
    B & If the true treatment effect structure is duration-specific,     \begin{itemize}
        \item Based on linear mixed models with an assumed constant treatment effect structure, are the proposed point and variance estimators robust to model misspecification?
        \item Based on linear mixed models with an assumed duration-specific treatment effect structure, are the proposed point and variance estimators robust to model misspecification?
    \end{itemize} & \vspace{-0.2in}\begin{itemize}
        \item Continuous outcomes
        \item True treatment effect increases over treatment duration
        \item  12 misspecified working models differed by treatment effect structure specification (constant or duration-specific), covariate adjustment, and random-effects specification
        \item Misspecification of working models:  moderate (scenario B1) or severe (scenario B2), similar to scenarios A1 and A2
    \end{itemize}\\
    \hline
    C$^{*}$ & For estimating marginal odds ratio estimands under the saturated treatment effect structure, \begin{itemize}
        \item Are the proposed linear mixed models and GEE estimators (subject to a g-computation step) robust under model misspecification?
        \item Does model-based inference with generalized linear mixed models target our marginal estimands? 
    \end{itemize}& \vspace{-0.2in}\begin{itemize}
        \item Binary outcomes
        \item True treatment effect varies by period and treatment duration
        \item  6 misspecified working models: linear mixed model, GEE, or generalized linear mixed models with or without covariate adjustment
    \end{itemize}\\
        \hline
    \end{tabular}
    \begin{tablenotes}
        \item $^{*}$Details provided in the Supplementary Material Section C.2.
    \end{tablenotes}
\end{table}

\subsection{Continuous outcomes}

\subsubsection{Simulation design A}
The first simulation includes data-generating distributions with a true constant treatment effect structure ($\Delta_j(d) \equiv\Delta)$ and demonstrates the robustness of linear mixed models under misspecification. We consider two scenarios (A1 and A2), and in each scenario, address two sample sizes---a small number of clusters $(I=30)$ or a large number of clusters $(I=100)$. For both scenarios, we set $N_i = 1000$ and $J=5$.

For scenario A1, we independently generate four covariates via $X_{ik1} \sim \mathcal{B}(0.5)$, $X_{ik2} \sim \mathcal{B}(0.8)$, $X_{ik3} = e_{i3}+f_{ik3}$ and $X_{ik4} = e_{i4}+f_{ik4}$, where $\mathcal{B}(p)$ represents the Bernoulli distribution with success probability $p$, $ e_{i3} \sim N(0,0.1)$, $ e_{i4} \sim N(0,0.1)$, $ f_{ik3} \sim N(0,0.4)$, and $ f_{ik4} \sim N(0,0.9)$.
We next generate the potential outcomes for $a=0,1$ as
\begin{align*}
    Y_{ijk}(a) &= \beta_{0j} + \beta_Z(X)a + \frac{3(j+1)}{2} X_{ik1} + X_{ik2} + \frac{6(j+1)}{J+1} X_{ik3}^2 + X_{ik4} + \alpha_i + \epsilon_{ijk},
\end{align*}
where $\beta_{0j} = 0.25+0.004j$ is the period effect, $\alpha_i \sim N(0, 0.1)$ is the cluster-level random effect, $\epsilon_{ijk} \sim N(0, 0.9)$ is the random error, and $\beta_Z(X) = 2 + \frac{1}{2}(X_{ik1} - \overline{X_{i\cdot 1}}) + (X_{ik3}^3 - \overline{X^3_{i\cdot 3}})$ is the covariates-specific intervention effect with $\overline{f(X_{i.p})}$ representing the average of $f(X_{ikp})$ over $k$ for any function $f$ and coviarate index $p$. Next, we randomize treatment adoption time $Z_i$ such that $I/J$ new clusters start receiving treatment at each period, and define $Y_{ijk} = I\{Z_i\le j\}Y_{ijk}(1) + I\{Z_i> j\}Y_{ijk}(0)$. Finally, we randomly sample $N_{ij}$ individuals for period $j$ in cluster $i$ (with $S_{ijk}=1$), and $N_{ij}$ is a random draw from integers in $[5,50]$. 

Scenario A2 differs from A1 in that we set $\beta_{jZ}(X) = 2 + (j+1)(X_{ik1} -\overline{X}_{i\cdot 1})/2 + (j+1)(X_{ik3}^3 - \overline{X^3}_{i\cdot 3})/(J+1) + \beta_i$ with $\beta_{i} \sim N(0,0.25)$ to build in period-varying treatment effect heterogeneity by covariates and a random cluster-specific treatment effect; we also replace the normal random intercept $\alpha_i$ by a Gamma random intercept $\widetilde{\alpha}_i$ plus a Gamma cluster-period random intercept, and replace the normal random error $\epsilon_{ijk}$ by a Poisson random error $\widetilde{\epsilon}_{ijk}$. 
These modifications introduce additional complications to the data-generating distribution, in order to assess the robustness of linear mixed models under simultaneous covariate and random-effects distribution misspecification. 

For each scenario, we generate $1000$ data replicates and fit linear mixed model~\eqref{eq: mixed model} with $TE_{ij}= \beta_Z I\{Z_i \le j\}$ under 6 specifications---combinations of covariate adjustment (no adjustment, $\bbeta_{\bX}$ set to be zero; partial adjustment of linear terms, $\beta_{X2},\beta_{X4}$ set to be zero; or full adjustment of linear terms, no constraint on $\bbeta_{\bX}$) and random effect specifications (simple exchangeable or nested exchangeable structure). For each estimator, we report bias, empirical standard error, and coverage probability of 95\% confidence intervals via normal approximation (based on both the model-based and sandwich variance estimators).

\subsubsection{Simulation design B}
The second simulation contains two scenarios (B1 and B2). The data-generating distribution of B1 resembles A1 but includes a duration-specific treatment effect structure. Specifically, the potential outcomes are generated via
\begin{align*}
    Y_{ijk}(z) &= \beta_{0j} + I\{z>0\}\beta_{Zd}(X) + \frac{3(j+1)}{2} X_{ik1} + X_{ik2} + \frac{6(j+1)}{J+1} X_{ik3}^2 + X_{ik4} + \alpha_i + \epsilon_{ijk},
\end{align*}
{for $z=0,\dots, j$ and $j=1,\dots,J$, where}
$\beta_{Zd}(X) = (1+d)\left\{\frac{1}{2} + \frac{1}{8} (X_{ik1} -\overline{X}_{i\cdot 1}) + \frac{1}{4} (X_{ik3}^3 - \overline{X^3}_{i\cdot 3})\right\}$ with $d = j-z+1$. Then we set {$Y_{ijk}=\sum_{z=1}^{j} I\{Z_i = z\}Y_{ijk}(z) + I\{Z_i > j\}Y_{ijk}(0)$}. Scenario B2 resembles A2 but includes a duration-specific treatment effect structure.


For each scenario, we generate $1000$ data replicates, and fit 12 linear mixed models for each data replicate. Among the 12 models, The first 6 are the same as in the first simulation study, and the remaining models modify the first 6 by replacing $TE_{ij} = \beta_{Z} I\{Z_i\le j\}$ with $TE_{ij} =  \sum_{d=1}^{j} \beta_{Zd} I\{Z_i =j-d+1\}$ in the working model specification. 
Since the true treatment effect is duration-specific, we aim to estimate $\Delta(d)$ for each $d$ and $\Delta^{D\text{-avg}}$, and investigate the same performance metrics as in the first simulation study.

\subsubsection{Simulation results}

\begin{figure}
    \centering
    \includegraphics[width=0.8\textwidth]{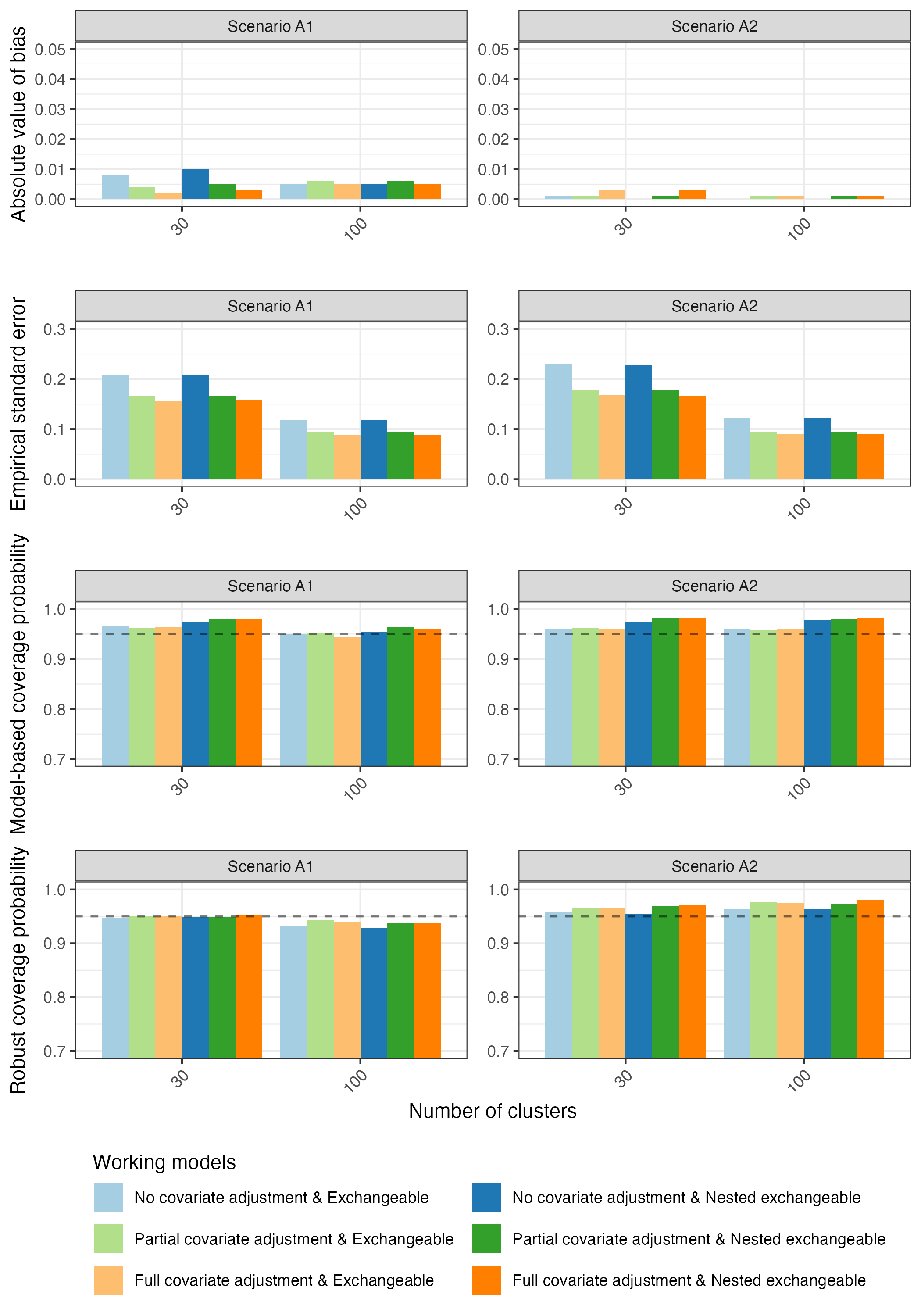}
    \vspace{0.2in}
    \caption{Simulation results for Design A. 
    The 6 working models vary by covariate adjustment (no, partial, or full adjustment) and working correlation structures (exchangeable or nested exchangeable). The performance metrics are absolute values of bias (row 1), empirical standard error (row 2), empirical coverage probability of 95\% model-based confidence intervals (row 3), and 95\% robust confidence intervals (row 4).}
    \label{fig:sim-A}
\end{figure}

Figure~\ref{fig:sim-A} presents the results for the first simulation. 
For both scenarios,
all estimators have negligible bias and nominal coverage for the true $\Delta$ using the sandwich variance estimator, {thereby supporting our theoretical results}. 
Interestingly, the model-based inference also leads to near-nominal coverage; 
however, this observation does not carry to the next simulation with a more complicated data-generating distribution. In addition, the simulations show that baseline covariate adjustment can improve precision: compared to no covariate adjustment, partial and full covariate adjustment by linear mixed models reduce the empirical variance by $\sim38\%$, and $\sim 44\%$, respectively.

\begin{figure}
    \centering
    \includegraphics[width=0.8\textwidth]{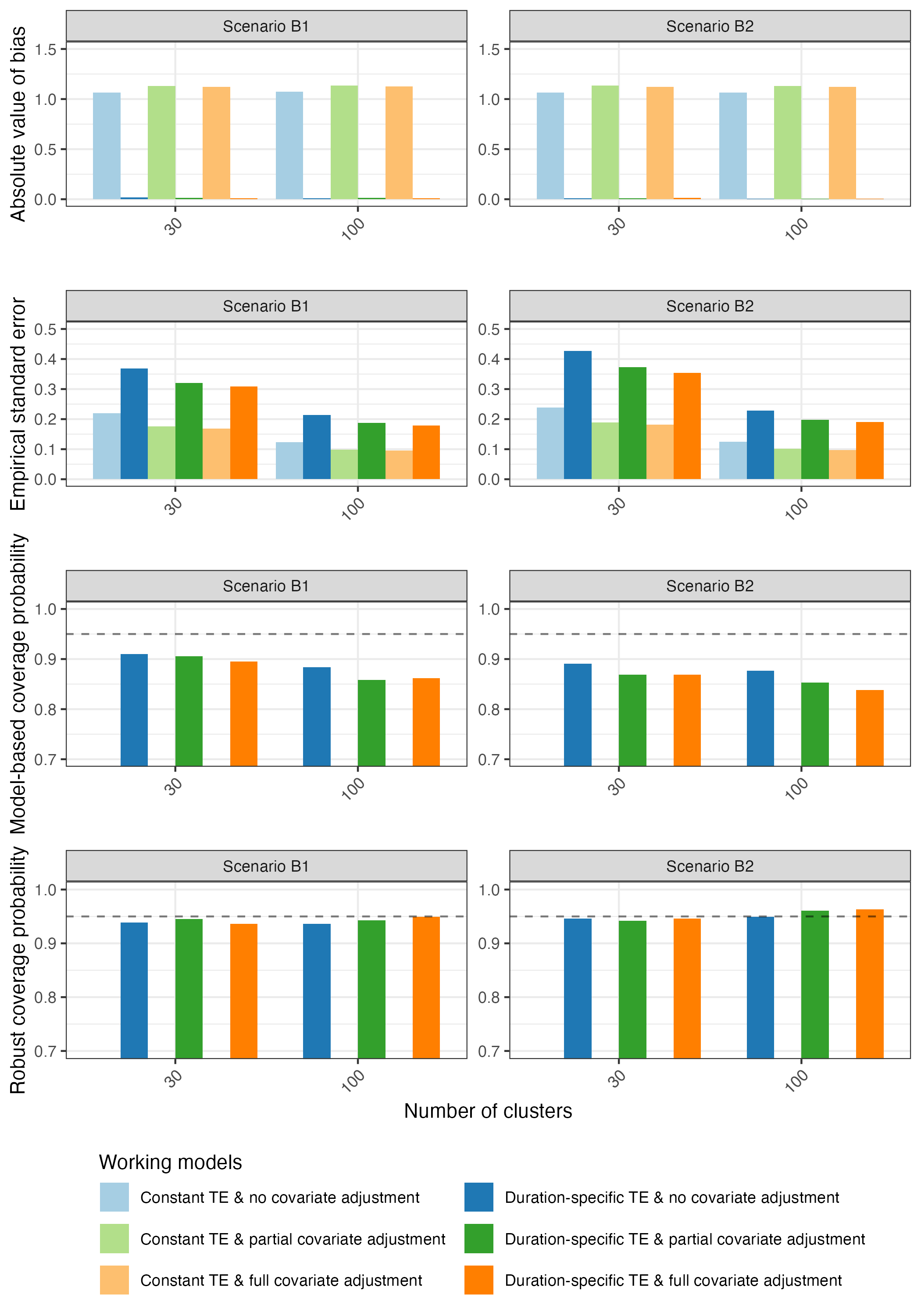}
    \vspace{0.2in}
    \caption{Simulation results for Design B. 
    The 6 working models vary by treatment effect structure specifications (constant or duraction specific) and covariate adjustment (no, partial, or full adjustment). All 6 working models use a simple exchangeable correlation structure. The performance metrics are absolute values of bias (row 1), empirical standard error (row 2), empirical coverage probability of 95\% model-based confidence intervals (row 3), and 95\% robust confidence intervals (row 4).}
    \label{fig:sim-B}
\end{figure}

Figure~\ref{fig:sim-B} presents the results for
the second simulation study. For brevity, we present the results for estimating $\Delta^{D\text{-avg}}=\sum_{d=1}^5 \Delta(d)/5 = 2$ under the simple exchangeable correlation structure and leave the full results for all $\Delta(d)$ to Web Tables 1-8 in the Supplementary Material. When estimating $\Delta^{D\text{-avg}}$, Figure~\ref{fig:sim-B} shows that assuming a constant treatment effect structure leads to severely biased estimates, similar to earlier findings in \citet{kenny2022analysis} and \citet{maleyeff2022assessing} in the absence of covariates. However, as long as a correct duration-specific treatment effect structure is considered in the working model, each estimator returns negligible bias and nominal coverage based on the sandwich variance estimator. Importantly, inference based on the model-based variance estimator is no longer valid, resulting in 5\%-12\% under-coverage. Finally, similar to the first simulation, we confirm that adjusting for prognostic covariates leads to substantial efficiency improvement.


\subsection{Binary outcomes}
We conduct additional simulations with binary outcomes to further evaluate our model-robust estimators for the treatment effect estimands defined on the odds ratio scale, and to compare with conventional estimators under the generalized linear mixed model. We simulate data from a logistic mixed model with a duration-specific treatment effect on the logit scale, akin to the one considered in design B. While the treatment effect given covariates is duration-specific on the logit scale, the implied true marginal treatment effect structure on the marginal odds ratio scale is saturated, as defined in \eqref{eq:gen-estimand}. We therefore examine working models with a saturated treatment effect structure, and fit models with no covariate adjustment and with linear covariate adjustment. Under this simulation design, for our model-robust estimator via either linear mixed models or GEE, the bias is negligible, the average estimated standard error roughly matches the empirical standard error, and the coverage probabilities are nominal, demonstrating the robustness feature for addressing ratio estimands. In contrast, model-based estimators via the generalized linear mixed model leads to relatively large bias, underestimation of the true empirical standard error, and lower than nominal coverage probability, especially when the number of clusters is large (30\% under coverage).  We provide details about the data-generation distribution, the true parameter values, and the full simulation results (Web Tables 9-10) in the Supplementary Material.

\section{A stepped wedge trial data example}\label{sec:data}

SMARThealth India \citep{peiris2019smarthealth} is a SW-CRT that evaluated the effect of a community health-worker-managed mobile health intervention on reducing cardiovascular disease in rural India. 
In this study, 18 primary health centers (clusters) were equally randomized to start intervention in 3 steps, with each step taking 6 months and outcomes measured for approximately $120$ participants per cluster; these participants comprised about 15\% of the high-risk cohort identified at baseline (source population) and were randomly sampled in each period, leading to a cross-sectional design. We focus on the continuous outcome of systolic blood pressure (SBP), an important risk factor for cardiovascular disease. The baseline covariates include age, sex, baseline SBP, and an indicator of baseline SBP $<140$ (indicating attainment of target SBP levels). We apply the linear mixed models~\eqref{eq: mixed model} with different specifications to estimate the marginal average treatment effect estimands.

\begin{figure}
    \centering
    \includegraphics[width=0.7\textwidth]{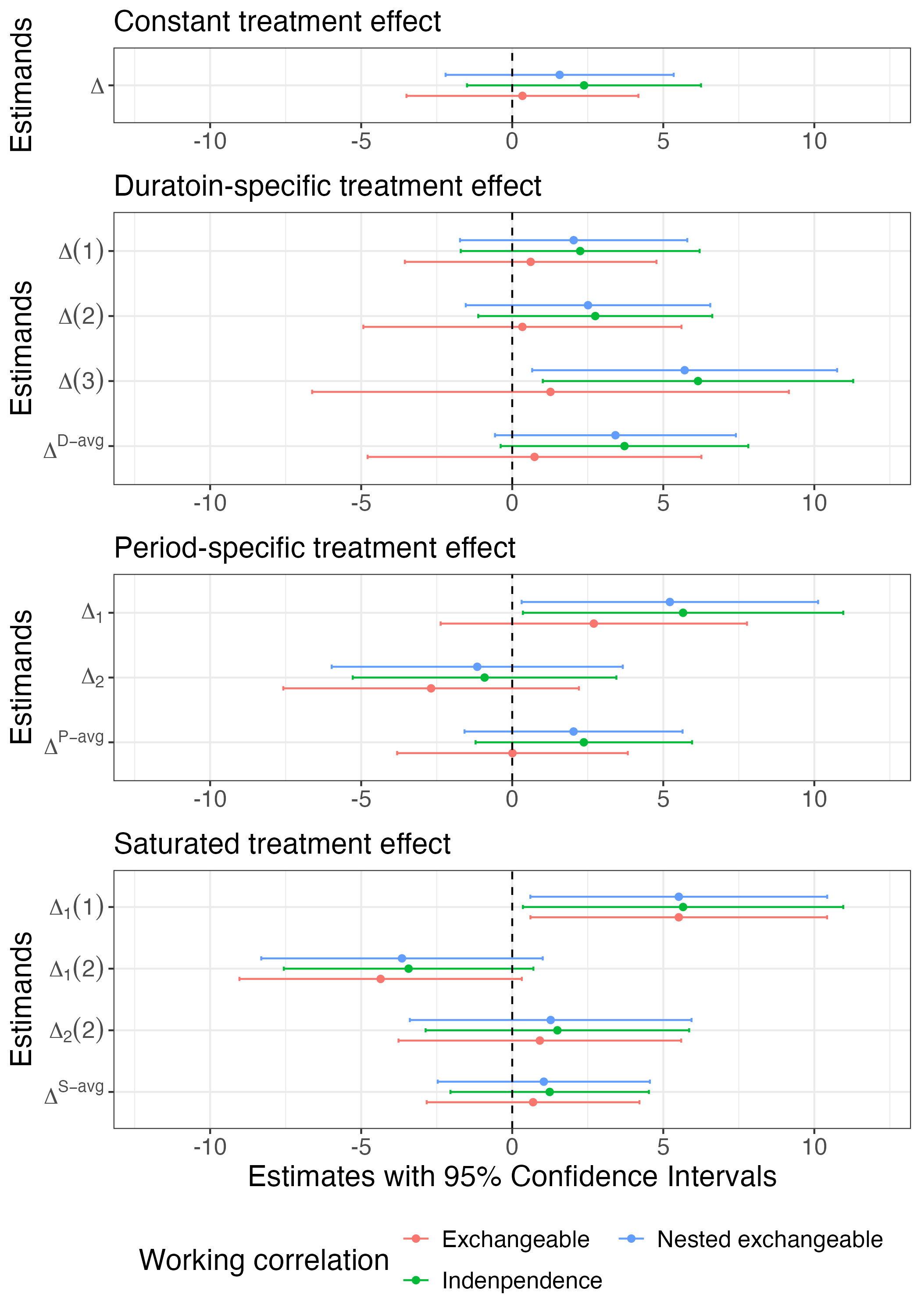}
    \vspace{0.2in}
    \caption{Point estimates and 95\% robust confidence intervals based on the analysis of SMARThealth India SW-CRT with unadjusted linear mixed models. }
    \label{fig:data-analysis}
\end{figure}

Figure~\ref{fig:data-analysis} summarizes the analysis results under different working linear mixed models without covariate adjustment. For each of the four treatment effect structure specifications, we consider independence, exchangeable, and nested exchangeable random-effects structures. We observe that the independence and the nested exchangeable random-effects structures result in similar confidence intervals, but lead to larger point estimates and smaller variances compared to the exchangeable random-effects structure.
Across all four treatment effect structure specifications, the average treatment effects ($\Delta$, $\Delta^{D\text{-avg}}$, $\Delta^{P\text{-avg}}$, $\Delta^{S\text{-avg}}$) are all estimated to be larger than zero, suggesting that the mobile health intervention may increase the SBP, but none of their 95\% confidence intervals exclude zero; 
this is consistent with the primary study findings. However, some interesting observations emerge when we compare the results under different treatment effect structures. Based on the likelihood ratio tests, there is at most borderline evidence from the data to support the duration-specific treatment effect structure over the constant treatment effect structure 
(p-value $=0.08$, $0.09$, $0.04$ under each random-effects structure). Nevertheless, there is strong evidence in favor of treatment effect heterogeneity across calendar time (p-value $<0.01$ when comparing period-specific or saturated versus constant treatment effect structure). This heterogeneity is further substantiated when we inspect each component treatment effect estimand. That is, the point estimates for $\Delta_1$ and $\Delta_1(1)$ are substantially larger than those for the remaining component estimands, with confidence intervals excluding zero. This suggests that the mobile health intervention may elevate SBP during the first rollout period. As discussed in \citet{peiris2019smarthealth}, this is likely due to several reasons, including a strong heatwave in Andhra Pradesh during the first rollout period and the possibility that standard care improved over calendar time. The results under covariate adjustment (Web Figure 1) are generally similar but show larger variance estimates compared to no covariate adjustment. This is likely because the collected covariates carry only small prognostic values in this trial, and the number of clusters is relatively limited. For brevity, we only report those results in the Supplementary Material.

\vspace{-0.2in}
\section{Discussion}
\label{s:discuss}
{
We contributed a causal inference framework for SW-CRTs accommodating different target estimands of interests, and showed how to perform robust inference based on linear mixed models and GEEs. One of our key findings is that robust inference in SW-CRTs typically requires correctly specified treatment effect structures, which is not needed in a parallel-arm design with a linear mixed model \citep{wang2021mixed}. This stronger requirement originates from a unique feature of the stepped wedge design that treatment effect estimands may vary by two time axes: calendar time and exposure time. In summary, our results continue to reiterate the importance of prioritizing the treatment effect structure specification over other model components in SW-CRTs, for the purpose of estimand-aligned causal inference.}

{In practice, there is a trade-off between the complexity in treatment effect structure and the power to detect the treatment effect signal. Among the four treatment effect structures, the effective sample size for addressing each component estimand is the smallest under the saturated specification. Thus, it would require a much larger sample size to achieve sufficient power for testing $H_0:\Delta_j(d)=0$. We generally recommend focusing on summary estimands over each component estimand in a single trial, and to the extent possible, leveraging content knowledge to simply the treatment effect structure. Finally, adjusting for prognostic covariates in the analysis remains an important consideration to improve efficiency.} 

Our developments inspire several future directions. First, we have primarily focused on the cluster-average treatment effect estimands, whereas the individual-average treatment effect estimands may also be of interest \citep{kahan2023estimands}. Inferring this latter set of estimands generally requires knowledge of the source population size $N_i$ \citep{wang2024model}, and the corresponding model-robust methods will be developed in future work.
Second, to better operate our large-sample results to frequent practice with a small number of clusters in SW-CRTs, it is important to further examine the application of small-sample corrections to the sandwich variance estimator, as in \citet{ouyang2023maintaining} and \citet{li2018sample}, but for targeting marginal estimands. 
{Third, different individuals in a cluster may have different amounts of treatment duration depending on their precise recruitment time points. 
Our current causal framework handles the intention-to-treat estimands by emphasizing cluster-level treatment duration, and inspecting estimands based on individual treatment duration necessitates an alternative framework with additional assumptions.} Finally, the working models we considered for our theoretical development, though typical in practice, are not exhaustive. Our theory does not address generalized linear mixed models, nor the exponential decay correlation structure \citep{kasza2019inference}. The score functions under these working model specifications often do not have a tractable analytical form, posing challenges in deriving explicit asymptotic results for studying model robustness. {To empirically explore the performance of working linear mixed models with an exponential decay correlation structure, we repeated our simulations and found that this working correlation structure yields similar results in terms of bias and coverage (Web Tables 11-17 of the Supplementary Material) to our proposed working models, and thus remain empirically robust to misspecification. We encourage additional simulations under a wider range of data-generating distributions to continue informing the robustness features of alternative model specifications in SW-CRTs.}

\backmatter



\section*{Acknowledgements}
Research in this article was supported by Patient-Centered Outcomes Research Institute Awards\textsuperscript{\textregistered} (PCORI\textsuperscript{\textregistered} Awards ME-2020C3-21072 and ME-2022C2-27676) and the National Institute Of Allergy And Infectious Diseases of the National Institutes of Health under Award Number R00AI173395. The statements presented in this article are solely the responsibility of the authors and do not necessarily represent the official views of the National Institutes of Health or PCORI\textsuperscript{\textregistered}, its Board of Governors, or the Methodology Committee. We thank Dr. Rui Wang for helpful discussions based on an earlier version of this article.

\section*{Supplementary Material}
{Web Appendices, Tables, and Figures, and code referenced in Sections~\ref{sec:lmm}-\ref{s:discuss} are available with this paper at the Biometrics website on Oxford Academic.}


\section*{Data Availability}
The data in Section \ref{sec:data} are publicly available on the Harvard Dataverse platform and can be accessed at \url{https://doi.org/10.7910/DVN/NSKFK2}.




\bibliographystyle{biom} 
\bibliography{references.bib}





\label{lastpage}

\end{document}


\def\spacingset#1{\renewcommand{\baselinestretch}%
{#1}\small\normalsize} \spacingset{1}

\date{\vspace{-5ex}}

\maketitle
\spacingset{1.5}
\setcounter{page}{1}
\appendix

In Section~\ref{sec: reg}, we provide the regularity conditions for our theoretical results. Section~\ref{sec: proofs} provides proofs to the theoretical results in the main paper. Sections~\ref{sec: add-simulation} and \ref{sec: add-application} provide additional results for simulation and data application. {R code for simulation and data analysis is available at \url{https://github.com/BingkaiWang/SW-CRT}.}


\section{Regularity conditions}\label{sec: reg}
For the estimating equation $\bpsi$ of the linear mixed models or the GEE, we assume the following regularity conditions.

(1) $E[\bpsi(\bO,\btheta)]=0$ has a unique solution, denoted as $\underline{\btheta}$, in the interior of a compact subset of the Euclidean space. 
\vspace{5pt}

(2) The function $\btheta \mapsto \bpsi({o}, \btheta)$, together with its first and second derivatives, is dominated by a square-integrable function for every ${o}$ in the support of $\bO$.
\vspace{5pt}

(3) $E\left[\frac{\partial \bpsi(\bO,\btheta)}{\partial \btheta} \big|_{\btheta = \underline\btheta}\right]$ exists and is nonsingular. 
\vspace{5pt}

These regularity conditions are standard assumptions in \cite{vaart_1998} to ensure the estimating function $\bpsi$ is well-behaved to prove the asymptotic results. Of note, Regularity condition (1) does not imply any component of the working model is correctly specified. Instead, it solely requires the uniqueness of maxima in maximum likelihood or maximum quasi-likelihood estimation. This can be achieved by carefully designing $\bpsi$ and restricting parameter space to rule out degenerative solutions. As a specific example, if $\bpsi$ is the estimating equation for linear regression (based on ordinary least squares), then regularity condition (1) is equivalent to the invertibility of the covariance matrix of covariates.

\section{Proofs}\label{sec: proofs}
\subsection{Introducing an equivalent definition of potential outcomes}

In the main paper, we let $Y_{ijk}(z)$ denote the potential outcome of individual $k$ in cluster $i$ during period $j$ had the cluster been first treated in period $z$ for $1\le z\le j$. In order to simplify the notation in the proof of our theoretical results in Section B.3-B.5, we introduce an alternative definition of potential outcomes, and show it is equivalent to the potential outcome defined in the main paper. 

Let $\widetilde{Y}_{ijk}(d)$ denote the potential outcome of individual $k$ in cluster $i$ during period $j$ had the cluster been treated for $d$ periods already for $1\le d \le j$. In addition, $\widetilde{Y}_{ijk}(0)$ denotes the untreated potential outcome.
 We next show these two potential outcome definitions, $\widetilde{Y}_{ijk}(d)$ and $Y_{ijk}(z)$ are mathematically equivalent and can be used interchangeably in proving the subsequent technical results. 
 First, $\widetilde{Y}_{ijk}(0)$ and $Y_{ijk}(0)$ both are potential outcomes in the absence of treatment in period $j$. When $d>0$, $\widetilde{Y}_{ijk}(d)$ refers to the individual potential outcome in period $j$ had the cluster been treated for $d$ periods, which is equivalent to the individual potential outcome in period $j$ had the treatment started in period $j-d+1$, i.e., $Y_{ijk}(j-d+1)$. That is, when $1\le z\le j$, $Y_{ijk}(z)$ and $\widetilde{Y}_{ijk}(j-z+1)$ both refer to the individual potential outcome in period $j$ had the treatment started in period $z$. Here, the underlying assumption is that the treatment is irreversible and one-directional, which automatically holds in a stepped wedge design.
With the above elaboration, we arrive at $Y_{ijk}(0) = \widetilde{Y}_{ijk}(0)$ and $Y_{ijk}(z)=\widetilde{Y}_{ijk}(j-z+1)$, which establishes a bijection between two potential outcome definitions.

Denoting $d = j-z+1$, the causal consistency assumption can be alternatively stated as 
        \begin{align*}
        Y_{ijk} &= \sum_{z=1}^j I\{Z_i=z\} Y_{ijk}(z) + I\{Z_i>j\} Y_{ijk}(0) \\
        &=\sum_{z=1}^j I\{Z_i=z\} \widetilde{Y}_{ijk}(j-z+1) + I\{Z_i>j\} \widetilde{Y}_{ijk}(0) \\
        &= \sum_{z=1}^j I\{Z_i=j-d+1\} \widetilde{Y}_{ijk}(d) + I\{Z_i>j\} \widetilde{Y}_{ijk}(0).
    \end{align*}
    In addition, Assumption 1 in the main paper remains the same under the alternative definition of potential outcomes because the set $\{Y_{ijk}(z):k=1,\dots, N_i, 1\le z\le j\le J\}$ is equal to $\{\widetilde{Y}_{ijk}(d):k=1,\dots, N_i, 1\le d\le j\le J\}$.
    

    In the proofs of our main results, we will use the potential outcome definition $\widetilde{Y}_{ijk}(d)$ in place of $Y_{ijk}(d)$. This is because our treatment effect estimand is defined on $Y_{ijk}(j-d+1)$, which is equal to $\widetilde{Y}_{ijk}(d)$. Therefore, using $\widetilde{Y}_{ijk}(d)$ can much simplify the notation. 
    Finally, we further simplify the notation from $\widetilde{Y}_{ijk}(d)$ to $Y_{ijk}(d)$ in our proofs, which improves readability without loss of clarity.

\subsection{Lemmas}
\begin{lemma}\label{lemma: asymptotics-theta}
Let $\bO_1, \ldots, \bO_m$ be i.i.d. samples from a common distribution on $\bO$. Let $\bpsi(\bO,\btheta)$ be a known estimating equation with parameters $\btheta\in \mathrm{\Theta}$, a compact set of the Euclidean space. 
Let $\widehat\btheta$ be the solution to $\sum_{i=1}^m \bpsi(\bO_i,\btheta) = \bzero$.
We assume that $\bpsi$ satisfies 
\begin{enumerate}
    \item There exists a unique solution in the interior of $\mathrm{\Theta}$, denoted as $\underline{\btheta}$, to the equation $E[\bpsi(\bO, \btheta)] =\bzero$.
    \item The function $\btheta \mapsto \bpsi({o}, \btheta)$, together with its first and second derivatives, is dominated by a square-integrable function for every ${o}$ in the support of $\bO$.
    \item $E\left[\frac{\partial \bpsi(\bO,\btheta)}{\partial \btheta} \big|_{\btheta = \underline\btheta}\right]$ is invertible.   
\end{enumerate}
Then we have $\widehat{\btheta} \xrightarrow{P} \underline{\btheta}$ and $m^{1/2}(\widehat{\btheta}  - \underline{\btheta}) \xrightarrow{d} N(0,\bfV)$, where $\bfV = E[IF(\bO,\underline\btheta)IF(\bO,\underline\btheta)^\top]$ and $IF(\bO,\underline\btheta) =- E\left[\frac{\partial \bpsi(\bO,\btheta)}{\partial \btheta} \big|_{\btheta = \underline\btheta}\right]^{-1}\bpsi(\bO,\underline\btheta)$ is the influence function for $\widehat\btheta$.
Furthermore, the sandwich variance estimator $m^{-1} \sum_{i=1}^m \widehat{IF}(\bO_i,\widehat{\btheta})\widehat{IF}(\bO_i,\widehat{\btheta})^\top$ converges in probability to $\bfV$, where $\widehat{IF}(\bO_i,\widehat{\btheta}) = \left\{m^{-1} \sum_{i=1}^m \frac{\partial \bpsi(\bO_i,\btheta)}{\partial \btheta} \big|_{\btheta = \widehat\btheta}\right\}^{-1}\bpsi(\bO_i, \widehat{\btheta})$.
\end{lemma}

\begin{proof}[Proof of Lemma 1]
By condition 2 for $\bpsi$, Example 19.8 of \cite{vaart_1998} implies that $\{\bpsi(\bO,\btheta): \btheta \in \mathrm{\Theta}\}$ is P-Glivenko-Cantelli. Then, with condition 1, Theorem 5.9 of \cite{vaart_1998} shows that $\widehat{\btheta} \xrightarrow{P} \underline{\btheta}$. Next, we apply Theorem 5.41 of \cite{vaart_1998} to obtain asymptotic normality, for which our assumptions on $\bpsi$ ensure all conditions needed in Theorem 5.41 are satisfied. Then we have
\begin{align*}
    m^{1/2}(\widehat{\btheta}  - \underline{\btheta})
    &= m^{-1/2}\sum_{i=1}^mIF(\bO_i,\underline\btheta) + o_p(1),
\end{align*}
which implies the desired asymptotic normality by the Central Limit Theorem.

We next prove the consistency of the sandwich variance estimator. First, we prove that $m^{-1} \sum_{i=1}^m \frac{\partial \bpsi(\bO_i,\btheta)}{\partial \btheta} \big|_{\btheta = \widehat\btheta} \xrightarrow{P} E\left[\frac{\partial \bpsi(\bO,\btheta)}{\partial \btheta} \big|_{\btheta = \underline\btheta} \right]$. Denoting $\dot{\bpsi}_{ij}(\widehat\btheta)$ as the transpose of the $j$th row of $\frac{\partial \bpsi(\bO_i,\btheta)}{\partial \btheta}\big|_{\btheta = \widehat\btheta}$, we apply the multivariate Taylor expansion to get
\begin{align*}
    m^{-1} \sum_{i=1}^m \dot{\bpsi}_{ij}(\widehat\btheta) -     m^{-1} \sum_{i=1}^m \dot{\bpsi}_{ij}(\underline\btheta) =  m^{-1} \sum_{i=1}^m \ddot{\bpsi}_{ij}(\widetilde\btheta) (\widehat\btheta-\underline\btheta)
\end{align*}
for some $\widetilde{\btheta}$ on the line segment between $\widehat{\btheta}$ and $\underline\btheta$ and $ \ddot{\bpsi}_{ij}$ being the derivative of $\dot{\bpsi}_{ij}$. By condition 2 and $\widehat{\btheta} \xrightarrow{P} \btheta$, we have $m^{-1} \sum_{i=1}^m \ddot{\bpsi}_{ij}(\widetilde\btheta) = O_p(1)$. As a result, $\widehat{\btheta} -\btheta =o_p(1)$ implies that  $ m^{-1} \sum_{i=1}^m \dot{\bpsi}_{ij}(\widehat\btheta) -     m^{-1} \sum_{i=1}^m \dot{\bpsi}_{ij}(\underline\btheta)  = o_p(1)$. Then the first step is completed by the fact that $m^{-1} \sum_{i=1}^m \dot{\bpsi}_{ij}(\underline\btheta) = E[\dot{\bpsi}_{ij}(\underline\btheta) ] + o_p(1)$, which results from the Law of Large Numbers and condition 2. Next, we prove $m^{-1} \sum_{i=1}^m \bpsi(\bO_i,\widehat{\btheta})\bpsi(\bO_i,\widehat{\btheta})^\top \xrightarrow{P} E[\bpsi(\bO,\underline{\btheta})\bpsi(\bO,\underline{\btheta})^\top]$ following a similar procedure to the first step. Letting $\bpsi_{ij}(\btheta)$ be the $j$th entry of $\bpsi(\bO_i,\btheta)$, we apply the multivariate Taylor expansion and get
\begin{align*}
    &m^{-1} \sum_{i=1}^m \bpsi_{ij}(\widehat\btheta)\bpsi(\bO_i,\widehat\btheta) -   m^{-1} \sum_{i=1}^m \bpsi_{ij}(\underline\btheta)\bpsi(\bO_i,\underline\btheta)  \\
    &=  m^{-1} \sum_{i=1}^m \left\{\bpsi(\bO_i,\widetilde\btheta)\dot{\bpsi}_{ij}(\widetilde\btheta)^\top + \bpsi_{ij}(\widetilde\btheta) \frac{\partial \bpsi(\bO_i,\btheta)}{\partial \btheta} \Big|_{\btheta = \widetilde\btheta} \right\} (\widehat\btheta-\underline\btheta) \\
    &= O_p(1)o_p(1),
\end{align*}
which, combined with the Law of Large Numbers on $m^{-1} \sum_{i=1}^m \bpsi_{ij}(\underline\btheta)\bpsi(\bO_i,\underline\btheta)$, implies the desired result in this step. Finally, by the Continuous Mapping Theorem, we have
\begin{align*}
    &m^{-1} \sum_{i=1}^m \widehat{IF}(\bO_i,\widehat{\btheta})\widehat{IF}(\bO_i,\widehat{\btheta})^\top\\
    &= \left\{m^{-1} \sum_{i=1}^m \frac{\partial \bpsi(\bO_i,\btheta)}{\partial \btheta} \big|_{\btheta = \widehat\btheta}\right\}^{-1}m^{-1} \sum_{i=1}^m \bpsi(\bO_i,\widehat\btheta)\bpsi(\bO_i,\widehat\btheta)  \left\{m^{-1} \sum_{i=1}^m \frac{\partial \bpsi(\bO_i,\btheta)}{\partial \btheta} \big|_{\btheta = \widehat\btheta}\right\}^{-1} \\
    &= \left\{E\left[\frac{\partial \bpsi(\bO,\btheta)}{\partial \btheta} \big|_{\btheta = \underline\btheta}\right]+o_p(1)\right\}^{-1}\left\{E[\bpsi(\bO,\underline\btheta)\bpsi(\bO,\underline\btheta)^\top]+o_p(1)\right\}\left\{E\left[\frac{\partial \bpsi(\bO,\btheta)}{\partial \btheta} \big|_{\btheta = \underline\btheta}\right]+o_p(1)\right\}^{-1} \\
    &= \bfV+o_p(1).
\end{align*}
\end{proof}

\subsection{Proof of Theorem 1}\label{sec:proof-LMM}
\begin{proof}[Proof of Theorem~1]
We first introduce a few definitions. Recall that, in Section B.1, we define $Y_{ijk}(d)$ as the potential outcome of individual $k$ in cluster $i$ during period $j$ had the cluster been treated for $d$ periods already for $1\le d \le j$. In addition, $Y_{ijk}(0)$ denotes the untreated potential outcome.
Let
\begin{align*}
    \bY_{ij} &= (Y_{ij1}, \dots, Y_{ijN_i})^\top \in \mathbb{R}^{N_i},\\
    \bY_{ij}(d) &= (Y_{ij1}(d), \dots, Y_{ijN_i}(d))^\top \in \mathbb{R}^{N_i},\\
    \bY_i &= (\bY_{i1}^\top, \dots, \bY_{iJ}^\top)^\top \in \mathbb{R}^{JN_i},\\
    \bY_i(d) &= (\bY_{i1}(d)^\top, \dots, \bY_{iJ}(d)^\top)^\top \in \mathbb{R}^{JN_i},\\
    \Lambda_{Z_i}^d &= \left(\begin{array}{cccccc}
    0   \\
    &   \ddots   \\
   & & 0\\
    &   &  &  I\{Z_i = 1\} & & \\
    &   &  & & \ddots & \\
    &   &  & & & I\{Z_i = J-d+1\}
    \end{array}\right) \in \mathbb{R}^{J\times J}\textrm{ for } d = 1,\dots, J,\\
    \Delta_{Z_i} &= diag\{I\{Z_i\le j\}:j=1,\dots, J\} \in \mathbb{R}^{J\times J},
\end{align*}
$\bfI_{q} \in \mathbb{R}^{q\times q}$ be the identity matrix for any positive integer $q$, and $\otimes$ be the Kronecker product operator. 
We have $\Delta_{Z_i}  = \sum_{d=1}^J \Lambda_{Z_i}^d$.
Of note, $\bY_{ij}(d)$ for $d>j$ is not identifiable with the observed data. Here we still define  them for notation convenience, but these quantities will not be evaluated throughout the proof. An alternative approach is to simply define $\bY_{ij}(d) =\bzero$ (or an arbitrary constant quantity) for $d > j$. With the above definitions, we observe that 
\begin{align*}
    \bY_{i1} &= I\{Z_i> 1\} \bY_{i1}(0) + I\{Z_i=1\}\bY_{i1}(1) \\
    \bY_{i2} &= I\{Z_i> 2\} \bY_{i2}(0) + I\{Z_i=2\}\bY_{i2}(1) + I\{Z_i=1\}\bY_{i2}(2) \\
    &\vdots \\
    \bY_{iJ} &= I\{Z_i> J\} \bY_{iJ}(0) + I\{Z_i=J\}\bY_{iJ}(1) + I\{Z_i=J-1\}\bY_{iJ}(2) +\dots + I\{Z_i=1\}\bY_{iJ}(J).  
\end{align*}
Then
\begin{align*}
    \bY_i &= \sum_{d=1}^J  (\Lambda_{Z_i}^d \otimes \bfI_N) \bY_i(d) + \{(\bfI_J - \Delta_{Z_i}) \otimes \bfI_{N_i}\} \bY_i(0) \\
    &= \sum_{d=1}^J (\Lambda_{Z_i}^d \otimes \bfI_N) \{\bY_i(d) - \bY_i(0)\} + \{\bfI_J \otimes \bfI_{N_i}\} \bY_i(0). \numberthis \label{eq:Y}
\end{align*}
For the working linear mixed model, we focus on the case of nested exchangeable correlation structure, since the working independence assumption (ordinary least squares estimation) or the exchangeable correlation structure is just a special case with some variance parameters set to zero.

Furthermore, for notation convenience, we omit the subscript $i$ when taking expectation with respect to distribution $\mathcal{P}$. For example, $E[f(\bO_i)]$ is simplified as $E[f(\bO)]$, where $\bO_i$ is the observed data of cluster $i$ and $f$ is an arbitrary measurable function. This simplification is justified by Assumption A1, which assumes the complete data of each cluster are independent and identically distributed. Likewise, $(Y_{ijk}, S_{ijk}, N_{ij})$ are denoted as $(Y_{.jk}, S_{.jk}, N_{.j})$ when taking expectations. This simplification is used throughout the proof for all notation.

We start by proving the results for the constant treatment effect $\Delta$.  The linear mixed model can be re-written as
\begin{align*}
    \bY_i = (\bfI_J\otimes \bone_{N_i})\bbeta_0  +  (\Delta_{Z_i}\bone_J) \otimes \bone_{N_i} \beta_Z + \bone_{J} \otimes \bfX_i\bbeta_{\bX} + \alpha_i \bone_{N_iJ} + \boldsymbol\gamma_i \otimes \bone_{N_i} + \boldsymbol\varepsilon_i,
\end{align*}
where $\bbeta_0 = (\beta_{01},\dots, \beta_{0J})$, $\bone_q$ is a $q$-dimensional vector of ones, $\bfX_i = (\bX_{i1}, \dots \bX_{iN_i})^\top \in \mathbb{R}^{N_i \times p}$, $\boldsymbol\gamma_i = (\gamma_{i1}, \dots, \gamma_{iJ})^\top$, and $\boldsymbol\varepsilon_i = (\varepsilon_{i11}, \dots, \varepsilon_{i1N_i}, \dots, \varepsilon_{iJ1}, \dots, \varepsilon_{iJN_i})^\top$. Then the working model becomes $$\bY_i | Z_i,\bfX_i, N_i \sim N(\bfQ_i\bbeta, \bfSigma_i),$$
where $\bfQ_i = (\bfI_J\otimes \bone_{N_i}, (\Delta_{Z_i}\bone_J) \otimes \bone_{N_i}, \bone_{J} \otimes \bfX_i) \in \mathbb{R}^{N_iJ\times (J+1+p)}$, $\bbeta = (\bbeta_0, \beta_Z, \bbeta_{\bX})$, and $  \bfSigma_i = \tau^2 \bone_{N_iJ}\bone_{N_iJ}^\top + \kappa^2 \bfI_J \otimes \bone_{N_i}\bone_{N_i}^\top + \sigma^2 \bfI_{N_iJ}$.

Let $\bY_{ij}^o = \{Y_{ijk}: S_{ijk}=1\}$ be the observed outcome vector for cluster $i$ in period $j$ and  $\mathbf{D}_{ij} \in \mathbb{R}^{N_i\times N_{ij}}$ be a matrix such that $\bY_{ij}^o = \mathbf{D}_{ij}^\top \bY_{ij}$. Specifically, the $k$-th row of $\mathbf{D}_{ij}$ is a binary vector with the $\widetilde{k}$-th entry 1 and the rest 0, where $\widetilde{k}$ is the $k$-th smallest index such that $S_{ijk} =1$; that is, $\mathbf{D}_{ij}$ is a deterministic function of $(S_{ij1}, \dots, S_{ijN_i}, N_i)$. Letting $M_i = \sum_{j=1}^J N_{ij}$, $\bS_i= (S_{i11}, \dots, S_{iJN_i})$, and $\bfD_i = bdiag\{\bfD_{ij}:j=1,\dots, J\} \in \mathbb{R}^{N_iJ\times M_i}$ denoting the block diagonal matrix of $\bfD_{ij}$, and $\bY_i^o = (\bY_{i1}^o, \dots, \bY_{iJ}^o)$, we have $\bY_i^o = \bfD_i^\top \bY_i$, $\mathbf{D}_{i}^\top \bone_{N_iJ} = \bone_{M_i}$, $\mathbf{D}_{i}\bone_{M_i} = \bS_i$, $ \bfD_i\bfD_i^\top  = diag\{\bS_i\}$, and  $\bfD_i^\top \bfD_i = \bfI_{M_i}$.
Then the observed outcome follows $\bY_i^o|Z_i,\bfX_i^o, N_i, \bS_i \sim N(\bfD_i^\top \bfQ_i \bbeta, \bfD_i^\top \bfSigma_i \bfD_i)$, where $\bfX_i^o = \bfD_i^\top (\bone_{J} \otimes \bfX_i)$ is the observed covariate matrix for cluster $i$ across periods.

Denote $\btheta = (\bbeta, \tau^2, \kappa^2,\sigma^2)$ as the vector of unknown parameters. Based on the observed data, the log-likelihood function given $\{Z_i, \bfX_i^o, N_i, \bS_i\}$ is 
\begin{align*}
    &l(\btheta; \{\bY_i^o\}_{i=1}^I|\{Z_i, \bfX_i^o, N_i, \bS_i\}_{i=1}^I)\\
    &=  C -\frac{1}{2}\sum_{i=1}^I\left\{\log(|\mathbf{D}_i^\top\bfSigma_i \mathbf{D}_i|) + (\bY_i^o - \mathbf{D}_i^\top\bfQ_i\bbeta)^\top  (\mathbf{D}_i^\top\bfSigma_i \mathbf{D}_i)^{-1} (\bY_i^o - \mathbf{D}_i^\top\bfQ_i\bbeta) \right\} \\
    &=  C -\frac{1}{2}\sum_{i=1}^I\left\{\log(|\mathbf{D}_i^\top\bfSigma_i \mathbf{D}_{i}|) + (\bY_i - \bfQ_i\bbeta)^\top \mathbf{D}_i (\mathbf{D}_i^\top\bfSigma_i \mathbf{D}_i)^{-1}\mathbf{D}_i^\top (\bY_i - \bfQ_i\bbeta) \right\} 
\end{align*}
where $C$ is a constant independent of the parameters $\btheta$. 
 The derivative of the log-likelihood function is then
\begin{align*}
   & \frac{\partial l(\btheta; \{\bY_i\}_{i=1}^I|\{Z_i, \bfX_i^o, N_i, \bS_i\}_{i=1}^I)}{\partial \btheta}\\
   &= -\sum_{i=1}^I \left(\begin{array}{c}
    2\bfQ_i^{\top}\bfV_i(\bY_i - \bfQ_i\bbeta)  \\
    -tr(\bfV_i) + (\bY_i - \bfQ_i\bbeta)^\top \bfV_i^2(\bY_i - \bfQ_i\bbeta)\\
  -\bone_{N_i}^\top\bfV_i\bone_{N_i} + (\bY_i - \bfQ_i\bbeta)^\top \bfV_i\bone_{N_iJ}\bone_{N_iJ}^\top \bfV_i(\bY_i - \bfQ_i\bbeta)    \\
  -tr(\bfV_i \bfI_J \otimes \bone_{N_i}\bone_{N_i}^\top) + (\bY_i - \bfQ_i\bbeta)^\top \bfV_i(\bfI_J \otimes \bone_{N_i}\bone_{N_i}^\top)\bfV_i(\bY_i - \bfQ_i\bbeta)
   \end{array}\right),
\end{align*}
where $\bfV_i = \mathbf{D}_i (\mathbf{D}_i^\top\bfSigma_i \mathbf{D}_i)^{-1} \mathbf{D}_i^\top \in \mathbb{R}^{N_iJ\times N_iJ}$ and $tr(\bfV_i)$ is the trace of $\bfV_i$. 
We hence define the estimating function as
\begin{equation}\label{eq:psi}
  \bpsi(\bO; \btheta) =   \left(\begin{array}{c}
    2\bfQ^{\top}\bfV(\bY - \bfQ\bbeta)  \\
    -tr(\bfV) + (\bY - \bfQ\bbeta)^\top \bfV^2(\bY - \bfQ\bbeta)\\
  -\bone_N^\top\bfV\bone_N + (\bY - \bfQ\bbeta)^\top \bfV\bone_N\bone_N^\top \bfV(\bY - \bfQ\bbeta)  \\
    -tr(\bfV \bfI_J \otimes \bone_{N}\bone_{N}^\top) + (\bY - \bfQ\bbeta)^\top \bfV(\bfI_J \otimes \bone_{N}\bone_{N}^\top)\bfV(\bY - \bfQ\bbeta)
   \end{array}\right).
\end{equation}
The maximum likelihood estimator for $\btheta$ is defined as a solution to the estimating equation
$$\sum_{i=1}^n \bpsi(\bO_i;\btheta)=\bzero.$$ 
Recall that we focus on the nested exchangeable correlation structure, which includes two additional special cases. That is, for the independence working correlation structure, $\bpsi$ is modified by setting $\tau^2=\kappa^2=0$. For the exchangeable working correlation structure, $\bpsi$ is modified by setting $\kappa^2=0$.

For this estimating equation $\bpsi$, we prove the convergence and asymptotic normality of $\widehat\btheta$ by applying Lemma 1. 
In Lemma 1, the conditions for $\bpsi$ are assumed in the main paper as regularity conditions, which implies the desired results. In addition, the consistency of sandwich variance estimators is also implied. 

Denoting $\underline\btheta = (\underline\bbeta_0, \underline\beta_Z, \underline\bbeta_{\bX}^{\top}, \underline\sigma^2, \underline\tau^2, \underline\kappa^2)$ as the solution to $E[\bpsi(\bO;\btheta)] = \bzero$, we next prove $\underline\beta_Z = \Delta$, which will imply the robustness of $\widehat{\beta}_Z$ and completes the proof under the constant treatment effect structure specification. To proceed, the first two entries of $E[\bpsi(\bO;\btheta)] = \bzero$ are 
\begin{align}
   & E[(\bfI_J\otimes \bone_{N})^\top \underline\bfV (\bY - \bfQ\underline\bbeta)] = \bzero, \label{eq: est-1} \\
   & E[\{(\Delta_{Z}\bone_J) \otimes \bone_{N}\}^\top \underline\bfV (\bY - \bfQ\underline\bbeta)] = 0, \label{eq: est-2} 
\end{align}
where $\underline\bfV$ is equal to $\bfV$ with $(\sigma^2,\tau^2,\kappa^2)$ replaced by $(\underline\sigma^2, \underline\tau^2, \underline\kappa^2)$.
Define $\bdelta = E[\Delta_{Z}\bone_J] = (\pi_1^s,\dots, \pi_J^s)^\top$, where $\pi_j^s = \sum_{j'=1}^j \pi_{j'}$. By left-multiplying Equation~(\ref{eq: est-1}) by $-\delta^\top$ and adding it to Equation~(\ref{eq: est-2}), we get
\begin{align*}
    E[\{(\Delta_{Z}\bone_J - \bdelta) \otimes \bone_{N}\}^\top \underline\bfV (\bY - \bfQ\underline\bbeta)] = 0.
\end{align*}
Since algebra shows that 
\begin{align*}
    \underline\bfV = diag\left\{\frac{1}{\underline\sigma^2} \bfD_j\bfD_j^\top - \frac{\underline\kappa^2}{\underline\sigma^2(\underline\sigma^2+ N_{.j}\underline\kappa^2)}\bS_j\bS_j^\top: j = 1,\dots, J\right\} - \left(\frac{1}{\underline\tau^2}+\sum_{j=1}^J\frac{N_{.j}}{\underline\sigma^2 + N_{.j} \underline\kappa^2}\right)^{-1} \boldsymbol{q}\boldsymbol{q}^\top,
\end{align*}
where $\boldsymbol{q} =  \left(\frac{1}{\underline\sigma^2 + N_{.1}\underline\kappa^2} \bS_1^\top, \dots, \frac{1}{\underline\sigma^2 + N_{.J}\underline\kappa^2} \bS_J^\top \right)^\top$, then Assumption A3 implies that $\underline\bfV$ is independent of $\Delta_Z$, which implies that $E[\{(\Delta_{Z}\bone_J - \bdelta) \otimes \bone_{N}\}^\top \underline\bfV(\bfI_J\otimes \bone_{N})] = \{E[\Delta_{Z}\bone_J - \bdelta] \otimes \bone_{N}\}^\top E[\underline{\bfV}(\bfI_J\otimes \bone_{N})] = \bzero$ and, similarly, by Assumptions A2 and A3, $E[\{(\Delta_{Z}\bone_J - \bdelta) \otimes \bone_{N}\}^\top \underline\bfV(\bone_J\otimes \bfX)] = \bzero$ and $E[\{(\Delta_{Z}\bone_J - \bdelta) \otimes \bone_{N}\}^\top \underline\bfV(\bfI_J\otimes \bfI_N)\bY(0)] = \bzero$. Therefore, using formula~(\ref{eq:Y}) and the defintion of $\bfQ$, we get
\begin{align*}
    \underline{\beta}_Z &= \frac{E[\{(\Delta_{Z}\bone_J - \bdelta) \otimes \bone_{N}\}^\top \underline\bfV \bY]}{E[\{(\Delta_{Z}\bone_J - \bdelta) \otimes \bone_{N}\}^\top \underline\bfV (\Delta_{Z}\bone_J \otimes \bone_{N})]} \\
    &= \sum_{d=1}^J\frac{E[\{(\Delta_{Z}\bone_J - \bdelta) \otimes \bone_{N}\}^\top \underline\bfV (\Lambda_{Z}^d \otimes \bfI_N) \{\bY(d) - \bY(0)\}]}{E[\{(\Delta_{Z}\bone_J - \bdelta) \otimes \bone_{N}\}^\top \underline\bfV (\Delta_{Z}\bone_J \otimes \bone_{N})]}.
\end{align*}
Since $\underline\bfV$ is a function of $\bS$ and $N$, by Assumptions A2 and A3, $E[\bY(d) - \bY(0)|\bS, N, Z] = E[\bY(d) - \bY(0)|N]$. Since Assumption A1 implies that $E[\bY_{ij}(d)|N] = v_j(d,N)\bone_N$ for $v_j(d,N) = E[Y_{.jk}(d)|N]$, we get $E[\bY(d) - \bY(0)|N] = \{\bv(d,N)-\bv(0,N) \}\otimes \bone_N$, where $\bv(d,N) = (v_1(d,N), \dots, v_J(d,N))$. Therefore, 
\begin{align*}
    \underline{\beta}_Z &= \sum_{d=1}^J\frac{E[\{(\Delta_{Z}\bone_J - \bdelta) \otimes \bone_{N}\}^\top \underline\bfV (\Lambda_{Z}^d \otimes \bone_N) \{\bv(d,N)-\bv(0,N) \}]}{E[\{(\Delta_{Z}\bone_J - \bdelta) \otimes \bone_{N}\}^\top \underline\bfV (\Delta_{Z}\bone_J \otimes \bone_{N})]}
\end{align*}
To further simplify the above formula, we use the definition of $\underline\bfV$ and compute
\begin{align*}
    &\{(\Delta_{Z}\bone_J - \delta) \otimes \bone_{N}\}^\top \underline\bfV \{\Lambda_{Z}^d \otimes \bone_N\} \\
    &= \left((I\{Z\le 1\} - \pi_1^s) \bone_N^\top, \dots, (I\{Z\le J\} - \pi_J^s) \bone_N^\top\right) \underline\bfV\{\Lambda_{Z}^d \otimes \bone_N\}\\
    &= \bU_d^\top,
\end{align*}
where the $j$-th entry of $\bU_d$ is 
$$\frac{N_{.j}\lambda_j^d }{\underline\sigma^2+N_{.j} \underline\kappa^2}\left[(I\{Z\le j\}-\pi_j^s) -   \left(\frac{1}{\underline\tau^2}+\sum_{j=1}^J\frac{N_{.j'}}{\underline\sigma^2 + N_{.j'} \underline\kappa^2}\right)^{-1}\sum_{j'=1}^J \frac{N_{\cdot j'}}{\underline\sigma^2+N_{.j'} \underline\kappa^2} (I\{Z\le j'\} - \pi_{j'}^s)\right]$$ with $\lambda_j^d$ being the $j$-th diagonal entry of $\Lambda_Z^d$. 
From the above formula of $\bU_d$, we observe that it is a  a function of $Z$ and $N_{.j}$, then Assumption A3 implies that $\bU_d$ is independent of $N$ and hence
\begin{align*}
      \underline{\beta}_Z = \frac{\sum_{d=1}^J E[\bU_d^\top \{\bv(d,N)-\bv(0,N) \}]}{\sum_{d=1}^JE[\bU_d^\top\bone_J]} = \frac{\sum_{d=1}^J E[\bU_d^\top]}{\sum_{d=1}^J E[\bU_d^\top\bone_J]} E[\bv(d,N)-\bv(0,N)].
\end{align*}
Assuming a duration-invariant and period-invariant treatment effect, we have $E[v_j(d,N)-\bv_j(0,N)] = \Delta$ for all $d\le j$, which implies that the $d$-th to the $J$-th entries of $E[\bv(d,N)-\bv(0,N)]$ are a constant $\Delta$. From the expression of $\bU_d$, we also observe that its first to $(d-1)$-th entry is zero since $\lambda_j^d=0$ for $d>j$. Therefore, $E[\bU_d^\top]E[\bv(d,N)-\bv(0,N)] = \Delta$ for each $d$, and we get $\beta_Z = \Delta$, which completes the proof for consistency.

Under the duration-specific treatment effect setting, the linear mixed model becomes
\begin{align*}
    \bY_i = (\bfI_J\otimes \bone_{N_i})\bbeta_0  +  \mathbf{H}_{Z_i} \otimes \bone_{N_i} \bbeta^D_Z + \bone_{J} \otimes \bfX_i\beta_{\bX} + \alpha_i \bone_{N_iJ} + \boldsymbol{\gamma}_i \otimes \bone_N + \boldsymbol\varepsilon_i,
\end{align*}
where
\begin{align*}
 \mathbf{H}_{Z_i} = \left(\begin{array}{cccc}
    I\{Z_i=1\}  &  \\
    I\{Z_i=2\}  &  I\{Z_i=1\} \\
    \vdots & & \ddots \\
    I\{Z_i=J\} &  I\{Z_i=J-1\} &  \dots& I\{Z_i = 1\}
 \end{array}\right).
\end{align*}
Notice that the estimating equations under this setting are the same as those under the constant treatment effect setting, except that $\bfQ = (\bfI_J\otimes \bone_{N}, \mathbf{H}_Z \otimes \bone_{N}, \bone_{J} \otimes \bfX)$ and $\bbeta = (\bbeta_0, \bbeta_Z^D, \bbeta_{\bX})$. Then the asymptotic normality of $\widehat{\bbeta}_Z^D$ and consistency of variance estimators can be proved similarly using Lemma 1. Hence, the proof for this setting is completed if we can prove $\underline{\bbeta}_Z^D = \bDelta^D$. To this end, we observe that the first two components of $E[\bpsi(\bO;\underline\btheta)] = \bzero$ imply that 
\begin{align}
   & E[(\bfI_J\otimes \bone_{N})^\top \underline\bfV (\bY - \bfQ\underline\bbeta)] = \bzero, \label{eq: est-3} \\
   & E[(\mathbf{H}_{Z}\otimes \bone_{N})^\top \underline\bfV (\bY - \bfQ\underline\bbeta)] = \bzero.\label{eq: est-4} 
\end{align}
By left-multiplying Equation~(\ref{eq: est-3}) by $-E[\mathbf{H}_Z]^\top$ and adding it to Equation~(\ref{eq: est-4}), we get
\begin{align*}
    E[\{(\mathbf{H}_Z - E[\mathbf{H}_{Z}]) \otimes \bone_{N}\}^\top \underline\bfV (\bY - \bfQ\underline\bbeta)] = 0.
\end{align*}
By Assumptions A2 and A3, we have $E[\{(\mathbf{H}_{Z} - E[\mathbf{H}_{Z}]) \otimes \bone_{N}\}^\top \underline\bfV(\bfI_J\otimes \bone_{N})] = E[\{(\mathbf{H}_{Z} - E[\mathbf{H}_{Z}]) \otimes \bone_{N}\}^\top \underline\bfV(\bfI_J\otimes \bfX)] = E[\{(\mathbf{H}_{Z} - E[\mathbf{H}_{Z}]) \otimes \bone_{N}\}^\top \underline\bfV(\bfI_J\otimes \bfI_{N})\bY(0)] = \bzero$. Then, using the formula~(\ref{eq:Y}) and the form of $\bfQ$, we have
\begin{align*}
    \underline\bbeta^D_Z &= E[\{(\mathbf{H}_{Z} - E[\mathbf{H}_{Z}]) \otimes \bone_{N}\}^\top \underline\bfV(\mathbf{H}_{Z} \otimes \bone_{N})]^{-1} \\
    &\quad \sum_{d=1}^J E[\{(\mathbf{H}_{Z} - E[\mathbf{H}_{Z}]) \otimes \bone_{N}\}^\top \underline\bfV(\Lambda_{Z}^d \otimes \bone_N)] E[\bv(d,N)-\bv(0,N)].
\end{align*}
Under the duration-specific treatment effect setting, $E[v_j(d,N)-v_j(0,N)] = \Delta(d)$ for all $d\le j$, which implies that the $d$-th to the $J$-th entries of $E[\bv(d,N)-\bv(0,N)]$ are a constant $\Delta(d)$. Since the first $d-1$ rows of $\Lambda_{Z}^d$ are all zero, then $\Lambda_{Z}^dE[\bv(d,N)-\bv(0,N)] = \Delta(d)\Lambda_{Z}^d \bone_J$. Using the fact that $\sum_{d=1}^J \Lambda_{Z}^d \Delta(d)\bone_J = \mathbf{H}_Z \bDelta^{D}$, we arrive at the final result: $ \underline\bbeta^D_Z =  \bDelta^{D}$. 

For the period-specific treatment effect setting, we drop the data from period $J$ to avoid over-parameterization. To simplify the switch from $J$ periods to $J-1$ periods, we use the superscript ${}^*$ to denote all subsequent changes in notation. For example, $J^*=J-1$, $\bY_i^*=(\bY_{i1},\dots,\bY_{i,J-1})$, $\bbeta_0^* = (\beta_{01},\dots, \beta_{0,J-1})$, and $\mathbf{M}^*$ is a $J^*$-by-$J^*$ matrix consisting of the first $J^*$ columns and rows of a matrix $\mathbf{M} \in \mathbb{R}^{J\times J}$. 
As a result, the linear mixed model becomes
\begin{align*}
    \bY_i^* = (\bfI_{J^*}\otimes \bone_{N_i})\bbeta_0^*  +  \Delta_{Z_i}^* \otimes \bone_{N_i} \bbeta_Z^P + \bone_{J^*} \otimes \bfX_i\bbeta_{\bX} + \alpha_i \bone_{N_iJ^*} + \boldsymbol\gamma_i^* \otimes \bone_{N_i} + \boldsymbol\varepsilon_i^*, 
\end{align*}
where $\bbeta_Z^P = (\beta_{1Z}, \dots, \beta_{J^*Z})$ is the treatment effect coefficient vector.
Notice that the estimating equations under this setting remain the same as those under the constant treatment effect setting, except that $\bfQ^* = (\bfI_{J^*}\otimes \bone_{N}, \Delta_{Z}^* \otimes \bone_{N}, \bone_{J^*} \otimes \bfX)$ and $\bbeta^* = (\bbeta_0^*, \bbeta_Z^P, \bbeta_{\bX})$. Then the asymptotic normality of $\widehat{\bbeta}_Z^P$ and consistency of variance estimators can be proved similarly using Lemma 1. Hence, the proof for this setting is completed if we can prove $\underline{\bbeta}_Z^P = \bDelta^P$. To this end, we observe that the first two components of $E[\bpsi(\bO^*;\underline\btheta)] = \bzero$ imply that 
\begin{align*}
   & E[(\bfI_J^*\otimes \bone_{N})^\top \underline\bfV^* (\bY^* - \bfQ^*\underline\bbeta^*)] = \bzero,  \\
   & E[(\Delta_{Z}^* \otimes \bone_{N})^\top \underline\bfV^* (\bY^* - \bfQ^*\underline\bbeta^*)] = \bzero
\end{align*}
which implies $E[\{(\Delta_{Z}^* - E[\Delta_{Z}^*]) \otimes \bone_{N}\}^\top \underline\bfV^* (\bY^* - \bfQ^*\underline\bbeta^*)] = 0$ and, following the previous proofs, we have
\begin{align*}
    \underline\bbeta^P_Z &= E[\{(\Delta_{Z}^* - E[\Delta_{Z}^*]) \otimes \bone_{N}\}^\top \underline\bfV^*(\Delta_{Z}^* \otimes \bone_{N})]^{-1} \\
    &\quad \sum_{d=1}^{J^*} E[\{(\Delta_{Z}^* - E[\Delta_{Z}^*]) \otimes \bone_{N}\}^\top \underline\bfV^*(\Lambda_{Z}^d{}^* \otimes \bone_N)] E[\bv^*(d,N)-\bv^*(0,N)],
\end{align*}
where the $j$-th entry of $\bv^*(d,N)$ is $E[Y_{.jk}(d)|N]$.
If the treatment effect is duration invariant, then $E[\bv(d,N)-\bv(0,N)] = \bDelta^P$. Using the fact that $\Delta_Z^* = \sum_{d=1}^{J^*} \Lambda_Z^d{}^*$, we have $\underline\bbeta^P_Z = \bDelta^P$. 

Finally, under the saturated treatment effect setting, we also drop the data from period $J$ to avoid over-parameterization. Similar to the period-specific treatment effects, we use the superscript ${}^*$ to denote all subsequent changes in notation. Then the linear mixed model becomes
\begin{align*}
    \bY_i^* = (\bfI_{J^*}\otimes \bone_{N_i})\bbeta_0^*  +  \sum_{d=1}^{J^*}\widetilde{\Lambda}_{Z_i}^d{}^* \otimes \bone_{N_i} \bbeta_{d}^S + \bone_{J^*} \otimes \bfX_i\beta_{\bX} + \alpha_i \bone_{N_iJ^*} + \boldsymbol\gamma_i^* \otimes \bone_{N_i} + \boldsymbol\varepsilon_i^*, 
\end{align*}
where $\widetilde{\Lambda}_{Z_i}^d{}^* \in \mathbb{R}^{J^* \times (J^*-d+1)}$ is the $d$-th to the $J^*$-th column of ${\Lambda}_{Z_i}^d{}^*$ and $\bbeta_{d}^S = (\beta_{dZd}, \dots, \beta_{J^*Zd}) \in R^{J^*-d+1}$. Notice that the estimating equations under this setting remain the same as those under the period-specific treatment effect setting, except that $\bfQ^* = (\bfI_{J^*}\otimes \bone_{N}, \widetilde{\Lambda}_{Z_i}^1{}^* \otimes \bone_{N}, \dots, \widetilde{\Lambda}_{Z_i}^{J^*}{}^* \otimes \bone_{N}, \bone_{J^*} \otimes \bfX)$ and $\bbeta^* = (\bbeta_0, \bbeta_1^S, \dots, \bbeta_{J^*}^S, \bbeta_{\bX})$. 
Furthermore, we highlight that columns of $\bfQ^*$ are not co-linear, i.e., there exists no non-zero vector $\boldsymbol{c}$ satisfying $\bfQ^* \boldsymbol{c} = \bzero$.  Hence, the model does not suffer from over-parameterization; and the regularity condition (1) is not violated. Then using Lemma 1, the asymptotic normality of $\widehat{\bbeta}_1^S,\dots, \widehat{\bbeta}_{J^*}^S$ and consistency of variance estimators can be proved. Hence, the proof for this setting is completed if we can prove $\underline{\beta}_{jZd} = \Delta_j(d)$ for $d\le j$. 
To this end, we observe that the first $J^*+1$ components of $E[\bpsi(\bO^*;\underline\btheta)] = \bzero$ are, for each $d = 1,\dots, J^*$,
\begin{align*}
   & E[(\bfI_{J^*}\otimes \bone_{N})^\top \underline\bfV^* (\bY^* - \bfQ^*\underline\bbeta^*)] = \bzero, \numberthis\label{eq:ee-s} \\
   & E[(\widetilde\Lambda_{Z}^d{}^* \otimes \bone_{N})^\top \underline\bfV^* (\bY^* - \bfQ^*\underline\bbeta^*)] = \bzero,
\end{align*}
which implies, for $d =1,\dots, J^*$, $E[(\widetilde\Lambda_{Z}^d{}^*- E[\widetilde\Lambda_{Z}^d{}^*] \otimes \bone_{N})^\top \underline\bfV^* (\bY^* - \bfQ^*\underline\bbeta^*)] = \bzero$. 
Since $E[(\widetilde\Lambda_{Z}^d{}^*- E[\widetilde\Lambda_{Z}^d{}^*] \otimes \bone_{N})^\top \underline\bfV^* (\bfI_{J^*} \otimes \bfI_N)\bY^*(0)]=\bzero$ and $E[(\widetilde\Lambda_{Z}^d{}^*- E[\widetilde\Lambda_{Z}^d{}^*] \otimes \bone_{N})^\top \underline\bfV^* (\bfI_{J^*} \otimes \bfX)]=\bzero$, formula~(\ref{eq:Y}) (with period $J$ dropped) and the expression of $\bfQ^*$ implies 
\begin{align*}
    \sum_{d'=1}^{J^*} E[\{(\widetilde\Lambda_{Z}^d{}^* -E[\widetilde\Lambda_{Z}^d{}^*]) \otimes \bone_{N}\}^\top \underline\bfV^* (\{\Lambda_{Z}^{d'}{}^*E[\bv^*(d',N)-\bv^*(0,N)] - \widetilde\Lambda_{Z}^{d'}{}^*\}\underline\bbeta_{d'}^S \otimes \bone_N)] = \bzero.
\end{align*}
We further denote $\widetilde\bv^*(d,N)$ as the $d$-th to the $J^*$-th entries of $\bv^*(d,N)$ for $d > 0$, and $\widetilde\bv^{d*}(0,N)$ as the $d$-th to the $J^*$-th entries of $\bv^*(0,N)$, then $\Lambda_{Z}^{d'}{}^*E[\bv^*(d',N)-\bv^*(0,N)] = \widetilde\Lambda_{Z}^{d'*} E[\widetilde\bv^*(d',N)-\widetilde\bv^{d*}(0,N)]$, which implies, for $d = 1, \dots, J^*$
\begin{align*}
    \sum_{d'=1}^{J^*} E[\{(\widetilde\Lambda_{Z}^d{}^* -E[\widetilde\Lambda_{Z}^d{}^*]) \otimes \bone_{N}\}^\top \underline\bfV^* (\widetilde\Lambda_{Z}^{d'*} \otimes \bone_N)]E[\widetilde\bv{}^*(d',N)-\widetilde\bv^{d*}(0,N) - \underline\bbeta_{d'}^S ] = \bzero.
\end{align*}
This linear system has $J^*(J^*+1)/2$ equations and $J^*(J^*+1)/2$ parameters in $\underline{\bbeta}_d^S$. 
For this linear system with parameters $\underline\bbeta_{d}^S$, a straightforward solution is $\underline\bbeta_{d}^S = E[\widetilde\bv^*(d,N)-\bv^{d*}(0,N)] = (\Delta_d(d),\dots, \Delta_{J^*}(d))$. By the regularity condition that the solution is unique, we have the desired consistency result. In fact, the regularity condition is unnecessary here since the design matrix $E[\{(\widetilde\Lambda_{Z}^* -E[\widetilde\Lambda_{Z}^*]) \otimes \bone_{N}\}^\top \underline\bfV^* (\widetilde\Lambda_{Z}^* \otimes \bone_N)]$ is already invertible (which has a form of a covariance matrix and does not degenerate due to Assumption A2), where $\widetilde\Lambda_{Z}^* = (\widetilde\Lambda_{Z}^1{}^*, \widetilde\Lambda_{Z}^2{}^*,\dots, \widetilde\Lambda_{Z}^{J^*}{}^*) \in\bfR^{J^*\times J^*(J^*+1)/2} $.
\end{proof}

\subsection{Proof of Theorem 2}
\begin{lemma}
Under any of the conditions (I)-(III) in Theorems 2, $\bfZ_i^{-1/2} \bfR_i^{-1} \bfZ_i^{-1/2} = \bfZ_i^{-1} \bfR_i^{-1}$.
\end{lemma}
\begin{proof}[Proof of Lemma 2]
It suffices to show $\bfR_i^{-1} \bfZ_i^{-1/2} = \bfZ_i^{-1/2}\bfR_i^{-1}$.
For condition (I) that $\rho_1 = \rho_2 = 0$, we have $\bfR_i = \bfI_{M_i}$, which directly implies the desired result. For condition (II) that $v(Y_{ijk}) \equiv \sigma^2$, then $\bfZ_i = \sigma^2\bfI_{M_i}$ is a diagonal matrix, which also yields the desired result. For condition (III) that $\rho_1=0$ and $\bX_{ik} = \bX_{ik'}$, we have $\bfR_i^{-1} = \textrm{bdiag}\{\bfI_{N_{ij}}-(\rho_2^{-1}+N_{ij})^{-1}\bone_{N_{ij}}\bone_{N_{ij}}^\top: j=1,\dots, J\}$ and $\bfZ_i =\textrm{bdiag}\{v_{ij}\bfI_{N_{ij}}:j=1\dots, J\}$ for a variance function $v_{ij}$ common across subjects within each period since $v(Y_{ijk})$ is a function of $Z_i, \bX_{ik}$, which is constant across $k$. Then $\bfR_i^{-1} \bfZ_i^{-1/2} = \textrm{bdiag}\{v_{ij}^{-1/2}\bfI_{N_{ij}}-v_{ij}^{-1/2}(\rho_2^{-1}+N_{ij})^{-1}\bone_{N_{ij}}\bone_{N_{ij}}^\top: j=1,\dots, J\} =\bfZ_i^{-1/2} \bfR_i^{-1} $, which completes the proof. Finally, the above derivation also works if we replace $J$ with $J^*=J-1$, i.e., dropping the data from the last period for period-specific or saturated treatment effects.
\end{proof}

\begin{proof}[Proof of Theorem 2]
We inherit all notation from the proof of Theorem 1. We first focus on the proof with condition (I), (II), or (III), where the mean model $E[Y_{ijk}|Z_i,\bX_{ik}] = g^{-1}(\beta_{0j} + TE_{ij} + \bbeta_{\bX}^\top \bX_{ik})$  can be misspecified.
By Lemma 2, the estimating equations (4) become $\sum_{i=1}^I {\bfU}_i^\top\bfZ_i^{-1} \bfR_i^{-1} (\bY_i^o - \bmu_i^o) = \bzero$. The proof structure is similar to Theorem 1 except for the additional g-computation step. Therefore, we omit the proof regarding asymptotic normality and the consistency of the sandwich variance estimator, as it is a direct application of Lemma 1.

We start by proving the results under the constant treatment effect setting. Let $\bmu_i = \{g^{-1}(\beta_{0j} + \beta_Z I\{Z_i\le j\} + \bbeta_{\bX}^\top \bX_{ik}): k = 1,\dots, N_i, j = 1,\dots, J\}$, then $\bmu_i^o = \bfD_i^\top\bmu_i$. Recall that $\bfQ_i = (\bfI_J\otimes \bone_{N_i}, (\Delta_{Z_i}\bone_J) \otimes \bone_{N_i}, \bone_{J} \otimes \bfX_i) \in \mathbb{R}^{N_iJ\times (J+1+p)}$ is the design matrix and $\bY_i^o = \bfD_i^\top \bY_i$. Since a canonical link function is used, then ${\bfU}_i = \frac{\partial \bmu_i^o}{\partial \bbeta } = \bfD_i^\top\frac{\partial \bmu_i}{\partial \bbeta } = \bfD_i^\top\bfZ_i \bfQ_i$. The above estimating equations become $\sum_{i=1}^I \bfQ_i^\top\bfD_i\bfR_i^{-1} \bfD_i^\top(\bY_i - \bmu_i) = \bzero$. Denoting $\btheta = (\Delta_{\textrm{GEE-g}}, \bbeta, \rho_1,\rho_2)$ as the vector of the parameters, our estimator is a solution to the estimating equations $\sum_{i=1}^I \bpsi(\bO_i;\btheta) = \bzero$, where
\begin{equation}\label{eq: gee-constant-1}
    \bpsi(\bO_i;\btheta)= \left(\begin{array}{c}
    \sum_{j=1}^J \lambda_{ij} \{\Delta_{\textrm{GEE-g}} - \mu_j(\beta_Z)+\mu_j(0) \}\\
    M_i \mu_j(b) - \bS_i^\top \bh_{ij}(b), j =1,\dots, J, b \in\{\beta_Z, 0\}  \\
    \bfQ_i^\top\bfD_i\bfR_i^{-1} \bfD_i^\top(\bY_i - \bmu_i)    \\
    f(\bO_i, \rho_1, \rho_2)
    \end{array}\right)
\end{equation}
with $M_i = \sum_{j=1}^J N_{ij}$, $\bh_{ij}(b) = \{g^{-1}(\beta_{0j} + b + \bbeta_{\bX}^\top \bX_{ik}): k = 1,\dots, N_i\} \otimes \bone_J$, 
\begin{align*}
     \lambda_{ij} = \frac{\pi_j^s(1-\pi_j^s)}{1+N_{ij}\rho_2} - (\frac{1}{\rho_1} +\sum_{j'=1}^J \frac{N_{ij'}}{1+N_{ij'}\rho_2})^{-1}\sum_{j'=1}^J \frac{N_{ij'} (\pi_{j\wedge j'}^s-\pi_j^s\pi_{j'}^s)}{1+N_{ij'}\rho_2},
\end{align*}
and $f$ is a pre-specified estimating equation for estimating $\rho_1, \rho_2$.
By Assumption A1 and regularity conditions (1)-(3), Lemma 1 implies that $\btheta$ converges in probability to $\underline\btheta = (\underline\Delta_{\textrm{GEE-g}}, \underline\bbeta, \underline\rho_1, \underline\rho_2)$ that solves $E[\bpsi(\bO; \btheta)] = \bzero$, is asymptotically normal distributed, and has a consistent sandwich variance estimator $\widehat{W}_{\textrm{GEE-g}}$.
Therefore, the only remaining part that needs to be addressed is consistency, i.e., $\underline\Delta_{\textrm{GEE-g}} = \Delta$. Using the fact that $E[\bpsi(\boldsymbol{O}, \underline\btheta)] = \bzero$ and $\bfQ = (\bfI_J\otimes \bone_{N}, (\Delta_{Z}\bone_J) \otimes \bone_{N}, \bone_{J} \otimes \bfX) $, we have
\begin{align*}
   & E[(\bfI_J\otimes \bone_{N})^\top \bfD\underline{\bfR}^{-1} \bfD^\top (\bY - \underline\bmu)] = \bzero,  \\
   & E[\{(\Delta_{Z}\bone_J) \otimes \bone_{N}\}^\top \bfD\underline\bfR^{-1} \bfD^\top(\bY - \underline\bmu)] = 0,
\end{align*}
where $\underline\bfR$ is $\bfR$ with $(\rho_1,\rho_2)$ substituted by $(\underline\rho_1,\underline\rho_2)$, and $\underline\bmu = \{g^{-1}(\underline\beta_{0j} + \underline\beta_Z I\{Z\le j\} + \underline\bbeta_{\bX}^\top \bX_{.k}): k = 1,\dots, N, j = 1,\dots, J\}$. Recall $\delta = E[\Delta_{Z}\bone_J] = (\pi_1^s,\dots, \pi_J^s)^\top$. By left-multiplying the first equation by $-\delta^\top$ and adding it to the second equation, we get
\begin{align*}
    E[\{(\Delta_{Z}\bone_J - \delta) \otimes \bone_{N}\}^\top \bfD\underline\bfR^{-1} \bfD^\top(\bY - \underline\bmu)] = 0.
\end{align*}
By the assumption that the treatment effect is duration invariant, we can replace $\bY$ with $\Delta_Z \otimes \bfI_N \{\bY(1) - \bY(0)\} - \bfI_J \otimes \bfI_N \bY(0)$ in the above equation. Furthermore, we have $\underline\bmu = \Delta_Z \otimes \bfI_N \{\underline\bmu(\underline{\beta}_Z) - \underline\bmu(0)\} - \bfI_J \otimes \bfI_N \underline\bmu(0)$, where $\underline\bmu(b) = \{g^{-1}(\underline\beta_{0j} + b + \underline\bbeta_{\bX}^\top \bX_{ik}): k = 1,\dots, N_i, j = 1,\dots, J\}$, which implies
\begin{equation}\label{eq:gee-proof1}
    E[\{(\Delta_{Z}\bone_J - \delta) \otimes \bone_{N}\}^\top \bfD\underline\bfR^{-1} \bfD^\top \Delta_Z \otimes \bfI_N \{\bY(1) - \bY(0) - \underline\bmu(\underline{\beta}_Z) +\underline\bmu(0) \}] = 0.
\end{equation}
Since $\underline\bmu(b) $ is a function of $\bX$, Assumption A1 implies that $E[\underline\bmu(b)|N]$ can be written as $ E[\underline{\widetilde{\bmu}}(b)|N]\otimes \bone_{N}$ where $\underline{\widetilde{\bmu}}(b)=\{E[g^{-1}(\underline\beta_{0j} + b + \underline\bbeta_{\bX}^\top \bX)]: j = 1,\dots, J\}$. Similarly, $E[\bY(d)|N]= \bv(d, N)\otimes \bone_{N}$. By Assumptions A2 and A3, the above equation~(\ref{eq:gee-proof1}) becomes
\begin{equation*}
    E[\{(\Delta_{Z}\bone_J - \delta) \otimes \bone_{N}\}^\top \bfD\underline\bfR^{-1} \bfD^\top \Delta_Z \otimes \bone_N \{\bv(1,N)-\bv(0,N)-\underline{\widetilde{\bmu}}(\underline{\beta}_Z)+\underline{\widetilde{\bmu}}(0)\}] = 0.
\end{equation*}
To further simplify the above equation, we compute 
\begin{align*}
    \bfD^\top \underline\bfR^{-1}\bfD 
    &= bdiag\{\bfD_{.j}\bfD_{.j}^\top - \frac{\underline\rho_2}{1+N_{.j} \underline\rho_2}\bS_{.j}\bS_{.j}^\top: j =1,\dots, J\} \\
    & \quad - (\frac{1}{\underline\rho_1} +\sum_{j=1}^J \frac{N_{.j}}{1+N_{.j}\underline\rho_2})^{-1}\left(\begin{array}{c}
     \frac{1}{1+N_{.1}\underline\rho_2}\bS_{.1}      \\
        \vdots \\
    \frac{1}{1+N_{.J}\underline\rho_2}\bS_{.J}     
    \end{array}\right) (\frac{1}{1+N_{.1}\underline\rho_2}\bS_{.1}, \dots, \frac{1}{1+N_{.J}\underline\rho_2}\bS_{.J})^\top,
\end{align*}
and then $\{(\Delta_{Z}\bone_J - \delta) \otimes \bone_{N}\}^\top \bfD\bfR^{-1} \bfD^\top \Delta_Z \otimes \bone_N $ becomes
\begin{align*}
    (\Delta_{Z}\bone_J - \delta)^\top \mathbf{G} \Delta_Z - (\frac{1}{\underline\rho_1} +\sum_{j=1}^J \frac{N_{.j}}{1+N_{.j}\underline\rho_2})^{-1}  (\Delta_{Z}\bone_J - \delta)^\top \mathbf{G} \bone_J \bone_J^\top \mathbf{G} \Delta_Z,
\end{align*}
where $\mathbf{G} = diag\{\frac{N_{.j}}{1+N_{.j} \underline\rho_2}: j =1,\dots,J\}$. We observe that the above formula is a function of $N_{.j}$ and $Z$ and is independent of potential outcomes and covariates by Assumptions A2 and A3. Furthermore, the $j$-th component of the above formula has expectation $E[\lambda_{ij}]$.
Under the setting of constant treatment effects, $E[\bv(1,N) - \bv(0,N)]=\Delta \bone_J $ and
\begin{align*}
    \Delta &= E[\{(\Delta_{Z}\bone_J - \delta) \otimes \bone_{N}\}^\top \bfD\underline\bfR^{-1} \bfD^\top \Delta_Z\bone_J \otimes \bone_N ]^{-1}\\
    &\quad E[\{(\Delta_{Z}\bone_J - \delta) \otimes \bone_{N}\}^\top \bfD\underline\bfR^{-1} \bfD^\top \Delta_Z \otimes \bone_N] E[\underline{\widetilde{\bmu}}(\underline{\beta}_Z)-\underline{\widetilde{\bmu}}(0)] \\
    &= (\sum_{j=1}^J E[\lambda_{ij}])^{-1} \sum_{j=1}^J E[\lambda_{ij}] E\left\{g^{-1}(\underline\beta_{0j} + \underline\beta_Za + \underline\bbeta_{\bX}^\top \bX) - g^{-1}(\underline\beta_{0j} + \underline\bbeta_{\bX}^\top \bX)\right\}.
\end{align*}
To show $\underline\Delta_{\textrm{GEE-g}} = \Delta$. We recall that the first two entries of estimating equations~(\ref{eq: gee-constant-1}) imply
\begin{align*}
E[\sum_{j=1}^J \lambda_{ij} \{\underline\Delta_{\textrm{GEE-g}} - \underline\mu_j(\underline{\beta}_Z)+\underline\mu_j(0) \}] &= 0,\\
       E[ M_i \underline\mu_j(b) - \bS_i^\top \underline\bh_{ij}(b)] &=0.
\end{align*}
The second equation yeilds that $E[\underline\mu_j(b)] = E[M_i]^{-1}E[\bS_i^\top \bh_{ij}(b)] = E[g^{-1}(\underline\beta_{0j} + b + \underline\bbeta_{\bX}^\top \bX)]$ using Assumption A3. The first equation then implies that
\begin{align*}
    \underline\Delta_{\textrm{GEE-g}} &= (\sum_{j=1}^J E[\lambda_{ij}])^{-1} \sum_{j=1}^J E[\lambda_{ij}] E\{\mu_j(\underline{\beta}_Z)-\mu_j(0)\} \\
    &= (\sum_{j=1}^J E[\lambda_{ij}])^{-1} \sum_{j=1}^J E[\lambda_{ij}] E\left\{g^{-1}(\underline\beta_{0j} + \underline\beta_Z + \underline\bbeta_{\bX}^\top \bX) - g^{-1}(\underline\beta_{0j} + \underline\bbeta_{\bX}^\top \bX)\right\} \\
    &= \Delta,
\end{align*}
which completes the proof for consistency. 

Next, we prove the results under the duration-specific treatment effect setting in a similar way. The corresponding estimating equations become
\begin{equation}\label{eq: gee-duration-1}
    \bpsi(\bO_i;\btheta)= \left(\begin{array}{c}
  \sum_{d=1}^J  \bLambda_i(d) [\mathbf{H}_d\bDelta_{\textrm{GEE-g}}^P -  \{\mu_1(\beta_{Zd})-\mu_1(0),\dots, \mu_J(\beta_{Zd})-\mu_J(0)\}^\top] \\
    M_i \mu_j(b) - \bS_i^\top \bh_{ij}(b), j =1,\dots, J; b\in\{\beta_{Z1},\dots, \beta_{ZJ},0\}  \\
    \bfQ_i^\top\bfD_i\bfR_i^{-1} \bfD_i^\top(\bY_i - \bmu_i)    \\
    f(\bO_i, \rho_1, \rho_2)
    \end{array}\right),
\end{equation}
where
$\bfQ_i = (\bfI_J\otimes \bone_{N_i}, \mathbf{H}_{Z_i} \otimes \bone_{N_i}, \bone_{J} \otimes \bfX_i)$,  $\mu_j(b)$, $\bh_{ij}(b)$ are defined as in the constant treatment effect setting, and
\begin{equation}
    \bLambda_i(d) = \{(\mathbf{H}_{Z_i}-E[\mathbf{H}_{Z}])\otimes \bone_{N_i}\}^\top \bfD_i\bfR_i^{-1} \bfD_i^\top(\Lambda_{Z_i}^d \otimes \bone_{N_i}).
\end{equation}
Some algebra shows that $\bLambda_i(d)$ is a function of $N_{ij}$ and $Z_i$ but not $\bS_i$ or $N_i$.
Then $E[\bpsi(\boldsymbol{O}, \underline\btheta)] = \bzero$ implies that
\begin{align*}
   & E[(\bfI_J\otimes \bone_{N})^\top \bfD\underline\bfR^{-1} \bfD^\top (\bY - \underline\bmu)] = \bzero,  \\
   & E[(\mathbf{H}_{Z}\otimes \bone_{N})^\top \bfD\underline\bfR^{-1} \bfD^\top(\bY - \underline\bmu)] = \bzero.
\end{align*}
Left multiplying the first equation by $-E[\mathbf{H}_Z]^\top$ and adding it to the second equation, we get $E[\{(\mathbf{H}_{Z}-E[\mathbf{H}_{Z}])\otimes \bone_{N}\}^\top \bfD\underline\bfR^{-1} \bfD^\top(\bY - \underline\bmu)] = \bzero.$ For $\bmu$, we have $\bmu_i = \sum_{d=1}^J (\Lambda_{Z_i}^d \otimes \bfI_{N_i}) \{\bmu_i(\beta_{Zd}) - \bmu_i(0)\} + \{\bfI_J \otimes \bfI_{N_i}\} \bmu_i(0)$. Using formula~(\ref{eq:Y}), we get
$$\sum_{d=1}^J E[\{(\mathbf{H}_{Z}-E[\mathbf{H}_{Z}])\otimes \bone_{N}\}^\top \bfD\underline\bfR^{-1} \bfD^\top(\Lambda_{Z}^d \otimes \bfI_N)\{\bY(d) - \bY(0) - \underline\bmu(\underline\beta_{Zd}) + \underline\bmu(0)\}] = \bzero.$$
Since the duration-specific treatment effect implies that $E[\bY(d) - \bY(0)|N] = \{\bv(d,N)-\bv(0,N)\} \otimes \bone_N$, $\Lambda_{Z}^d E[\bv(d,N)-\bv(0,N)] =\Lambda_{Z}^d \Delta(d) \bone_J$, and $\sum_{d=1}^J \Lambda_{Z}^d \Delta(d)\bone_J = \sum_{d=1}^J \Lambda_{Z}^d \mathbf{H}_d \bDelta^{D}$, we have
\begin{align*}
    \bDelta^{D} &= E[\sum_{d=1}^J\{(\mathbf{H}_{Z}-E[\mathbf{H}_{Z}])\otimes \bone_{N}\}^\top \bfD\underline\bfR^{-1} \bfD^\top(\Lambda_{Z}^d\mathbf{H}_d\otimes \bone_N)]^{-1} \\
    & \sum_{d=1}^J E[\{(\mathbf{H}_{Z}-E[\mathbf{H}_{Z}])\otimes \bone_{N}\}^\top \bfD\underline\bfR^{-1} \bfD^\top(\Lambda_{Z}^d \otimes \bone_N)\{\underline{\widetilde\bmu}(\underline\beta_{Zd}) - \underline{\widetilde\bmu}(0)\}]\\
    &= E[\sum_{d=1}^J\bLambda(d) \mathbf{H}_d]^{-1} E[\sum_{d=1}^J \bLambda(d) \{\underline{\widetilde\bmu}(\underline\beta_{Zd}) - \underline{\widetilde\bmu}(0)\}].
\end{align*}
Recall that the first two entries of estimating equations~(\ref{eq: gee-duration-1}) imply that 
\begin{align*}
    E\left[\sum_{d=1}^J  \bLambda(d) [\mathbf{H}_d\underline{\bDelta}_{\textrm{GEE-g}}^P -  \{\underline\mu_1(\beta_{Zd})-\underline\mu_1(0),\dots, \underline\mu_J(\beta_{Zd})-\underline\mu_J(0)\}^\top]\right] &=\bzero\\
    E[M_i \mu_j(b) - \bS_i^\top \bh_{ij}(b)]&=0, 
\end{align*}
we have $E[\underline\mu_j(b)] = E[g^{-1}(\underline{\beta}_{0j} + b + \underline{\bbeta}_{\bX}^\top \bX)]$, and hence $ E[\sum_{d=1}^J  \bLambda(d) [\mathbf{H}_d\underline\bDelta_{\textrm{GEE-g}}^P -\underline{\widetilde\bmu}(\underline\beta_{Zd}) +\underline{\widetilde\bmu}(0)]]=\bzero$, yielding $\underline\bDelta_{\textrm{GEE-g}}^P = \bDelta^P$.

We next prove the result under the period-specific treatment effect setting. Since we need to drop the data from period $J$ to avoid over-parameterization, we use the superscript ${}^*$ to denote all subsequent changes in notation, as in the proof of Theorem 1. For example, $J^*=J-1$, $\bY_i^*=(\bY_{i1},\dots,\bY_{i,J-1})$, $\bbeta_0^* = (\beta_{01},\dots, \beta_{0,J-1})$, and $\mathbf{M}^*$ is a $J^*$-by-$J^*$ matrix consisting of the first $J^*$ columns and rows of a matrix $\mathbf{M} \in \mathbb{R}^{J\times J}$. 
As a result, the corresponding estimating equations are
\begin{equation}\label{eq: gee-period-1}
    \bpsi(\bO_i^*;\btheta)= \left(\begin{array}{c}
    \bDelta_{\textrm{GEE-g}}^P -  \{\mu_1^*(\beta_{1Z})-\mu_1^*(0),\dots, \mu_J^*(\beta_{J^*Z})-\mu_{J^*}^*(0)\}^\top \\
     M_i^* \mu_j^*(\beta_{j'Z}) - \bS_i^*{}^\top \bh_{ij}^*(\beta_{j'Z}), j =1,\dots, J^*; j'=j \textrm{ or } 0  \\
    \bfQ_i^*{}^\top\bfD_i^*\bfR_i^*{}^{-1} \bfD_i^*{}^\top(\bY_i^* - \bmu_i^*)    \\
    f(\bO_i^*, \rho_1, \rho_2)
    \end{array}\right),
\end{equation}
with $\bfQ^* = (\bfI_{J^*}\otimes \bone_{N}, \Delta_{Z}^* \otimes \bone_{N}, \bone_{J^*} \otimes \bfX)$. Then $E[\bpsi(\boldsymbol{O}^*, \underline\btheta)] = \bzero$ implies that
\begin{align*}
   & E[(\bfI_{J^*}\otimes \bone_{N})^\top \bfD^*\underline\bfR^*{}^{-1} \bfD^*{}^\top (\bY^* -  \underline\bmu^*)] = \bzero,  \\
   & E[(\bDelta_{Z}^*\otimes \bone_{N})^\top  \bfD^*\underline\bfR^*{}^{-1} \bfD^*{}^\top(\bY^* -  \underline\bmu^*)] = \bzero.
\end{align*}
Left multiplying the first equation by $-E[\bDelta_{Z}^*]^\top$ and adding it to the second equation, we get $E[\{(\bDelta_{Z}^*-E[\bDelta_{Z}^*])\otimes \bone_{N}\}^\top \bfD^*\underline\bfR^*{}^{-1} \bfD^*{}^\top(\bY^* - \underline\bmu^*)] = \bzero.$ Under the setting of period-specific treatment effects, we have $\bmu_i^* =  (\bDelta_{Z_i}^* \otimes \bfI_{N_i}) \{\bmu_i^{*P}(1) - \bmu_i^*(0)\} + \{\bfI_{J^*} \otimes \bfI_{N_i}\} \bmu_i^*(0)$, where $\bmu_i^{*P}(1) = \{g^{-1}(\underline\beta_{0j} + \underline\beta_{jZ} + \underline\bbeta_{\bX}^\top \bX_{ik}): k = 1,\dots, N_i, j = 1,\dots, J^*\}$. Using formula~(\ref{eq:Y}) with $J$ replaced by $J^*$, we get
$$E[\{(\bDelta_{Z}^*-E[\bDelta_{Z}^*])\otimes \bone_{N}\}^\top \bfD^*\underline\bfR^*{}^{-1} \bfD^*{}^\top(\bDelta_{Z}^* \otimes \bfI_N)\{\bY(1)^* - \bY(0)^* - \bmu^{*P}(1) + \bmu^*(0)\}] = \bzero,$$
and hence
$$E[\{(\bDelta_{Z}^*-E[\bDelta_{Z}^*])\otimes \bone_{N}\}^\top \bfD^*\underline\bfR^*{}^{-1} \bfD^*{}^\top(\bDelta_{Z}^* \otimes \bone_N)]E[\bv^*(1,N) - \bv^*(0,N)- \underline{\widetilde\bmu}^{*P}(1) + \underline{\widetilde\bmu}^*(0)] = \bzero,$$
implying $E[\underline{\widetilde\bmu}^{*P}(1) -\underline{\widetilde\bmu}^*(0)] = E[\bv^*(1,N) - \bv^*(0,N)] = \bDelta^P$. 
Recall the first two entries of estimating equations~(\ref{eq: gee-period-1}) imply $E[\underline{\mu}_j^*(\underline{\beta}_{jZ})] = E[g^{-1}(\underline\beta_{0j} + \underline\beta_{jZ} + \underline\bbeta_{\bX}^\top \bX_{ik})]$, $E[\underline{\mu}_j^*(0)] = E[g^{-1}(\underline\beta_{0j} + \underline\bbeta_{\bX}^\top \bX_{ik})]$ and $\underline\bDelta_{\textrm{GEE-g}}^P =  \{\underline\mu_1^*(\beta_{1Z})-\underline\mu_1^*(0),\dots, \underline\mu_{J^*}^*(\beta_{J^*Z})-\underline\mu_{J^*}^*(0)\}^\top = \bDelta^P$, which completes the proof for consistency.

Finally, we prove the results under the saturated treatment effect setting. Like the period-specific treatment effect setting, we drop period $J$ to avoid over-parameterization, and we thus use superscript ${}^*$ to notate this change. The corresponding estimating equations are
\begin{equation}\label{eq: gee-saturated-1}
    \bpsi(\bO_i^*;\btheta)= \left(\begin{array}{c}
    \bDelta_{\textrm{GEE-g}}^S -  \{\mu_1^*(\beta_{1Z1})-\mu_1^*(0),\dots, \mu_{J^*}^*(\beta_{J^*ZJ^*})-\mu_{J^*}^*(0)\}^\top \\
     M_i^* \mu_j^*(\beta_{j'Zd}) - \bS_i^*{}^\top \bh_{ij}^*(\beta_{j'Zd}), 1\le d \le j \le J^*; j'=j \textrm{ or } 0  \\
    \bfQ_i^*{}^\top\bfD_i^*\bfR_i^*{}^{-1} \bfD_i^*{}^\top(\bY_i^* - \bmu_i^*)    \\
    f(\bO_i^*, \rho_1, \rho_2)
    \end{array}\right),
\end{equation}
with $\bfQ^* = (\bfI_{J^*}\otimes \bone_{N}, \widetilde\Lambda_{Z}^1{}^* \otimes \bone_{N},\dots, \widetilde\Lambda_{Z}^{J^*}{}^* \otimes \bone_{N}, \bone_{J^*} \otimes \bfX)$. Following a similar proof, we have, for each $d = 1,\dots, J$,
\begin{align*}
   & E[(\bfI_{J^*}\otimes \bone_{N})^\top \bfD^*\underline\bfR^*{}^{-1} \bfD^*{}^\top (\bY^* - \bmu^*)] = \bzero,\\
   & E[(\widetilde\Lambda_{Z}^d{}^* \otimes \bone_{N})^\top \bfD^*\underline\bfR^*{}^{-1} \bfD^*{}^\top(\bY^* - \bmu^*)] = \bzero,
\end{align*}
which implies, for $d =1,\dots, J^*$,
\begin{align*}
    &\sum_{d'=1}^{J^*} E[\{(\widetilde\Lambda_{Z}^d{}^* -E[\widetilde\Lambda_{Z}^d{}^*]) \otimes \bone_{N}\}^\top \bfD^*\underline\bfR^*{}^{-1} \bfD^*{}^\top(\widetilde\Lambda_{Z}^{d'*} \otimes \bone_N)]\\
    &\quad E[\widetilde\bv^*(d',N)-\widetilde\bv^*(0,N) -  \underline{\widetilde\bmu}_j^{S*}(d') + \underline{\widetilde\bmu}_j^{S,d*}(0)] = \bzero,
\end{align*}
where $\underline{\widetilde\bmu}_j^{S*}(d) = \{g^{-1}(\underline\beta_{0j} + \underline\beta_{jZd} + \underline\bbeta_{\bX}^\top \bX_{ik}):  j = d,\dots, J^*\} \in \mathbf{R}^{J^*-d+1}$ and $\underline{\widetilde\bmu}_j^{S,d*}(0)=\{g^{-1}(\underline\beta_{0j} + \underline\bbeta_{\bX}^\top \bX_{ik}):  j = d,\dots, J^*\} \in \mathbf{R}^{J^*-d+1}$.
For this linear system, $E[\widetilde\bv^*(d',N)-\widetilde\bv^*(0,N)]  =  E[\underline{\widetilde\bmu}_j^{S*}(d') - \underline{\widetilde\bmu}_j^{S*}(0)]$ for all $d$ is a direct solution. By the regularity condition that the solution is unique, we have, for $1\le d\le j \le J^*$, $$E[g^{-1}(\underline\beta_{0j} + \underline\beta_{jZd} + \underline\bbeta_{\bX}^\top \bX_{ik})] -E[g^{-1}(\underline\beta_{0j} + \underline\bbeta_{\bX}^\top \bX_{ik})] = E[Y_{ijk}(d)-Y_{ijk}(0)] = \Delta_j(d).$$
Recall that the second entry of the estimating equations~(\ref{eq: gee-saturated-1}) imply that \\$\underline\mu_j^*(\underline\beta_{jZd}) = E[M_i^*]^{-1}E[\bS_i^*{}^\top \underline{\bh}_{ij}^*(\underline\beta_{jZd})] = E[g^{-1}(\underline\beta_{0j} + \underline\beta_{jZd} + \underline\bbeta_{\bX}^\top \bX_{ik})]$ and, similarly, $\underline\mu_j^*(0) = E[g^{-1}(\underline\beta_{0j} + \underline\bbeta_{\bX}^\top \bX_{ik})]$. Therefore, the first entry of the the estimating equations~(\ref{eq: gee-saturated-1}) imply that $\underline\bDelta_{\textrm{GEE-g}}^S = \bDelta^S$. In fact, the regularity condition (1) is unnecessary here since the design matrix $E[\{(\widetilde\Lambda_{Z}^* -E[\widetilde\Lambda_{Z}^*]) \otimes \bone_{N}\}^\top \bfD^*\underline\bfR^*{}^{-1} \bfD^*{}^\top (\widetilde\Lambda_{Z}^* \otimes \bone_N)]$ is already invertible (which has a form a covariance matrix and does not degenerate due to Assumption A2), where $\widetilde\Lambda_{Z}^* = (\widetilde\Lambda_{Z}^1{}^*, \widetilde\Lambda_{Z}^2{}^*,\dots, \widetilde\Lambda_{Z}^{J^*}{}^*) \in\bfR^{J^*\times J^*(J^*+1)/2} $.

For condition (IV) that the mean model $E[Y_{ijk}|Z_i,\bX_{ik}] = g^{-1}(\beta_{0j} + TE_{ij} + \bbeta_{\bX}^\top \bX_{ik})$ is correctly specified for some parameters $\bbeta^*$, the classical GEE theory \citep{liang1986longitudinal} implies that the probability limit of $\widehat{\bbeta}$, $\underline\bbeta$, is equal to $\bbeta^*$. Therefore, the probability limit of $\widehat{\mu}_j(\widehat{b})$, $E[g^{-1}(\underline\beta_{0j} + \underline{b} + \underline{\bbeta}_{\bX}^\top \bX_{ik})]$, is the corresponding the expectation of potential outcomes, and the desired results are implied. 
\end{proof}

\subsection{Proof of Corollary 1}
\begin{proof}[Proof of Corollary 1]
In this proof, we inherit all notation defined in the proof of Theorem 2. 
For estimating the treatment effect estimands on other scales, the estimating equations become
\begin{equation}\label{eq: gee-saturated-2}
    \bpsi(\bO_i^*;\btheta)= \left(\begin{array}{c}
    \bPhi_{\textrm{GEE-g}}^S -  \{f(\mu_1^*(\beta_{1Z1}),\mu_1^*(0)),\dots, f(\mu_{J^*}^*(\beta_{JZJ}),\mu_{J^*}^*(0))\}^\top \\
     M_i^* \mu_j^*(\beta_{j'Zd}) - \bS_i^*{}^\top \bh_{ij}^*(\beta_{j'Zd}), 1\le d \le j \le J^*; j'=j \textrm{ or } 0  \\
    \bfQ_i^*{}^\top\bfD_i^*\bfR_i^*{}^{-1} \bfD_i^*{}^\top(\bY_i^* - \bmu_i^*)    \\
    f(\bO_i^*, \rho_1, \rho_2)
    \end{array}\right),
\end{equation}
which is equal to estimating equations~(\ref{eq: gee-saturated-1}) if $f(x,y) = x -y$ and the difference is only on the first entry of the estimating equations.  Therefore, we only need to prove consistency.

For GEE with the saturated treatment effect structure specification, Equations~(\ref{eq:Y}) and the estimating equations~(\ref{eq: gee-saturated-2}) imply that
\begin{align*}
    & E\left[(\bfI_{J^*}\otimes \bone_{N})^\top \bfD^*\underline\bfR^*{}^{-1} \bfD^*{}^\top \left[\sum_{d=1}^{J^*} (\Lambda_{Z}^d{}^* \otimes \bfI_N) \{\bY^*(d) - \bY^*(0)\} + (\bfI_{J^*} \otimes \bfI_{N}) \bY^*(0) \right.\right. \\
    &\quad \left.\left.
    -\sum_{d=1}^{J^*} (\Lambda_{Z}^d{}^* \otimes \bfI_N) \{\bmu^S{}^*(d) - \bmu^*(0)\} - (\bfI_{J^*} \otimes \bfI_{N}) \bmu^*(0)\right]\right] = \bzero,
\end{align*}
where $\bmu^S{}^*(d) =  \{g^{-1}(\underline\beta_{0j} + \underline\beta_{jZd} + \underline\bbeta_{\bX}^\top \bX_{ik}): k= 1,\dots, N_i,  j = d,\dots, J^*\}$. 
Following the same proof in Theorem 2, we then have
\begin{align*}
    & E\left[(\bfI_{J^*}\otimes \bone_{N})^\top \bfD^*\underline\bfR^*{}^{-1} \bfD^*{}^\top \left[\sum_{d=1}^{J^*} (\widetilde\Lambda_{Z}^{d*} \otimes \bone_N) \{\bv^*(d,N) - \bv^*(0,N)\} + (\bfI_{J^*} \otimes \bone_{N}) \bv^*(0,N) \right.\right. \\
    &\quad \left.\left.
    -\sum_{d=1}^{J^*} (\widetilde\Lambda_{Z}^{d*} \otimes \bone_N) \{\underline{\widetilde\bmu}_j^{S*}(d') - \underline{\widetilde\bmu}_j^{S,d*}(0)\} - (\bfI_{J^*} \otimes \bone_{N}) \underline{\widetilde{\bmu}}^*(0)\right]\right] = \bzero,
\end{align*}
where $\underline{\widetilde\bmu}^*(0) =  \{g^{-1}(\underline\beta_{0j}+ \underline\bbeta_{\bX}^\top \bX_{ik}):  j = d,\dots, J^*\}$.
In the proof of Theorem~2, we have shown that, under the setting of saturated treatment effect, $E[\widetilde\bv^*(d',N)-\widetilde\bv^*(0,N)]  =  E[\underline{\widetilde\bmu}_j^{S*}(d') - \underline{\widetilde\bmu}_j^{S,d*}(0)]$. Hence,
\begin{align*}
    E[(\bfI_{J^*}\otimes \bone_{N})^\top \bfD^*\underline\bfR^*{}^{-1} \bfD^*{}^\top (\bfI_{J^*}\otimes \bone_{N})]E[\bv^*(0,N)
   -\underline{\widetilde\bmu}^*(0)] = \bzero.
\end{align*}
Since $E[(\bfI_{J^*}\otimes \bone_{N})^\top \bfV (\bfI_{J^*}\otimes \bone_{N})]$ is a positive-definite diagonal matrix, we have $E[\bv^*(0,N)
   -\underline{\widetilde\bmu}^*(0)] = 0$ and hence $E[Y_{ijk}(0)]=E[g^{-1}(\underline\beta_{0j}+ \underline\bbeta_{\bX}^\top \bX_{ik})]$. Combined with the result that $E[\widetilde\bv^*(d',N)-\widetilde\bv^*(0,N)]  =  E[\underline{\widetilde\bmu}_j^S{}^*(d') - \underline{\widetilde\bmu}_j^S{}^*(0)]$, we also obtain $E[Y_{ijk}(d)]=E[g^{-1}(\underline\beta_{0j}+ \underline\beta_{jZd}+\underline\bbeta_{\bX}^\top \bX_{ik})]$. Since the second entry of the estimating equations~(\ref{eq: gee-saturated-2}) imply that $\underline\mu_j^*(\underline\beta_{jZd}) = E[M_i^*]^{-1}E[\bS_i^*{}^\top \underline{\bh}_{ij}^*(\underline\beta_{jZd})] = E[g^{-1}(\underline\beta_{0j} + \underline\beta_{jZd} + \underline\bbeta_{\bX}^\top \bX_{ik})]$ and, similarly, $\underline\mu_j^*(0) = E[g^{-1}(\underline\beta_{0j} + \underline\bbeta_{\bX}^\top \bX_{ik})]$, we get $\underline\mu_j^*(0) = E[Y_{ijk}(0)]$ and $\underline\mu_j^*(\underline{\beta_{jZd}}) = E[Y_{ijk}(d)]$. Based on the first entry of the estimating equations~(\ref{eq: gee-saturated-2}), we have $\Phi^S_{GEE-g,j}(d) = f\{\underline\mu_j^*(\underline{\beta_{jZd}}), \underline\mu_j^*(0)\} = f(E[Y_{ijk}(d)], E[Y_{ijk}(0)]) = \Phi_j(d)$, which completes the proof for consistency.

\end{proof}

\section{Additional simulation results}\label{sec: add-simulation}
\subsection{Complete results under simulation design B}
\begin{itemize}
    \item Web Tables \ref{tb:Tablew1}-\ref{tb:Tablew4} present the results for continuous outcomes, from Simulation Scenario B1.
    \begin{itemize}
        \item Web Table \ref{tb:Tablew1} refers to $I=30$ under the simple exchangeable correlation structure.
        \item Web Table \ref{tb:Tablew2} refers to $I=30$ under the nested exchangeable correlation structure.
        \item Web Table \ref{tb:Tablew3} refers to $I=100$ under the simple exchangeable correlation structure.
        \item Web Table \ref{tb:Tablew4} refers to $I=100$ under the nested exchangeable correlation structure.
    \end{itemize}
    \item Web Tables \ref{tb:Tablew5}-\ref{tb:Tablew8} present the results for continuous outcomes, from Simulation Scenario B2.
    \begin{itemize}
        \item Web Table \ref{tb:Tablew5} refers to $I=30$ under the simple exchangeable correlation structure.
        \item Web Table \ref{tb:Tablew6} refers to $I=30$ under the nested exchangeable correlation structure.
        \item Web Table \ref{tb:Tablew7} refers to $I=100$ under the simple exchangeable correlation structure.
        \item Web Table \ref{tb:Tablew8} refers to $I=100$ under the nested exchangeable correlation structure.
    \end{itemize}
\end{itemize}

\begin{table}[htbp]
\footnotesize
\caption{Simulation results for continuous outcomes under Simulation Scenario B1 with $I=30$. The working linear mixed model includes a simple exchangeable correlation structure.}\label{tb:Tablew1}
\centering
\resizebox{\textwidth}{!}{%
\begin{threeparttable}
\begin{tabular}{lclcccccccc}
\toprule
\midrule
Scenario & $I$ & Model\tnote{a} & TE\tnote{b} & Bias\tnote{c} & ESE\tnote{d} & RE\tnote{e} & ASE\textsubscript{MB}\tnote{f} & ASE\textsubscript{ROB}\tnote{g} & ECP\textsubscript{MB}\tnote{h} & ECP\textsubscript{ROB}\tnote{i} \\
\midrule
B1 & 30 & B(i) & $\Delta$ & -1.067 & 0.220 & 1.000 & 0.214 & 0.212 & 0.002 & 0.003 \\
& & B(ii) & $\Delta^{D\text{-avg}}$ & 0.019 & 0.369 & 1.000 & 0.298 & 0.339 & 0.910 & 0.939 \\
& & & $\Delta(1)$ & 0.010 & 0.223 & 1.000 & 0.228 & 0.208 & 0.969 & 0.956 \\
& & & $\Delta(2)$ & 0.016 & 0.288 & 1.000 & 0.277 & 0.282 & 0.950 & 0.952 \\
& & & $\Delta(3)$ & 0.017 & 0.427 & 1.000 & 0.339 & 0.384 & 0.914 & 0.934 \\
& & & $\Delta(4)$ & 0.020 & 0.556 & 1.000 & 0.422 & 0.521 & 0.887 & 0.952 \\
& & & $\Delta(5)$ & 0.034 & 0.841 & 1.000 & 0.562 & 0.732 & 0.847 & 0.920 \\
& & B(iii) & $\Delta$ & -1.131 & 0.176 & 1.567 & 0.170 & 0.172 & 0.000 & 0.000 \\
& & B(iv) & $\Delta^{D\text{-avg}}$ & 0.013 & 0.320 & 1.331 & 0.248 & 0.290 & 0.906 & 0.945 \\
& & & $\Delta(1)$ & 0.007 & 0.176 & 1.605 & 0.183 & 0.166 & 0.975 & 0.946 \\
& & & $\Delta(2)$ & 0.008 & 0.242 & 1.408 & 0.224 & 0.229 & 0.948 & 0.945 \\
& & & $\Delta(3)$ & 0.008 & 0.349 & 1.495 & 0.277 & 0.317 & 0.905 & 0.936 \\
& & & $\Delta(4)$ & 0.014 & 0.472 & 1.390 & 0.346 & 0.433 & 0.881 & 0.942 \\
& & & $\Delta(5)$ & 0.025 & 0.705 & 1.424 & 0.459 & 0.605 & 0.852 & 0.926 \\
& & B(v) & $\Delta$ & -1.121 & 0.168 & 1.709 & 0.162 & 0.166 & 0.000 & 0.000 \\
& & B(vi) & $\Delta^{D\text{-avg}}$ & 0.010 & 0.309 & 1.428 & 0.231 & 0.279 & 0.895 & 0.936 \\
& & & $\Delta(1)$ & 0.005 & 0.166 & 1.808 & 0.172 & 0.156 & 0.976 & 0.959 \\
& & & $\Delta(2)$ & 0.006 & 0.232 & 1.539 & 0.211 & 0.217 & 0.945 & 0.949 \\
& & & $\Delta(3)$ & 0.006 & 0.338 & 1.592 & 0.259 & 0.305 & 0.899 & 0.932 \\
& & & $\Delta(4)$ & 0.010 & 0.459 & 1.471 & 0.324 & 0.419 & 0.884 & 0.947 \\
& & & $\Delta(5)$ & 0.022 & 0.686 & 1.502 & 0.430 & 0.588 & 0.837 & 0.920 \\
\bottomrule
\end{tabular}\smallskip
\begin{tablenotes}\linespread{1}\scriptsize
\item[a] TE: treatment effect. $\Delta = 2$; $\Delta^{D\text{-avg}}$ = $\sum_{d=1}^{5} \Delta(d)/5 = 2$; $\Delta(1) = 1$; $\Delta(2) = 1.5$; $\Delta(3) = 2$; $\Delta(4) = 2.5$; $\Delta(5) = 3$. \smallskip
\item[b] Model: six working linear mixed-models under the simple exchangeable correlation structure, which are combinations of covariate adjustment settings and treatment effect settings. B(i), no adjustment with constant TE; B(ii), no adjustment with duration-specific TE; B(iii), partial adjustment with constant TE; B(iv), partial adjustment with duration-specific TE; B(v), full adjustment with constant TE; B(vi), full adjustment with duration-specific TE. \smallskip
\item[c] Bias: bias as the average of (estimate - estimand) over simulations.\smallskip
\item[d] ESE: empirical standard error (SE).\smallskip
\item[e] RE: relative effeciency.\smallskip
\item[f] ASE\textsubscript{MB}: averaged model-based SE.\smallskip
\item[g] ASE\textsubscript{ROB}: averaged robust SE.\smallskip
\item[h] ECP\textsubscript{MB}: empirical coverage probability based on the model-based SE.\smallskip
\item[i] ECP\textsubscript{ROB}: empirical coverage probability based on the robust SE.
\end{tablenotes}
\end{threeparttable}
}
\end{table}

\begin{table}[htbp]
\footnotesize
\caption{Simulation results for continuous outcomes under Simulation Scenario B1 with $I=30$. The working linear mixed model includes a nested exchangeable correlation structure.}\label{tb:Tablew2}
\centering
\resizebox{\textwidth}{!}{%
\begin{threeparttable}
\begin{tabular}{lclcccccccc}
\toprule
\midrule
Scenario & $I$ & Model\tnote{a} & TE\tnote{b} & Bias\tnote{c} & ESE\tnote{d} & RE\tnote{e} & ASE\textsubscript{MB}\tnote{f} & ASE\textsubscript{ROB}\tnote{g} & ECP\textsubscript{MB}\tnote{h} & ECP\textsubscript{ROB}\tnote{i} \\
\midrule
B1 & 30 & B(i) & $\Delta$ & -1.028 & 0.219 & 1.000 & 0.234 & 0.210 & 0.007 & 0.008 \\
& & B(ii) & $\Delta^{D\text{-avg}}$ & 0.021 & 0.372 & 1.000 & 0.307 & 0.342 & 0.914 & 0.939 \\
& & & $\Delta(1)$ & 0.011 & 0.223 & 1.000 & 0.237 & 0.209 & 0.974 & 0.954 \\
& & & $\Delta(2)$ & 0.017 & 0.289 & 1.000 & 0.287 & 0.283 & 0.958 & 0.951 \\
& & & $\Delta(3)$ & 0.019 & 0.430 & 1.000 & 0.351 & 0.387 & 0.922 & 0.934 \\
& & & $\Delta(4)$ & 0.022 & 0.558 & 1.000 & 0.436 & 0.526 & 0.899 & 0.950 \\
& & & $\Delta(5)$ & 0.035 & 0.845 & 1.000 & 0.581 & 0.739 & 0.857 & 0.918 \\
& & B(iii) & $\Delta$ & -1.086 & 0.176 & 1.545 & 0.198 & 0.171 & 0.000 & 0.000 \\
& & B(iv) & $\Delta^{D\text{-avg}}$ & 0.013 & 0.323 & 1.327 & 0.261 & 0.292 & 0.923 & 0.950 \\
& & & $\Delta(1)$ & 0.008 & 0.177 & 1.595 & 0.194 & 0.166 & 0.982 & 0.943 \\
& & & $\Delta(2)$ & 0.009 & 0.244 & 1.399 & 0.238 & 0.230 & 0.962 & 0.943 \\
& & & $\Delta(3)$ & 0.009 & 0.353 & 1.486 & 0.293 & 0.319 & 0.918 & 0.938 \\
& & & $\Delta(4)$ & 0.014 & 0.474 & 1.388 & 0.366 & 0.435 & 0.902 & 0.944 \\
& & & $\Delta(5)$ & 0.024 & 0.708 & 1.425 & 0.486 & 0.610 & 0.870 & 0.930 \\
& & B(v) & $\Delta$ & -1.067 & 0.167 & 1.714 & 0.190 & 0.165 & 0.000 & 0.000 \\
& & B(vi) & $\Delta^{D\text{-avg}}$ & 0.010 & 0.312 & 1.419 & 0.244 & 0.280 & 0.908 & 0.933 \\
& & & $\Delta(1)$ & 0.006 & 0.167 & 1.793 & 0.184 & 0.156 & 0.982 & 0.958 \\
& & & $\Delta(2)$ & 0.006 & 0.233 & 1.529 & 0.225 & 0.218 & 0.958 & 0.946 \\
& & & $\Delta(3)$ & 0.007 & 0.342 & 1.576 & 0.276 & 0.307 & 0.913 & 0.933 \\
& & & $\Delta(4)$ & 0.010 & 0.461 & 1.465 & 0.344 & 0.421 & 0.903 & 0.952 \\
& & & $\Delta(5)$ & 0.020 & 0.690 & 1.500 & 0.457 & 0.593 & 0.854 & 0.922 \\
\bottomrule
\end{tabular}\smallskip
\begin{tablenotes}\linespread{1}\scriptsize
\item[a] TE: treatment effect. $\Delta = 2$; $\Delta^{D\text{-avg}}$ = $\sum_{d=1}^{5} \Delta(d)/5 = 2$; $\Delta(1) = 1$; $\Delta(2) = 1.5$; $\Delta(3) = 2$; $\Delta(4) = 2.5$; $\Delta(5) = 3$. \smallskip
\item[b] Model: six working linear mixed-models under the nested exchangeable correlation structure, which are combinations of covariate adjustment settings and treatment effect settings. B(i), no adjustment with constant TE; B(ii), no adjustment with duration-specific TE; B(iii), partial adjustment with constant TE; B(iv), partial adjustment with duration-specific TE; B(v), full adjustment with constant TE; B(vi), full adjustment with duration-specific TE. \smallskip
\item[c] Bias: bias as the average of (estimate - estimand) over simulations.\smallskip
\item[d] ESE: empirical standard error (SE).\smallskip
\item[e] RE: relative effeciency.\smallskip
\item[f] ASE\textsubscript{MB}: averaged model-based SE.\smallskip
\item[g] ASE\textsubscript{ROB}: averaged robust SE.\smallskip
\item[h] ECP\textsubscript{MB}: empirical coverage probability based on the model-based SE.\smallskip
\item[i] ECP\textsubscript{ROB}: empirical coverage probability based on the robust SE.
\end{tablenotes}
\end{threeparttable}
}
\end{table}

\begin{table}[htbp]
\footnotesize
\caption{Simulation results for continuous outcomes under Simulation Scenario B1 with $I=100$. The working linear mixed model includes a simple exchangeable correlation structure.}\label{tb:Tablew3}
\centering
\resizebox{\textwidth}{!}{%
\begin{threeparttable}
\begin{tabular}{lclcccccccc}
\toprule
\midrule
Scenario & $I$ & Model\tnote{a} & TE\tnote{b} & Bias\tnote{c} & ESE\tnote{d} & RE\tnote{e} & ASE\textsubscript{MB}\tnote{f} & ASE\textsubscript{ROB}\tnote{g} & ECP\textsubscript{MB}\tnote{h} & ECP\textsubscript{ROB}\tnote{i} \\
\midrule
B1 & 100 & B(i) & $\Delta$ & -1.074 & 0.123 & 1.000 & 0.117 & 0.119 & 0.000 & 0.000 \\
& & B(ii) & $\Delta^{D\text{-avg}}$ & 0.010 & 0.214 & 1.000 & 0.165 & 0.199 & 0.884 & 0.936 \\
& & & $\Delta(1)$ & 0.005 & 0.125 & 1.000 & 0.125 & 0.118 & 0.956 & 0.937 \\
& & & $\Delta(2)$ & 0.009 & 0.172 & 1.000 & 0.153 & 0.164 & 0.917 & 0.933 \\
& & & $\Delta(3)$ & 0.006 & 0.234 & 1.000 & 0.187 & 0.223 & 0.886 & 0.946 \\
& & & $\Delta(4)$ & 0.012 & 0.322 & 1.000 & 0.233 & 0.303 & 0.863 & 0.942 \\
& & & $\Delta(5)$ & 0.015 & 0.471 & 1.000 & 0.309 & 0.439 & 0.811 & 0.937 \\
& & B(iii) & $\Delta$ & -1.134 & 0.099 & 1.538 & 0.094 & 0.097 & 0.000 & 0.000 \\
& & B(iv) & $\Delta^{D\text{-avg}}$ & 0.012 & 0.187 & 1.304 & 0.137 & 0.174 & 0.858 & 0.943 \\
& & & $\Delta(1)$ & 0.007 & 0.100 & 1.555 & 0.101 & 0.095 & 0.949 & 0.938 \\
& & & $\Delta(2)$ & 0.009 & 0.140 & 1.498 & 0.124 & 0.135 & 0.918 & 0.947 \\
& & & $\Delta(3)$ & 0.009 & 0.201 & 1.355 & 0.153 & 0.188 & 0.877 & 0.942 \\
& & & $\Delta(4)$ & 0.015 & 0.273 & 1.392 & 0.192 & 0.258 & 0.852 & 0.943 \\
& & & $\Delta(5)$ & 0.021 & 0.397 & 1.410 & 0.253 & 0.369 & 0.799 & 0.942 \\
& & B(v) & $\Delta$ & -1.126 & 0.096 & 1.652 & 0.089 & 0.094 & 0.000 & 0.000 \\
& & B(vi) & $\Delta^{D\text{-avg}}$ & 0.010 & 0.179 & 1.430 & 0.129 & 0.168 & 0.862 & 0.949 \\
& & & $\Delta(1)$ & 0.006 & 0.094 & 1.764 & 0.095 & 0.090 & 0.956 & 0.940 \\
& & & $\Delta(2)$ & 0.007 & 0.133 & 1.665 & 0.117 & 0.129 & 0.916 & 0.943 \\
& & & $\Delta(3)$ & 0.008 & 0.192 & 1.479 & 0.144 & 0.181 & 0.861 & 0.945 \\
& & & $\Delta(4)$ & 0.013 & 0.261 & 1.514 & 0.180 & 0.251 & 0.843 & 0.946 \\
& & & $\Delta(5)$ & 0.016 & 0.386 & 1.490 & 0.238 & 0.360 & 0.780 & 0.940 \\
\bottomrule
\end{tabular}\smallskip
\begin{tablenotes}\linespread{1}\scriptsize
\item[a] TE: treatment effect. $\Delta = 2$; $\Delta^{D\text{-avg}}$ = $\sum_{d=1}^{5} \Delta(d)/5 = 2$; $\Delta(1) = 1$; $\Delta(2) = 1.5$; $\Delta(3) = 2$; $\Delta(4) = 2.5$; $\Delta(5) = 3$. \smallskip
\item[b] Model: six working linear mixed-models under the simple exchangeable correlation structure, which are combinations of covariate adjustment settings and treatment effect settings. B(i), no adjustment with constant TE; B(ii), no adjustment with duration-specific TE; B(iii), partial adjustment with constant TE; B(iv), partial adjustment with duration-specific TE; B(v), full adjustment with constant TE; B(vi), full adjustment with duration-specific TE. \smallskip
\item[c] Bias: bias as the average of (estimate - estimand) over simulations.\smallskip
\item[d] ESE: empirical standard error (SE).\smallskip
\item[e] RE: relative effeciency.\smallskip
\item[f] ASE\textsubscript{MB}: averaged model-based SE.\smallskip
\item[g] ASE\textsubscript{ROB}: averaged robust SE.\smallskip
\item[h] ECP\textsubscript{MB}: empirical coverage probability based on the model-based SE.\smallskip
\item[i] ECP\textsubscript{ROB}: empirical coverage probability based on the robust SE.
\end{tablenotes}
\end{threeparttable}
}
\end{table}

\begin{table}[htbp]
\footnotesize
\caption{Simulation results for continuous outcomes under Simulation Scenario B1 with $I=100$. The working linear mixed model includes a nested exchangeable correlation structure.}\label{tb:Tablew4}
\centering
\resizebox{\textwidth}{!}{%
\begin{threeparttable}
\begin{tabular}{lclcccccccc}
\toprule
\midrule
Scenario & $I$ & Model\tnote{a} & TE\tnote{b} & Bias\tnote{c} & ESE\tnote{d} & RE\tnote{e} & ASE\textsubscript{MB}\tnote{f} & ASE\textsubscript{ROB}\tnote{g} & ECP\textsubscript{MB}\tnote{h} & ECP\textsubscript{ROB}\tnote{i} \\
\midrule
B1 & 100 & B(i) & $\Delta$ & -1.038 & 0.122 & 1.000 & 0.129 & 0.119 & 0.000 & 0.000 \\
& & B(ii) & $\Delta^{D\text{-avg}}$ & 0.010 & 0.215 & 1.000 & 0.170 & 0.199 & 0.894 & 0.936 \\
& & & $\Delta(1)$ & 0.005 & 0.125 & 1.000 & 0.130 & 0.118 & 0.960 & 0.939 \\
& & & $\Delta(2)$ & 0.009 & 0.172 & 1.000 & 0.158 & 0.164 & 0.923 & 0.936 \\
& & & $\Delta(3)$ & 0.006 & 0.234 & 1.000 & 0.193 & 0.224 & 0.896 & 0.947 \\
& & & $\Delta(4)$ & 0.012 & 0.323 & 1.000 & 0.241 & 0.304 & 0.871 & 0.943 \\
& & & $\Delta(5)$ & 0.016 & 0.472 & 1.000 & 0.319 & 0.441 & 0.819 & 0.937 \\
& & B(iii) & $\Delta$ & -1.093 & 0.098 & 1.531 & 0.109 & 0.097 & 0.000 & 0.000 \\
& & B(iv) & $\Delta^{D\text{-avg}}$ & 0.012 & 0.189 & 1.301 & 0.145 & 0.175 & 0.874 & 0.943 \\
& & & $\Delta(1)$ & 0.007 & 0.100 & 1.558 & 0.107 & 0.095 & 0.959 & 0.935 \\
& & & $\Delta(2)$ & 0.009 & 0.141 & 1.493 & 0.132 & 0.136 & 0.931 & 0.944 \\
& & & $\Delta(3)$ & 0.010 & 0.202 & 1.347 & 0.162 & 0.188 & 0.896 & 0.944 \\
& & & $\Delta(4)$ & 0.016 & 0.274 & 1.387 & 0.203 & 0.259 & 0.873 & 0.938 \\
& & & $\Delta(5)$ & 0.021 & 0.398 & 1.410 & 0.268 & 0.371 & 0.818 & 0.945 \\
& & B(v) & $\Delta$ & -1.077 & 0.095 & 1.657 & 0.105 & 0.093 & 0.000 & 0.000 \\
& & B(vi) & $\Delta^{D\text{-avg}}$ & 0.010 & 0.180 & 1.428 & 0.136 & 0.169 & 0.884 & 0.951 \\
& & & $\Delta(1)$ & 0.006 & 0.094 & 1.772 & 0.102 & 0.089 & 0.966 & 0.942 \\
& & & $\Delta(2)$ & 0.007 & 0.134 & 1.663 & 0.125 & 0.130 & 0.936 & 0.945 \\
& & & $\Delta(3)$ & 0.009 & 0.193 & 1.469 & 0.154 & 0.182 & 0.893 & 0.945 \\
& & & $\Delta(4)$ & 0.013 & 0.263 & 1.509 & 0.192 & 0.252 & 0.859 & 0.943 \\
& & & $\Delta(5)$ & 0.017 & 0.387 & 1.492 & 0.254 & 0.362 & 0.805 & 0.940 \\
\bottomrule
\end{tabular}\smallskip
\begin{tablenotes}\linespread{1}\scriptsize
\item[a] TE: treatment effect. $\Delta = 2$; $\Delta^{D\text{-avg}}$ = $\sum_{d=1}^{5} \Delta(d)/5 = 2$; $\Delta(1) = 1$; $\Delta(2) = 1.5$; $\Delta(3) = 2$; $\Delta(4) = 2.5$; $\Delta(5) = 3$. \smallskip
\item[b] Model: six working linear mixed-models under the nested exchangeable correlation structure, which are combinations of covariate adjustment settings and treatment effect settings. B(i), no adjustment with constant TE; B(ii), no adjustment with duration-specific TE; B(iii), partial adjustment with constant TE; B(iv), partial adjustment with duration-specific TE; B(v), full adjustment with constant TE; B(vi), full adjustment with duration-specific TE. \smallskip
\item[c] Bias: bias as the average of (estimate - estimand) over simulations.\smallskip
\item[d] ESE: empirical standard error (SE).\smallskip
\item[e] RE: relative effeciency.\smallskip
\item[f] ASE\textsubscript{MB}: averaged model-based SE.\smallskip
\item[g] ASE\textsubscript{ROB}: averaged robust SE.\smallskip
\item[h] ECP\textsubscript{MB}: empirical coverage probability based on the model-based SE.\smallskip
\item[i] ECP\textsubscript{ROB}: empirical coverage probability based on the robust SE.
\end{tablenotes}
\end{threeparttable}
}
\end{table}

\begin{table}[htbp]
\footnotesize
\caption{Simulation results for continuous outcomes under Simulation Scenario B2 with $I=30$. The working linear mixed model includes a simple exchangeable correlation structure.}\label{tb:Tablew5}
\centering
\resizebox{\textwidth}{!}{%
\begin{threeparttable}
\begin{tabular}{lclcccccccc}
\toprule
\midrule
Scenario & $I$ & Model\tnote{a} & TE\tnote{b} & Bias\tnote{c} & ESE\tnote{d} & RE\tnote{e} & ASE\textsubscript{MB}\tnote{f} & ASE\textsubscript{ROB}\tnote{g} & ECP\textsubscript{MB}\tnote{h} & ECP\textsubscript{ROB}\tnote{i} \\
\midrule
B2 & 30 & B(i) & $\Delta$ & -1.065 & 0.239 & 1.000 & 0.233 & 0.244 & 0.008 & 0.012 \\
& & B(ii) & $\Delta^{D\text{-avg}}$ & -0.009 & 0.427 & 1.000 & 0.322 & 0.394 & 0.891 & 0.946 \\
& & & $\Delta(1)$ & 0.004 & 0.239 & 1.000 & 0.249 & 0.241 & 0.973 & 0.965 \\
& & & $\Delta(2)$ & -0.008 & 0.341 & 1.000 & 0.301 & 0.326 & 0.941 & 0.949 \\
& & & $\Delta(3)$ & -0.022 & 0.483 & 1.000 & 0.368 & 0.443 & 0.887 & 0.941 \\
& & & $\Delta(4)$ & -0.016 & 0.653 & 1.000 & 0.458 & 0.606 & 0.863 & 0.950 \\
& & & $\Delta(5)$ & -0.003 & 0.968 & 1.000 & 0.609 & 0.870 & 0.814 & 0.930 \\
& & B(iii) & $\Delta$ & -1.133 & 0.189 & 1.594 & 0.180 & 0.205 & 0.000 & 0.000 \\
& & B(iv) & $\Delta^{D\text{-avg}}$ & -0.008 & 0.373 & 1.313 & 0.261 & 0.343 & 0.869 & 0.942 \\
& & & $\Delta(1)$ & 0.002 & 0.191 & 1.565 & 0.193 & 0.200 & 0.962 & 0.963 \\
& & & $\Delta(2)$ & -0.008 & 0.281 & 1.471 & 0.237 & 0.270 & 0.934 & 0.952 \\
& & & $\Delta(3)$ & -0.016 & 0.402 & 1.442 & 0.292 & 0.368 & 0.880 & 0.947 \\
& & & $\Delta(4)$ & -0.010 & 0.553 & 1.396 & 0.366 & 0.504 & 0.834 & 0.939 \\
& & & $\Delta(5)$ & -0.007 & 0.787 & 1.514 & 0.485 & 0.709 & 0.801 & 0.919 \\
& & B(v) & $\Delta$ & -1.124 & 0.181 & 1.745 & 0.172 & 0.200 & 0.000 & 0.000 \\
& & B(vi) & $\Delta^{D\text{-avg}}$ & -0.012 & 0.354 & 1.454 & 0.246 & 0.333 & 0.869 & 0.946 \\
& & & $\Delta(1)$ & 0.000 & 0.179 & 1.787 & 0.184 & 0.192 & 0.964 & 0.970 \\
& & & $\Delta(2)$ & -0.010 & 0.265 & 1.653 & 0.224 & 0.260 & 0.933 & 0.961 \\
& & & $\Delta(3)$ & -0.020 & 0.383 & 1.591 & 0.276 & 0.357 & 0.876 & 0.950 \\
& & & $\Delta(4)$ & -0.017 & 0.532 & 1.507 & 0.345 & 0.491 & 0.838 & 0.946 \\
& & & $\Delta(5)$ & -0.014 & 0.762 & 1.615 & 0.458 & 0.695 & 0.795 & 0.930 \\
\bottomrule
\end{tabular}\smallskip
\begin{tablenotes}\linespread{1}\scriptsize
\item[a] TE: treatment effect. $\Delta = 2$; $\Delta^{D\text{-avg}}$ = $\sum_{d=1}^{5} \Delta(d)/5 = 2$; $\Delta(1) = 1$; $\Delta(2) = 1.5$; $\Delta(3) = 2$; $\Delta(4) = 2.5$; $\Delta(5) = 3$. \smallskip
\item[b] Model: six working linear mixed-models under the simple exchangeable correlation structure, which are combinations of covariate adjustment settings and treatment effect settings. B(i), no adjustment with constant TE; B(ii), no adjustment with duration-specific TE; B(iii), partial adjustment with constant TE; B(iv), partial adjustment with duration-specific TE; B(v), full adjustment with constant TE; B(vi), full adjustment with duration-specific TE. \smallskip
\item[c] Bias: bias as the average of (estimate - estimand) over simulations.\smallskip
\item[d] ESE: empirical standard error (SE).\smallskip
\item[e] RE: relative effeciency.\smallskip
\item[f] ASE\textsubscript{MB}: averaged model-based SE.\smallskip
\item[g] ASE\textsubscript{ROB}: averaged robust SE.\smallskip
\item[h] ECP\textsubscript{MB}: empirical coverage probability based on the model-based SE.\smallskip
\item[i] ECP\textsubscript{ROB}: empirical coverage probability based on the robust SE.
\end{tablenotes}
\end{threeparttable}
}
\end{table}

\begin{table}[htbp]
\footnotesize
\caption{Simulation results for continuous outcomes under Simulation Scenario B2 with $I=30$. The working linear mixed model includes a nested exchangeable correlation structure.}\label{tb:Tablew6}
\centering
\resizebox{\textwidth}{!}{%
\begin{threeparttable}
\begin{tabular}{lclcccccccc}
\toprule
\midrule
Scenario & $I$ & Model\tnote{a} & TE\tnote{b} & Bias\tnote{c} & ESE\tnote{d} & RE\tnote{e} & ASE\textsubscript{MB}\tnote{f} & ASE\textsubscript{ROB}\tnote{g} & ECP\textsubscript{MB}\tnote{h} & ECP\textsubscript{ROB}\tnote{i} \\
\midrule
B2 & 30 & B(i) & $\Delta$ & -1.009 & 0.236 & 1.000 & 0.262 & 0.241 & 0.025 & 0.026 \\
& & B(ii) & $\Delta^{D\text{-avg}}$ & -0.007 & 0.428 & 1.000 & 0.340 & 0.397 & 0.906 & 0.944 \\
& & & $\Delta(1)$ & 0.005 & 0.238 & 1.000 & 0.268 & 0.242 & 0.981 & 0.963 \\
& & & $\Delta(2)$ & -0.007 & 0.342 & 1.000 & 0.323 & 0.327 & 0.956 & 0.951 \\
& & & $\Delta(3)$ & -0.020 & 0.484 & 1.000 & 0.392 & 0.446 & 0.906 & 0.943 \\
& & & $\Delta(4)$ & -0.012 & 0.654 & 1.000 & 0.486 & 0.611 & 0.885 & 0.947 \\
& & & $\Delta(5)$ & -0.001 & 0.975 & 1.000 & 0.648 & 0.882 & 0.835 & 0.932 \\
& & B(iii) & $\Delta$ & -1.069 & 0.186 & 1.609 & 0.220 & 0.203 & 0.001 & 0.000 \\
& & B(iv) & $\Delta^{D\text{-avg}}$ & -0.007 & 0.374 & 1.306 & 0.290 & 0.344 & 0.906 & 0.940 \\
& & & $\Delta(1)$ & 0.003 & 0.190 & 1.573 & 0.220 & 0.200 & 0.984 & 0.967 \\
& & & $\Delta(2)$ & -0.007 & 0.282 & 1.468 & 0.268 & 0.270 & 0.960 & 0.954 \\
& & & $\Delta(3)$ & -0.015 & 0.403 & 1.444 & 0.329 & 0.370 & 0.919 & 0.944 \\
& & & $\Delta(4)$ & -0.008 & 0.554 & 1.395 & 0.409 & 0.506 & 0.887 & 0.940 \\
& & & $\Delta(5)$ & -0.006 & 0.795 & 1.501 & 0.544 & 0.719 & 0.840 & 0.927 \\
& & B(v) & $\Delta$ & -1.048 & 0.177 & 1.776 & 0.213 & 0.198 & 0.001 & 0.000 \\
& & B(vi) & $\Delta^{D\text{-avg}}$ & -0.011 & 0.355 & 1.450 & 0.274 & 0.334 & 0.904 & 0.943 \\
& & & $\Delta(1)$ & 0.000 & 0.177 & 1.806 & 0.211 & 0.191 & 0.986 & 0.973 \\
& & & $\Delta(2)$ & -0.009 & 0.266 & 1.656 & 0.256 & 0.260 & 0.963 & 0.961 \\
& & & $\Delta(3)$ & -0.019 & 0.382 & 1.605 & 0.312 & 0.358 & 0.924 & 0.955 \\
& & & $\Delta(4)$ & -0.015 & 0.532 & 1.509 & 0.388 & 0.493 & 0.885 & 0.951 \\
& & & $\Delta(5)$ & -0.013 & 0.770 & 1.600 & 0.516 & 0.705 & 0.837 & 0.934 \\
\bottomrule
\end{tabular}\smallskip
\begin{tablenotes}\linespread{1}\scriptsize
\item[a] TE: treatment effect. $\Delta = 2$; $\Delta^{D\text{-avg}}$ = $\sum_{d=1}^{5} \Delta(d)/5 = 2$; $\Delta(1) = 1$; $\Delta(2) = 1.5$; $\Delta(3) = 2$; $\Delta(4) = 2.5$; $\Delta(5) = 3$. \smallskip
\item[b] Model: six working linear mixed-models under the nested exchangeable correlation structure, which are combinations of covariate adjustment settings and treatment effect settings. B(i), no adjustment with constant TE; B(ii), no adjustment with duration-specific TE; B(iii), partial adjustment with constant TE; B(iv), partial adjustment with duration-specific TE; B(v), full adjustment with constant TE; B(vi), full adjustment with duration-specific TE. \smallskip
\item[c] Bias: bias as the average of (estimate - estimand) over simulations.\smallskip
\item[d] ESE: empirical standard error (SE).\smallskip
\item[e] RE: relative effeciency.\smallskip
\item[f] ASE\textsubscript{MB}: averaged model-based SE.\smallskip
\item[g] ASE\textsubscript{ROB}: averaged robust SE.\smallskip
\item[h] ECP\textsubscript{MB}: empirical coverage probability based on the model-based SE.\smallskip
\item[i] ECP\textsubscript{ROB}: empirical coverage probability based on the robust SE.
\end{tablenotes}
\end{threeparttable}
}
\end{table}

\begin{table}[htbp]
\footnotesize
\caption{Simulation results for continuous outcomes under Simulation Scenario B2 with $I=100$. The working linear mixed model includes a simple exchangeable correlation structure.}\label{tb:Tablew7}
\centering
\resizebox{\textwidth}{!}{%
\begin{threeparttable}
\begin{tabular}{lclcccccccc}
\toprule
\midrule
Scenario & $I$ & Model\tnote{a} & TE\tnote{b} & Bias\tnote{c} & ESE\tnote{d} & RE\tnote{e} & ASE\textsubscript{MB}\tnote{f} & ASE\textsubscript{ROB}\tnote{g} & ECP\textsubscript{MB}\tnote{h} & ECP\textsubscript{ROB}\tnote{i} \\
\midrule
B2 & 100 & B(i) & $\Delta$ & -1.066 & 0.125 & 1.000 & 0.128 & 0.138 & 0.000 & 0.000 \\
& & B(ii) & $\Delta^{D\text{-avg}}$ & 0.004 & 0.228 & 1.000 & 0.179 & 0.227 & 0.877 & 0.949 \\
& & & $\Delta(1)$ & 0.000 & 0.128 & 1.000 & 0.137 & 0.137 & 0.967 & 0.964 \\
& & & $\Delta(2)$ & 0.006 & 0.179 & 1.000 & 0.166 & 0.186 & 0.932 & 0.955 \\
& & & $\Delta(3)$ & -0.002 & 0.246 & 1.000 & 0.203 & 0.254 & 0.895 & 0.961 \\
& & & $\Delta(4)$ & 0.001 & 0.356 & 1.000 & 0.253 & 0.350 & 0.847 & 0.950 \\
& & & $\Delta(5)$ & 0.014 & 0.538 & 1.000 & 0.336 & 0.512 & 0.786 & 0.936 \\
& & B(iii) & $\Delta$ & -1.131 & 0.101 & 1.540 & 0.099 & 0.117 & 0.000 & 0.000 \\
& & B(iv) & $\Delta^{D\text{-avg}}$ & 0.005 & 0.198 & 1.330 & 0.145 & 0.202 & 0.853 & 0.961 \\
& & & $\Delta(1)$ & 0.003 & 0.102 & 1.578 & 0.106 & 0.115 & 0.963 & 0.978 \\
& & & $\Delta(2)$ & 0.005 & 0.146 & 1.502 & 0.131 & 0.156 & 0.924 & 0.971 \\
& & & $\Delta(3)$ & 0.001 & 0.206 & 1.430 & 0.162 & 0.215 & 0.880 & 0.966 \\
& & & $\Delta(4)$ & 0.002 & 0.296 & 1.452 & 0.202 & 0.296 & 0.829 & 0.957 \\
& & & $\Delta(5)$ & 0.014 & 0.439 & 1.504 & 0.268 & 0.424 & 0.771 & 0.949 \\
& & B(v) & $\Delta$ & -1.122 & 0.097 & 1.645 & 0.095 & 0.114 & 0.000 & 0.000 \\
& & B(vi) & $\Delta^{D\text{-avg}}$ & 0.005 & 0.191 & 1.426 & 0.137 & 0.196 & 0.838 & 0.963 \\
& & & $\Delta(1)$ & 0.003 & 0.097 & 1.745 & 0.101 & 0.110 & 0.969 & 0.984 \\
& & & $\Delta(2)$ & 0.004 & 0.140 & 1.631 & 0.124 & 0.151 & 0.928 & 0.975 \\
& & & $\Delta(3)$ & 0.002 & 0.200 & 1.519 & 0.153 & 0.209 & 0.860 & 0.966 \\
& & & $\Delta(4)$ & 0.003 & 0.286 & 1.545 & 0.192 & 0.289 & 0.811 & 0.957 \\
& & & $\Delta(5)$ & 0.012 & 0.427 & 1.585 & 0.254 & 0.416 & 0.746 & 0.944 \\
\bottomrule
\end{tabular}\smallskip
\begin{tablenotes}\linespread{1}\scriptsize
\item[a] TE: treatment effect. $\Delta = 2$; $\Delta^{D\text{-avg}}$ = $\sum_{d=1}^{5} \Delta(d)/5 = 2$; $\Delta(1) = 1$; $\Delta(2) = 1.5$; $\Delta(3) = 2$; $\Delta(4) = 2.5$; $\Delta(5) = 3$. \smallskip
\item[b] Model: six working linear mixed-models under the simple exchangeable correlation structure, which are combinations of covariate adjustment settings and treatment effect settings. B(i), no adjustment with constant TE; B(ii), no adjustment with duration-specific TE; B(iii), partial adjustment with constant TE; B(iv), partial adjustment with duration-specific TE; B(v), full adjustment with constant TE; B(vi), full adjustment with duration-specific TE. \smallskip
\item[c] Bias: bias as the average of (estimate - estimand) over simulations.\smallskip
\item[d] ESE: empirical standard error (SE).\smallskip
\item[e] RE: relative effeciency.\smallskip
\item[f] ASE\textsubscript{MB}: averaged model-based SE.\smallskip
\item[g] ASE\textsubscript{ROB}: averaged robust SE.\smallskip
\item[h] ECP\textsubscript{MB}: empirical coverage probability based on the model-based SE.\smallskip
\item[i] ECP\textsubscript{ROB}: empirical coverage probability based on the robust SE.
\end{tablenotes}
\end{threeparttable}
}
\end{table}

\begin{table}[htbp]
\footnotesize
\caption{Simulation results for continuous outcomes under Simulation Scenario B2 with $I=100$. The working linear mixed model includes a nested exchangeable correlation structure.}\label{tb:Tablew8}
\centering
\resizebox{\textwidth}{!}{%
\begin{threeparttable}
\begin{tabular}{lclcccccccc}
\toprule
\midrule
Scenario & $I$ & Model\tnote{a} & TE\tnote{b} & Bias\tnote{c} & ESE\tnote{d} & RE\tnote{e} & ASE\textsubscript{MB}\tnote{f} & ASE\textsubscript{ROB}\tnote{g} & ECP\textsubscript{MB}\tnote{h} & ECP\textsubscript{ROB}\tnote{i} \\
\midrule
B2 & 100 & B(i) & $\Delta$ & -1.013 & 0.123 & 1.000 & 0.144 & 0.137 & 0.000 & 0.000 \\
& & B(ii) & $\Delta^{D\text{-avg}}$ & 0.003 & 0.230 & 1.000 & 0.189 & 0.228 & 0.897 & 0.944 \\
& & & $\Delta(1)$ & 0.000 & 0.129 & 1.000 & 0.147 & 0.137 & 0.978 & 0.961 \\
& & & $\Delta(2)$ & 0.006 & 0.179 & 1.000 & 0.178 & 0.186 & 0.951 & 0.956 \\
& & & $\Delta(3)$ & -0.003 & 0.247 & 1.000 & 0.217 & 0.255 & 0.914 & 0.960 \\
& & & $\Delta(4)$ & 0.000 & 0.358 & 1.000 & 0.269 & 0.353 & 0.869 & 0.953 \\
& & & $\Delta(5)$ & 0.013 & 0.541 & 1.000 & 0.358 & 0.516 & 0.819 & 0.935 \\
& & B(iii) & $\Delta$ & -1.070 & 0.099 & 1.542 & 0.122 & 0.115 & 0.000 & 0.000 \\
& & B(iv) & $\Delta^{D\text{-avg}}$ & 0.004 & 0.199 & 1.330 & 0.162 & 0.203 & 0.903 & 0.962 \\
& & & $\Delta(1)$ & 0.003 & 0.102 & 1.603 & 0.122 & 0.114 & 0.989 & 0.982 \\
& & & $\Delta(2)$ & 0.004 & 0.146 & 1.501 & 0.149 & 0.156 & 0.966 & 0.971 \\
& & & $\Delta(3)$ & 0.000 & 0.207 & 1.423 & 0.183 & 0.215 & 0.922 & 0.967 \\
& & & $\Delta(4)$ & 0.001 & 0.298 & 1.451 & 0.228 & 0.298 & 0.868 & 0.958 \\
& & & $\Delta(5)$ & 0.013 & 0.442 & 1.503 & 0.302 & 0.427 & 0.828 & 0.945 \\
& & B(v) & $\Delta$ & -1.050 & 0.096 & 1.648 & 0.118 & 0.112 & 0.000 & 0.000 \\
& & B(vi) & $\Delta^{D\text{-avg}}$ & 0.004 & 0.192 & 1.427 & 0.154 & 0.197 & 0.891 & 0.966 \\
& & & $\Delta(1)$ & 0.003 & 0.096 & 1.785 & 0.117 & 0.109 & 0.989 & 0.981 \\
& & & $\Delta(2)$ & 0.004 & 0.141 & 1.627 & 0.142 & 0.150 & 0.964 & 0.973 \\
& & & $\Delta(3)$ & 0.001 & 0.201 & 1.517 & 0.174 & 0.209 & 0.926 & 0.965 \\
& & & $\Delta(4)$ & 0.002 & 0.289 & 1.542 & 0.217 & 0.291 & 0.852 & 0.953 \\
& & & $\Delta(5)$ & 0.011 & 0.430 & 1.581 & 0.288 & 0.419 & 0.816 & 0.943 \\
\bottomrule
\end{tabular}\smallskip
\begin{tablenotes}\linespread{1}\scriptsize
\item[a] TE: treatment effect. $\Delta = 2$; $\Delta^{D\text{-avg}}$ = $\sum_{d=1}^{5} \Delta(d)/5 = 2$; $\Delta(1) = 1$; $\Delta(2) = 1.5$; $\Delta(3) = 2$; $\Delta(4) = 2.5$; $\Delta(5) = 3$. \smallskip
\item[b] Model: six working linear mixed-models under the nested exchangeable correlation structure, which are combinations of covariate adjustment settings and treatment effect settings. B(i), no adjustment with constant TE; B(ii), no adjustment with duration-specific TE; B(iii), partial adjustment with constant TE; B(iv), partial adjustment with duration-specific TE; B(v), full adjustment with constant TE; B(vi), full adjustment with duration-specific TE. \smallskip
\item[c] Bias: bias as the average of (estimate - estimand) over simulations.\smallskip
\item[d] ESE: empirical standard error (SE).\smallskip
\item[e] RE: relative effeciency.\smallskip
\item[f] ASE\textsubscript{MB}: averaged model-based SE.\smallskip
\item[g] ASE\textsubscript{ROB}: averaged robust SE.\smallskip
\item[h] ECP\textsubscript{MB}: empirical coverage probability based on the model-based SE.\smallskip
\item[i] ECP\textsubscript{ROB}: empirical coverage probability based on the robust SE.
\end{tablenotes}
\end{threeparttable}
}
\end{table}

\clearpage

\subsection{Simulation design C with results}
In this simulation, we set $I \in \{30, 100, 300, 1000\}$, $N_i = 5000$, and $J = 3$. We consider two covariates, a binary covariate $X_{ik1} \sim \mathcal{B}(0.05)$ and a continuous covariate $X_{ik2} = e_{i2} + f_{ik2}$ with $e_{ik} \sim N(0,0.025)$ and $f_{ik2} \sim N(0,0.1)$. Each potential outcome $Y_{ijk}(z)$ is independently sampled from $\mathcal{B}(\mu_{ijk}(d))$, where $d=I\{z>0\}(j-z+1)$ and $\mu_{ijk}(z)$ follows
\begin{align*}
    \text{logit}(\mu_{ijk}(d)) &= \beta_{0j} + \beta_{Zd}(X) + \left(X_{ik1} + \frac{j}{J} X_{ik2}^2\right) + \alpha_i + \gamma_{ij},
\end{align*}
where
$\beta_{0j} = 0.08+0.02j$ is the period effect, 
$\alpha_i \sim N(0, 0.015)$ is the cluster-level random effect, $\gamma_{ij} \sim N(0, 0.01)$ is the cluster-period random effect, and $\beta_{Zd}(X) = I\{d>0\}\left\{0.72 + 0.18d + \frac{1}{4}(X_{ik1} - \overline{X}_{i\cdot 1}) + \frac{d}{10}(X_{ik2}^2 - \overline{X^2}_{i\cdot 2}) + \beta_i\right\}$ encodes a duration-specific treatment effect on the logit scale.
Within $\beta_{Zd}(X)$, $\beta_i \sim N(0, 0.25)$ under scenario C1 and $\beta_i \sim N(0, 0.04)$ under scenario C2 represents a random intervention effect. Using $N_{ij}$ and $Z_i$ generated as in simulation design A, we generate the observed outcome for the set of enrolled individuals as $Y_{ijk}=\sum_{z=1}^{j} I\{Z_i = z\}Y_{ijk}(d) + I\{Z_i > j\}Y_{ijk}(0)$. 

We generate $1000$ data replicates to estimate the saturated treatment effect on the odds ratio scale, i.e., the marginal estimand of the form
$$\Phi_j(d) = \frac{E[Y_{ijk}(j-d+1)]}{1-E[Y_{ijk}(j-d+1)]}\Big/\frac{E[Y_{ijk}(0)]}{1-E[Y_{ijk}(0)]},~~~\text{for}~~~(j,d) \in \{(1,1), (2,1), (2,2)\}.$$ 
Given each simulated replicate, we fit 6 models. The first two are linear mixed models with the saturated treatment effect structure and nested exchangeable correlation structure. The next two models are GEE with independence correlation structure and the saturated treatment effect structure. The last two models are generalized linear mixed models (GLMMs) with the saturated treatment effect structure and nested exchangeable random-effect structure, with a logit link function (following common practice and approximating the data-generating distribution to some extent). For each of the three types of models, we specify the first one without covariate adjustment and the other one adjusting for $(X_{ik1},X_{ik2})$ as linear terms. For linear mixed models and GEE, we estimate $\Phi_j(d)$ and compute the variance estimators as described in Section~6 of the main paper. For GLMMs, we use $e^{\widehat{\beta}_{jZd}}$ with model-based variance estimators to make inferences about $\Phi_j(d)$ as a comparison. Although in theory $e^{\widehat{\beta}_{jZd}}$ does not always target marginal estimand $\Phi_j(d)$ defined with potential outcomes, inference with $e^{\widehat{\beta}_{jZd}}$ and its model-based variance estimator is common practice in published trials; thus our goal to illustrate the performance characteristics of this model-based approach when the actual target estimand is a marginal, model-free one. To simplify the presentation of results, we only report bias, the ratio of averaged standard error estimators over empirical standard errors, and the coverage probability of 95\%
confidence intervals via normal approximations.

Web Table \ref{tb:Tablew9} presents the results of the third simulation study for estimators without covariate adjustment; the performance of estimators with covariate adjustment is similar, and is provided in Web Table \ref{tb:Tablew10}.
For both mixed models and GEE, the bias is negligible, the average estimated standard error roughly matches the true empirical standard error, and the coverage probabilities are nominal, demonstrating the feasibility of using these two working models to accurately capture the marginal treatment effect estimands on the odds ratio scale, even if the working models are misspecified. 
In contrast, GLMM estimators show relatively large biases and underestimation of the empirical standard error, leading to low coverage probability, especially when the number of clusters is large (e.g., 30\% under coverage). Beyond this setting, we have also conducted additional simulations to estimate the marginal treatment effect estimands defined on the risk difference or relative risk scales for binary outcomes. The patterns are generally similar and the results are thus omitted for brevity. 


\begin{table}[htbp]
\caption{Simulation scenarios and results for binary outcomes under Simulation Design C without covariate adjustment. The working linear mixed model includes a nested exchangeable correlation structure and the GEE is fitted under working independence.}\label{tb:Tablew9}
\centering
\resizebox{\textwidth}{!}{%
\begin{threeparttable}
\begin{tabular}{lcccccccccccc}
\toprule
\midrule
\multirow{2}[3]{*}{Scenario} & \multirow{2}[3]{*}{$I$} & \multirow{2}[3]{*}{TE\tnote{a}} & \multirow{2}[3]{*}{Estimand} & \multicolumn{3}{c}{Linear Mixed Model} & \multicolumn{3}{c}{GEE} & \multicolumn{3}{c}{GLMM\tnote{b}} \\
\cmidrule(lr){5-7} \cmidrule(lr){8-10} \cmidrule(lr){11-13}
& & & & Bias\tnote{c} & ASE/ESE\tnote{d} & ECP\tnote{e} & Bias\tnote{c} & ASE/ESE\tnote{d} & ECP\tnote{e} & Bias\tnote{c} & ASE/ESE\tnote{d} & ECP\tnote{e} \\
\midrule
C1 & 30 & $\Phi_1(1)$ & 2.299 & 0.061 & 0.891 & 0.915 & 0.060 & 0.884 & 0.913 & 0.153 & 0.801 & 0.899 \\
& & $\Phi_2(1)$ & 2.292 & 0.075 & 0.901 & 0.915 & 0.069 & 0.897 & 0.921 & 0.133 & 0.775 & 0.899 \\
& & $\Phi_2(2)$ & 2.716 & 0.089 & 0.897 & 0.910 & 0.092 & 0.876 & 0.916 & 0.216 & 0.858 & 0.928 \\
& 100 & $\Phi_1(1)$ & 2.299 & 0.024 & 0.989 & 0.952 & 0.023 & 0.982 & 0.953 & 0.127 & 0.883 & 0.911 \\
& & $\Phi_2(1)$ & 2.292 & 0.008 & 0.951 & 0.935 & 0.004 & 0.984 & 0.937 & 0.079 & 0.821 & 0.887 \\
& & $\Phi_2(2)$ & 2.716 & 0.010 & 0.961 & 0.934 & 0.009 & 0.974 & 0.930 & 0.151 & 0.913 & 0.917 \\
& 300 & $\Phi_1(1)$ & 2.299 & 0.008 & 0.977 & 0.941 & 0.007 & 0.973 & 0.939 & 0.112 & 0.882 & 0.866 \\
& & $\Phi_2(1)$ & 2.292 & 0.006 & 0.989 & 0.952 & 0.005 & 0.997 & 0.949 & 0.080 & 0.858 & 0.886 \\
& & $\Phi_2(2)$ & 2.716 & 0.004 & 0.991 & 0.939 & 0.002 & 0.989 & 0.941 & 0.147 & 0.946 & 0.881 \\
& 1000 & $\Phi_1(1)$ & 2.299 & 0.002 & 1.007 & 0.952 & 0.002 & 1.012 & 0.950 & 0.107 & 0.905 & 0.674 \\
& & $\Phi_2(1)$ & 2.292 & 0.002 & 0.926 & 0.932 & 0.000 & 0.929 & 0.927 & 0.076 & 0.801 & 0.771 \\
& & $\Phi_2(2)$ & 2.716 & 0.005 & 1.004 & 0.952 & 0.005 & 1.007 & 0.952 & 0.151 & 0.958 & 0.681 \\\midrule
C2 & 30 & $\Phi_1(1)$ & 2.340 & 0.032 & 0.924 & 0.909 & 0.032 & 0.914 & 0.907 & 0.079 & 0.858 & 0.911 \\
& & $\Phi_2(1)$ & 2.311 & 0.045 & 0.904 & 0.920 & 0.040 & 0.920 & 0.924 & 0.077 & 0.749 & 0.879 \\
& & $\Phi_2(2)$ & 2.732 & 0.076 & 0.858 & 0.910 & 0.074 & 0.858 & 0.910 & 0.139 & 0.792 & 0.910 \\
& 100 & $\Phi_1(1)$ & 2.340 & 0.029 & 0.979 & 0.942 & 0.029 & 0.975 & 0.947 & 0.080 & 0.892 & 0.924 \\
& & $\Phi_2(1)$ & 2.311 & 0.004 & 0.996 & 0.946 & 0.000 & 1.020 & 0.940 & 0.041 & 0.779 & 0.889 \\
& & $\Phi_2(2)$ & 2.732 & 0.009 & 0.991 & 0.949 & 0.007 & 0.996 & 0.941 & 0.076 & 0.882 & 0.916 \\
& 300 & $\Phi_1(1)$ & 2.340 & 0.003 & 0.986 & 0.948 & 0.003 & 0.982 & 0.947 & 0.054 & 0.906 & 0.903 \\
& & $\Phi_2(1)$ & 2.311 & 0.002 & 0.998 & 0.946 & 0.002 & 0.981 & 0.937 & 0.039 & 0.778 & 0.877 \\
& & $\Phi_2(2)$ & 2.732 & -0.002 & 0.974 & 0.944 & -0.002 & 0.971 & 0.938 & 0.066 & 0.870 & 0.888 \\
& 1000 & $\Phi_1(1)$ & 2.340 & 0.002 & 0.999 & 0.941 & 0.002 & 1.006 & 0.950 & 0.054 & 0.914 & 0.869 \\
& & $\Phi_2(1)$ & 2.311 & 0.001 & 0.924 & 0.928 & -0.001 & 0.933 & 0.933 & 0.037 & 0.718 & 0.815 \\
& & $\Phi_2(2)$ & 2.732 & 0.005 & 1.027 & 0.947 & 0.004 & 1.026 & 0.954 & 0.074 & 0.910 & 0.856 \\
\bottomrule
\end{tabular}\smallskip
\begin{tablenotes}\linespread{1}\small
\item[a] TE: treatment effect. \smallskip
\item[b] GLMM: Generalized linear mixed model. \smallskip
\item[c] Bias: bias as the average of (estimate - estimand) over simulations. \smallskip
\item[d] ASE/ESE: averaged standard error (SE) divided by empirical SE.\smallskip
\item[e] ECP: empirical coverage probability.
\end{tablenotes}
\end{threeparttable}
}
\end{table}

\begin{table}[htbp]
\caption{Simulation scenarios and results for binary outcomes under Simulation Design C with covariate adjustment. The working linear mixed model includes a nested exchangeable correlation structure and the GEE is fitted under working independence.}\label{tb:Tablew10}
\centering
\resizebox{\textwidth}{!}{%
\begin{threeparttable}
\begin{tabular}{lcccccccccccc}
\toprule
\midrule
\multirow{2}[3]{*}{Scenario} & \multirow{2}[3]{*}{$I$} & \multirow{2}[3]{*}{TE\tnote{a}} & \multirow{2}[3]{*}{Estimand} & \multicolumn{3}{c}{Linear Mixed Model} & \multicolumn{3}{c}{GEE} & \multicolumn{3}{c}{GLMM\tnote{b}} \\
\cmidrule(lr){5-7} \cmidrule(lr){8-10} \cmidrule(lr){11-13}
& & & & Bias\tnote{c} & ASE/ESE\tnote{d} & ECP\tnote{e} & Bias\tnote{c} & ASE/ESE\tnote{d} & ECP\tnote{e} & Bias\tnote{c} & ASE/ESE\tnote{d} & ECP\tnote{e} \\
\midrule
C1 & 30 & $\Phi_1(1)$ & 2.299 & 0.060 & 0.893 & 0.913 & 0.059 & 0.880 & 0.910 & 0.170 & 0.801 & 0.906 \\
& & $\Phi_2(1)$ & 2.291 & 0.074 & 0.894 & 0.918 & 0.068 & 0.897 & 0.924 & 0.149 & 0.777 & 0.898 \\
& & $\Phi_2(2)$ & 2.716 & 0.090 & 0.897 & 0.914 & 0.093 & 0.873 & 0.913 & 0.240 & 0.860 & 0.935 \\
& 100 & $\Phi_1(1)$ & 2.299 & 0.024 & 0.991 & 0.956 & 0.023 & 0.981 & 0.952 & 0.143 & 0.883 & 0.909 \\
& & $\Phi_2(1)$ & 2.291 & 0.008 & 0.952 & 0.931 & 0.004 & 0.983 & 0.939 & 0.094 & 0.821 & 0.887 \\
& & $\Phi_2(2)$ & 2.716 & 0.010 & 0.964 & 0.930 & 0.009 & 0.973 & 0.930 & 0.172 & 0.914 & 0.913 \\
& 300 & $\Phi_1(1)$ & 2.299 & 0.008 & 0.977 & 0.941 & 0.008 & 0.972 & 0.942 & 0.127 & 0.881 & 0.846 \\
& & $\Phi_2(1)$ & 2.291 & 0.007 & 0.989 & 0.955 & 0.006 & 0.995 & 0.949 & 0.095 & 0.857 & 0.875 \\
& & $\Phi_2(2)$ & 2.716 & 0.004 & 0.994 & 0.942 & 0.003 & 0.989 & 0.943 & 0.169 & 0.949 & 0.862 \\
& 1000 & $\Phi_1(1)$ & 2.299 & 0.002 & 1.006 & 0.954 & 0.002 & 1.009 & 0.949 & 0.122 & 0.905 & 0.625 \\
& & $\Phi_2(1)$ & 2.291 & 0.002 & 0.929 & 0.928 & 0.000 & 0.930 & 0.934 & 0.090 & 0.803 & 0.739 \\
& & $\Phi_2(2)$ & 2.716 & 0.005 & 1.007 & 0.951 & 0.005 & 1.008 & 0.953 & 0.171 & 0.960 & 0.611 \\\midrule
C2 & 30 & $\Phi_1(1)$ & 2.340 & 0.036 & 0.925 & 0.910 & 0.036 & 0.911 & 0.904 & 0.113 & 0.856 & 0.902 \\
& & $\Phi_2(1)$ & 2.311 & 0.044 & 0.904 & 0.919 & 0.040 & 0.914 & 0.927 & 0.103 & 0.747 & 0.884 \\
& & $\Phi_2(2)$ & 2.732 & 0.082 & 0.860 & 0.911 & 0.079 & 0.854 & 0.910 & 0.181 & 0.791 & 0.909 \\
& 100 & $\Phi_1(1)$ & 2.340 & 0.030 & 0.983 & 0.947 & 0.030 & 0.977 & 0.947 & 0.110 & 0.894 & 0.922 \\
& & $\Phi_2(1)$ & 2.311 & 0.004 & 1.000 & 0.945 & 0.001 & 1.017 & 0.940 & 0.067 & 0.779 & 0.885 \\
& & $\Phi_2(2)$ & 2.732 & 0.010 & 0.995 & 0.945 & 0.008 & 0.996 & 0.946 & 0.112 & 0.884 & 0.916 \\
& 300 & $\Phi_1(1)$ & 2.340 & 0.003 & 0.987 & 0.950 & 0.003 & 0.982 & 0.944 & 0.083 & 0.906 & 0.877 \\
& & $\Phi_2(1)$ & 2.311 & 0.002 & 0.998 & 0.948 & 0.002 & 0.980 & 0.940 & 0.064 & 0.776 & 0.840 \\
& & $\Phi_2(2)$ & 2.732 & -0.001 & 0.975 & 0.945 & -0.002 & 0.969 & 0.938 & 0.100 & 0.871 & 0.882 \\
& 1000 & $\Phi_1(1)$ & 2.340 & 0.001 & 1.000 & 0.944 & 0.002 & 1.006 & 0.947 & 0.081 & 0.914 & 0.764 \\
& & $\Phi_2(1)$ & 2.311 & 0.001 & 0.928 & 0.931 & 0.000 & 0.936 & 0.937 & 0.062 & 0.720 & 0.750 \\
& & $\Phi_2(2)$ & 2.732 & 0.005 & 1.030 & 0.946 & 0.005 & 1.026 & 0.955 & 0.107 & 0.912 & 0.771 \\
\bottomrule
\end{tabular}\smallskip
\begin{tablenotes}\linespread{1}\small
\item[a] TE: treatment effect. \smallskip
\item[b] GLMM: Generalized linear mixed model. \smallskip
\item[c] Bias: bias as the average of (estimate - estimand) over simulations. \smallskip
\item[d] ASE/ESE: averaged standard error (SE) divided by empirical SE.\smallskip
\item[e] ECP: empirical coverage probability.
\end{tablenotes}
\end{threeparttable}
}
\end{table}

\clearpage 

\subsection{Additional simulations when the working linear mixed models include a discrete-time exponential decay correlation structure}

In Section 4, our general theoretical development focuses on linear mixed models with independence, exchangeable, or nested exchangeable correlation structure, and we have not addressed other possible correlation structure specifications. This is primarily because the score functions under other types of working correlation structure specifications often do not have a tractable analytical form, posing challenges in deriving explicit asymptotic results for studying model robustness. This is the case with the linear mixed models coupled with the discrete-time exponential decay correlation structure proposed in \citet{kasza2019impact} and \citet{kasza2019inference}. Therefore, we empirically explore, under the same data-generating distributions outlined in Section 7 of the main paper, whether linear mixed models coupled with the discrete-time exponential decay correlation structure can still recover the marginal treatment effect estimands (defined under the potential outcomes framework in Sections 3 and 6) even when the working models are misspecified. We caution that this set of simulations is by no means exhaustive, and only generates preliminary insights on robustness features for more complicated model specifications. We hope this set of simulations can inspire more future simulation studies to address robustness properties of model misspecification in SW-CRTs. 

To further elaborate, recall that the linear mixed model that is commonly used in analyzing SW-CRTs with continuous outcomes is given in Section 4 as:
\begin{equation*}
    Y_{ijk} = \beta_{0j} + TE_{ij}  + \bbeta_{\bX}^\top \bX_{ik} + RE_{ij} + \varepsilon_{ijk}.
\end{equation*}
Under the discrete-time exponential decay correlation structure, the collection of random effects in cluster $i$ is assumed to follow $(RE_{i1},\ldots,RE_{iJ})\sim N(0,\tau^2{\bM})$, and $\bM$ is a first-order autoregressive matrix parameterized by a decay parameter $r$:
\begin{equation}\label{eq:MAR1}
{\bM}=\left(
       \begin{array}{ccccc}
         1 & r & r^2 & \ldots & r^{J-1} \\
         r & 1 & r & \ldots & r^{J-2} \\
         \vdots & \vdots & \vdots & \ddots & \vdots \\
         r^{J-1} & r^{J-2} & r^{J-3} & \ldots & 1 \\
       \end{array}
     \right).
\end{equation}
Clearly this structure includes the simple exchangeable correlation structure as a special case by setting $r=1$, but does not have a clear nesting relationship with the nested exchangeable correlation structure. We follow the implementation in \citet{ouyang2023maintaining} to estimate all model parameters (via the \texttt{glmmTMB} R package), and extract the component in $TE_{ij}$ as in Section 4 to estimate the marginal estimands under each treatment effect structure specification. Because the likelihood function is more complicated under the exponential decay structure, we propose a GEE-style approximate sandwich variance estimator that does not further the uncertainty due to estimating the variance and correlation parameters, in a similar spirit to \citet{li2020design} and \citet{li2022marginal}. In brief and resuming notation from Section \ref{sec:proof-LMM} to the extent possible, we denote the estimators for the conditional mean structure (period effects, treatment effects, and covariate effects) as $\widehat\bbeta$ and those for the remaining variance and correlation parameters as $\widehat\btheta_{-\bbeta}$, and the approximate sandwich variance estimator for $\bbeta$ (sufficient for deriving estimators for our proposed marginal estimands) is given by
\begin{equation*}
   \widehat{\bfV}_{\bbeta} = \left\{ \sum_{i=1}^I \frac{\partial \bpsi_1(\bO_i,\widehat\bbeta,\widehat\btheta_{-\bbeta})}{\partial \bbeta}  \right\}^{-1} 
    \left\{\sum_{i=1}^I \bpsi_1(\bO_i,\widehat\bbeta,\widehat\btheta_{-\bbeta})\bpsi_1(\bO_i,\widehat\bbeta,\widehat\btheta_{-\bbeta})^\top\right\}
    \left\{ \sum_{i=1}^I \frac{\partial \bpsi_1(\bO_i,\widehat\bbeta,\widehat\btheta_{-\bbeta})}{\partial \bbeta}  \right\}^{-1},
\end{equation*}
where $\bpsi_1(\bO_i,\bbeta,\btheta_{-\bbeta})=2\bfQ_i^{\top}\bfV_i(\bY_i - \bfQ_i\bbeta)$, $\bfV_i = \mathbf{D}_i (\mathbf{D}_i^\top\bfSigma_i \mathbf{D}_i)^{-1} \mathbf{D}_i^\top \in \mathbb{R}^{N_iJ\times N_iJ}$ and $\bfSigma_i$ now is the marginal covariance matrix for $\bY_i|Z_i,\bX_i,N_i$ under the discrete-time exponential decay correlation structure (involving $\bM$ and decay parameter $r$). The R code to implement the point and variance estimators is available at \url{https://github.com/BingkaiWang/SW-CRT}.

Below is a summary of tables that present the simulation results, in parallel to the main simulations we have reported in Section 6 of the main paper. 

\begin{itemize}
    \item Web Table \ref{tb:Tablew11} presents the results for continuous outcomes, from Simulation Scenarios A1 and A2.
    \item Web Tables \ref{tb:Tablew12}-\ref{tb:Tablew13} present the results for continuous outcomes, from Simulation Scenario B1.
    \begin{itemize}
        \item Web Table \ref{tb:Tablew12} refers to $I=30$.
        \item Web Table \ref{tb:Tablew13} refers to $I=100$.
    \end{itemize}
    \item Web Tables \ref{tb:Tablew14}-\ref{tb:Tablew15} present the results for continuous outcomes, from Simulation Scenario B2.
    \begin{itemize}
        \item Web Table \ref{tb:Tablew14} refers to $I=30$.
        \item Web Table \ref{tb:Tablew15} refers to $I=100$.
    \end{itemize}
    \item Web Tables \ref{tb:Tablew16}-\ref{tb:Tablew17} present the results for binary outcomes, from Simulation Scenarios C1 and C2.
    \begin{itemize}
        \item Web Table \ref{tb:Tablew16} refers to results without covariate adjustment.
        \item Web Table \ref{tb:Tablew17} refers to results with covariate adjustment.
    \end{itemize}
\end{itemize}

Overall, our main finding is that using a linear mixed model with a discrete-time exponential decay correlation correlation structure yields similar results in terms of bias and coverage to our proposed working linear mixed models with an exchangeable or nested exchangeable correlation structure (despite slight under-coverage with $I=30$ due to small number of clusters). The relative efficiency by comparing estimators from different working correlation models is generally close to $1$ regardless of simulation scenarios and treatment effect structure settings. Thus the linear mixed models with a discrete-time exponential decay correlation structure remains empirically robust to misspecification within the settings we examined in our simulations. Finally, when estimating marginal odds ratio estimands with binary outcomes, using a g-computation estimator from linear mixed models with a discrete-time exponential decay correlation correlation structure yields small bias and close to nominal coverage, even when the working model is misspecified. In contrast, the model-based coefficient estimator from generalized linear mixed model with a logistic link and a discrete-time exponential decay correlation correlation structure is biased to the marginal estimands, and this bias does not vanish as the number of clusters $I$ increases. This bias also leads to under-coverage (Web Tables 16 and 17).
\begin{table}[htbp]
\caption{Simulation scenarios and results for continuous outcomes under Simulation Scenario A. The working linear mixed model includes a discrete-time exponential decay correlation structure.}\label{tb:Tablew11}
\centering
\resizebox{\textwidth}{!}{%
\begin{threeparttable}
\begin{tabular}{lclccccccccccc}
\toprule
\midrule
\multirow{2}[3]{*}{Scenario} & \multirow{2}[3]{*}{$I$} & \multirow{2}[3]{*}{Model\tnote{a}} & \multirow{2}[3]{*}{Bias\tnote{b}} & \multirow{2}[3]{*}{ESE\tnote{c}} & \multirow{2}[3]{*}{RE\tnote{d}} & \multirow{2}[3]{*}{ASE\textsubscript{MB}\tnote{e}} & \multirow{2}[3]{*}{ASE\textsubscript{ROB}\tnote{f}} & \multirow{2}[3]{*}{ECP\textsubscript{MB}\tnote{g}} & \multirow{2}[3]{*}{ECP\textsubscript{ROB}\tnote{h}} & \multicolumn{2}{c}{ED/SE\tnote{i}} & \multicolumn{2}{c}{ED/NE\tnote{j}} \\
\cmidrule(lr){11-12} \cmidrule(lr){13-14}
& & & & & & & & & & Bias & ESE & Bias & ESE \\
\midrule
A1 & 30 & A(vii) & 0.009 & 0.206 & 1.000 & 0.226 & 0.196 & 0.978 & 0.943 & 1.031 & 0.993 & 0.897 & 0.994 \\
& & A(viii) & 0.005 & 0.162 & 1.616 & 0.185 & 0.155 & 0.983 & 0.951 & 1.295 & 0.976 & 1.018 & 0.975 \\
& & A(ix) & 0.003 & 0.154 & 1.784 & 0.176 & 0.146 & 0.986 & 0.953 & 1.507 & 0.979 & 1.038 & 0.979 \\
& 100 & A(vii) & 0.005 & 0.118 & 1.000 & 0.124 & 0.112 & 0.961 & 0.926 & 0.996 & 0.995 & 0.997 & 0.993 \\
& & A(viii) & 0.006 & 0.093 & 1.593 & 0.102 & 0.089 & 0.967 & 0.938 & 1.010 & 0.989 & 1.037 & 0.988 \\
& & A(ix) & 0.005 & 0.087 & 1.822 & 0.097 & 0.083 & 0.971 & 0.943 & 1.035 & 0.983 & 1.059 & 0.983 \\\midrule
A2 & 30 & A(vii) & 0.001 & 0.228 & 1.000 & 0.256 & 0.229 & 0.978 & 0.958 & 1.161 & 0.989 & 3.884 & 0.993 \\
& & A(viii) & 0.001 & 0.175 & 1.690 & 0.206 & 0.188 & 0.988 & 0.969 & 1.331 & 0.980 & 1.315 & 0.984 \\
& & A(ix) & -0.001 & 0.164 & 1.926 & 0.198 & 0.180 & 0.985 & 0.971 & 0.437 & 0.979 & 0.414 & 0.988 \\
& 100 & A(vii) & 0.000 & 0.121 & 1.000 & 0.140 & 0.131 & 0.979 & 0.966 & 13.809 & 0.994 & 16.435 & 0.997 \\
& & A(viii) & 0.002 & 0.093 & 1.670 & 0.113 & 0.108 & 0.984 & 0.978 & 1.257 & 0.982 & 1.301 & 0.990 \\
& & A(ix) & 0.002 & 0.089 & 1.824 & 0.109 & 0.103 & 0.985 & 0.977 & 1.301 & 0.984 & 1.319 & 0.993 \\
\bottomrule
\end{tabular}\smallskip
\begin{tablenotes}\small
\item[a] Model: three working linear mixed-models under the exponential decay correlation structure, with different covariate adjustment settings. A (vii), no adjustment; A (viii), partial adjustment; A (ix), full adjustment.\smallskip
\item[b] Bias: averaged bias (= estimate - estimand, where estimand = 2 across all scenarios).\smallskip
\item[c] ESE: empirical standard error (SE).\smallskip
\item[d] RE: relative effeciency.\smallskip
\item[e] ASE\textsubscript{MB}: averaged model-based SE.\smallskip
\item[f] ASE\textsubscript{ROB}: averaged robust sandwich SE.\smallskip
\item[g] ECP\textsubscript{MB}: empirical coverage probability based on the model-based SE.\smallskip
\item[h] ECP\textsubscript{ROB}: empirical coverage probability based on the robust sandwich SE.\smallskip
\item[i] ED: exponential decay correlation structure. SE: simple exchangeable correlation structure. ED/SE: ratio of Bias or ESE between the models under ED and SE.\smallskip
\item[j] ED: exponential decay correlation structure. NE: nested exchangeable correlation structure. ED/NE: ratio of Bias or ESE between the models under ED and NE.
\end{tablenotes}
\end{threeparttable}
}
\end{table}

\begin{table}[htbp]
\footnotesize
\caption{Simulation results for continuous outcomes under Simulation Scenario B1 with $I=30$. The working linear mixed model includes a discrete-time exponential decay correlation structure.}\label{tb:Tablew12}
\centering
\resizebox{\textwidth}{!}{%
\begin{threeparttable}
\begin{tabular}{lclcccccccccccc}
\toprule
\midrule
\multirow{2}[3]{*}{Scenario} & \multirow{2}[3]{*}{$I$} & \multirow{2}[3]{*}{Model\tnote{a}} & \multirow{2}[3]{*}{TE\tnote{b}} & \multirow{2}[3]{*}{Bias\tnote{c}} & \multirow{2}[3]{*}{ESE\tnote{d}} & \multirow{2}[3]{*}{RE\tnote{e}} & \multirow{2}[3]{*}{ASE\textsubscript{MB}\tnote{f}} & \multirow{2}[3]{*}{ASE\textsubscript{ROB}\tnote{g}} & \multirow{2}[3]{*}{ECP\textsubscript{MB}\tnote{h}} & \multirow{2}[3]{*}{ECP\textsubscript{ROB}\tnote{i}} & \multicolumn{2}{c}{ED/SE\tnote{j}} & \multicolumn{2}{c}{ED/NE\tnote{k}} \\
\cmidrule(lr){12-13} \cmidrule(lr){14-15}
& & & & & & & & & & & Bias & ESE & Bias & ESE \\
\midrule
B1 & 30 & B(i) & $\Delta$ & -1.056 & 0.217 & 1.000 & 0.245 & 0.197 & 0.005 & 0.001 & 0.989 & 0.986 & 1.027 & 0.992 \\
& & B(ii) & $\Delta^{D\text{-avg}}$ & 0.020 & 0.373 & 1.000 & 0.322 & 0.341 & 0.932 & 0.939 & 1.036 & 1.010 & 0.971 & 1.003 \\
& & & $\Delta(1)$ & 0.010 & 0.222 & 1.000 & 0.238 & 0.207 & 0.973 & 0.952 & 1.006 & 0.998 & 0.916 & 0.994 \\
& & & $\Delta(2)$ & 0.016 & 0.287 & 1.000 & 0.296 & 0.281 & 0.967 & 0.951 & 0.998 & 0.996 & 0.934 & 0.994 \\
& & & $\Delta(3)$ & 0.018 & 0.428 & 1.000 & 0.366 & 0.385 & 0.931 & 0.934 & 1.084 & 1.003 & 0.971 & 0.997 \\
& & & $\Delta(4)$ & 0.021 & 0.560 & 1.000 & 0.457 & 0.525 & 0.915 & 0.949 & 1.043 & 1.007 & 0.967 & 1.004 \\
& & & $\Delta(5)$ & 0.036 & 0.850 & 1.000 & 0.608 & 0.740 & 0.862 & 0.917 & 1.035 & 1.010 & 1.009 & 1.006 \\
& & B(iii) & $\Delta$ & -1.122 & 0.169 & 1.655 & 0.202 & 0.158 & 0.000 & 0.000 & 0.992 & 0.960 & 1.033 & 0.959 \\
& & B(iv) & $\Delta^{D\text{-avg}}$ & 0.012 & 0.321 & 1.345 & 0.281 & 0.291 & 0.939 & 0.942 & 0.995 & 1.004 & 0.976 & 0.996 \\
& & & $\Delta(1)$ & 0.007 & 0.174 & 1.631 & 0.195 & 0.165 & 0.982 & 0.942 & 1.030 & 0.990 & 0.928 & 0.984 \\
& & & $\Delta(2)$ & 0.009 & 0.238 & 1.449 & 0.249 & 0.226 & 0.976 & 0.947 & 1.046 & 0.982 & 1.006 & 0.976 \\
& & & $\Delta(3)$ & 0.010 & 0.347 & 1.523 & 0.313 & 0.315 & 0.945 & 0.936 & 1.213 & 0.994 & 1.070 & 0.984 \\
& & & $\Delta(4)$ & 0.013 & 0.473 & 1.402 & 0.395 & 0.434 & 0.924 & 0.943 & 0.961 & 1.003 & 0.951 & 0.999 \\
& & & $\Delta(5)$ & 0.023 & 0.710 & 1.433 & 0.524 & 0.611 & 0.886 & 0.931 & 0.916 & 1.007 & 0.959 & 1.003 \\
& & B(v) & $\Delta$ & -1.110 & 0.162 & 1.800 & 0.195 & 0.149 & 0.000 & 0.000 & 0.991 & 0.961 & 1.040 & 0.968 \\
& & B(vi) & $\Delta^{D\text{-avg}}$ & 0.009 & 0.311 & 1.438 & 0.264 & 0.279 & 0.930 & 0.938 & 0.950 & 1.006 & 0.943 & 0.996 \\
& & & $\Delta(1)$ & 0.005 & 0.164 & 1.825 & 0.185 & 0.155 & 0.985 & 0.963 & 1.015 & 0.993 & 0.924 & 0.986 \\
& & & $\Delta(2)$ & 0.006 & 0.228 & 1.582 & 0.235 & 0.214 & 0.969 & 0.953 & 1.027 & 0.983 & 0.985 & 0.977 \\
& & & $\Delta(3)$ & 0.008 & 0.337 & 1.616 & 0.295 & 0.303 & 0.938 & 0.936 & 1.258 & 0.996 & 1.063 & 0.984 \\
& & & $\Delta(4)$ & 0.009 & 0.460 & 1.481 & 0.371 & 0.420 & 0.931 & 0.947 & 0.895 & 1.004 & 0.904 & 0.998 \\
& & & $\Delta(5)$ & 0.018 & 0.691 & 1.514 & 0.492 & 0.594 & 0.879 & 0.920 & 0.853 & 1.007 & 0.913 & 1.001 \\
\bottomrule
\end{tabular}\smallskip
\begin{tablenotes}\linespread{1}\scriptsize
\item[a] TE: treatment effect. $\Delta = 2$; $\Delta^{D\text{-avg}}$ = $\sum_{d=1}^{5} \Delta(d)/5 = 2$; $\Delta(1) = 1$; $\Delta(2) = 1.5$; $\Delta(3) = 2$; $\Delta(4) = 2.5$; $\Delta(5) = 3$. \smallskip
\item[b] Model: six working linear mixed-models under the exponential decay correlation structure, which are combinations of covariate adjustment settings and treatment effect settings. B(i), no adjustment with constant TE; B(ii), no adjustment with duration-specific TE; B(iii), partial adjustment with constant TE; B(iv), partial adjustment with duration-specific TE; B(v), full adjustment with constant TE; B(vi), full adjustment with duration-specific TE. \smallskip
\item[c] Bias: bias as the average of (estimate - estimand) over simulations.\smallskip
\item[d] ESE: empirical standard error (SE).\smallskip
\item[e] RE: relative effeciency.\smallskip
\item[f] ASE\textsubscript{MB}: averaged model-based SE.\smallskip
\item[g] ASE\textsubscript{ROB}: averaged robust SE.\smallskip
\item[h] ECP\textsubscript{MB}: empirical coverage probability based on the model-based SE.\smallskip
\item[i] ECP\textsubscript{ROB}: empirical coverage probability based on the robust SE.\smallskip
\item[j] ED: exponential decay correlation structure. SE: simple exchangeable correlation structure. ED/SE: ratio of Bias or ESE between the models under ED and SE.\smallskip
\item[k] ED: exponential decay correlation structure. NE: nested exchangeable correlation structure. ED/NE: ratio of Bias or ESE between the models under ED and NE.
\end{tablenotes}
\end{threeparttable}
}
\end{table}

\begin{table}[htbp]
\footnotesize
\caption{Simulation results for continuous outcomes under Simulation Scenario B1 with $I=100$. The working linear mixed model includes a discrete-time exponential decay correlation structure.}\label{tb:Tablew13}
\centering
\resizebox{\textwidth}{!}{%
\begin{threeparttable}
\begin{tabular}{lclcccccccccccc}
\toprule
\midrule
\multirow{2}[3]{*}{Scenario} & \multirow{2}[3]{*}{$I$} & \multirow{2}[3]{*}{Model\tnote{a}} & \multirow{2}[3]{*}{TE\tnote{b}} & \multirow{2}[3]{*}{Bias\tnote{c}} & \multirow{2}[3]{*}{ESE\tnote{d}} & \multirow{2}[3]{*}{RE\tnote{e}} & \multirow{2}[3]{*}{ASE\textsubscript{MB}\tnote{f}} & \multirow{2}[3]{*}{ASE\textsubscript{ROB}\tnote{g}} & \multirow{2}[3]{*}{ECP\textsubscript{MB}\tnote{h}} & \multirow{2}[3]{*}{ECP\textsubscript{ROB}\tnote{i}} & \multicolumn{2}{c}{ED/SE\tnote{j}} & \multicolumn{2}{c}{ED/NE\tnote{k}} \\
\cmidrule(lr){12-13} \cmidrule(lr){14-15}
& & & & & & & & & & & Bias & ESE & Bias & ESE \\
\midrule
B1 & 100 & B(i) & $\Delta$ & -1.064 & 0.121 & 1.000 & 0.134 & 0.112 & 0.000 & 0.000 & 0.990 & 0.985 & 1.025 & 0.994 \\
& & B(ii) & $\Delta^{D\text{-avg}}$ & 0.010 & 0.216 & 1.000 & 0.179 & 0.200 & 0.910 & 0.939 & 1.012 & 1.007 & 0.998 & 1.003 \\
& & & $\Delta(1)$ & 0.005 & 0.125 & 1.000 & 0.131 & 0.118 & 0.966 & 0.938 & 0.951 & 1.000 & 0.959 & 0.997 \\
& & & $\Delta(2)$ & 0.009 & 0.172 & 1.000 & 0.163 & 0.163 & 0.936 & 0.934 & 1.004 & 0.998 & 1.002 & 0.995 \\
& & & $\Delta(3)$ & 0.007 & 0.234 & 1.000 & 0.203 & 0.223 & 0.908 & 0.943 & 1.066 & 1.000 & 1.034 & 0.997 \\
& & & $\Delta(4)$ & 0.012 & 0.325 & 1.000 & 0.254 & 0.305 & 0.891 & 0.944 & 0.975 & 1.010 & 0.962 & 1.005 \\
& & & $\Delta(5)$ & 0.016 & 0.474 & 1.000 & 0.336 & 0.444 & 0.843 & 0.939 & 1.045 & 1.006 & 1.019 & 1.004 \\
& & B(iii) & $\Delta$ & -1.127 & 0.096 & 1.576 & 0.111 & 0.089 & 0.000 & 0.000 & 0.994 & 0.973 & 1.031 & 0.980 \\
& & B(iv) & $\Delta^{D\text{-avg}}$ & 0.013 & 0.189 & 1.309 & 0.157 & 0.175 & 0.907 & 0.940 & 1.038 & 1.006 & 1.015 & 1.000 \\
& & & $\Delta(1)$ & 0.007 & 0.100 & 1.566 & 0.108 & 0.094 & 0.965 & 0.934 & 0.963 & 0.996 & 0.978 & 0.995 \\
& & & $\Delta(2)$ & 0.009 & 0.139 & 1.523 & 0.138 & 0.133 & 0.950 & 0.938 & 1.030 & 0.990 & 1.038 & 0.985 \\
& & & $\Delta(3)$ & 0.010 & 0.200 & 1.369 & 0.174 & 0.186 & 0.923 & 0.943 & 1.069 & 0.995 & 1.043 & 0.989 \\
& & & $\Delta(4)$ & 0.016 & 0.275 & 1.397 & 0.220 & 0.259 & 0.897 & 0.941 & 1.031 & 1.008 & 0.995 & 1.001 \\
& & & $\Delta(5)$ & 0.022 & 0.399 & 1.409 & 0.291 & 0.374 & 0.849 & 0.945 & 1.058 & 1.006 & 1.020 & 1.004 \\
& & B(v) & $\Delta$ & -1.118 & 0.092 & 1.722 & 0.107 & 0.084 & 0.000 & 0.000 & 0.993 & 0.965 & 1.038 & 0.975 \\
& & B(vi) & $\Delta^{D\text{-avg}}$ & 0.011 & 0.180 & 1.437 & 0.148 & 0.169 & 0.907 & 0.943 & 1.060 & 1.005 & 1.033 & 1.000 \\
& & & $\Delta(1)$ & 0.006 & 0.094 & 1.785 & 0.102 & 0.089 & 0.969 & 0.940 & 0.974 & 0.994 & 0.992 & 0.994 \\
& & & $\Delta(2)$ & 0.007 & 0.131 & 1.708 & 0.131 & 0.127 & 0.951 & 0.947 & 1.055 & 0.986 & 1.068 & 0.982 \\
& & & $\Delta(3)$ & 0.009 & 0.191 & 1.495 & 0.165 & 0.180 & 0.909 & 0.948 & 1.106 & 0.995 & 1.071 & 0.988 \\
& & & $\Delta(4)$ & 0.013 & 0.264 & 1.517 & 0.208 & 0.252 & 0.897 & 0.945 & 1.042 & 1.009 & 1.005 & 1.002 \\
& & & $\Delta(5)$ & 0.017 & 0.388 & 1.491 & 0.275 & 0.365 & 0.834 & 0.943 & 1.084 & 1.006 & 1.035 & 1.004 \\
\bottomrule
\end{tabular}\smallskip
\begin{tablenotes}\linespread{1}\scriptsize
\item[a] TE: treatment effect. $\Delta = 2$; $\Delta^{D\text{-avg}}$ = $\sum_{d=1}^{5} \Delta(d)/5 = 2$; $\Delta(1) = 1$; $\Delta(2) = 1.5$; $\Delta(3) = 2$; $\Delta(4) = 2.5$; $\Delta(5) = 3$. \smallskip
\item[b] Model: six working linear mixed-models under the exponential decay correlation structure, which are combinations of covariate adjustment settings and treatment effect settings. B(i), no adjustment with constant TE; B(ii), no adjustment with duration-specific TE; B(iii), partial adjustment with constant TE; B(iv), partial adjustment with duration-specific TE; B(v), full adjustment with constant TE; B(vi), full adjustment with duration-specific TE. \smallskip
\item[c] Bias: bias as the average of (estimate - estimand) over simulations.\smallskip
\item[d] ESE: empirical standard error (SE).\smallskip
\item[e] RE: relative effeciency.\smallskip
\item[f] ASE\textsubscript{MB}: averaged model-based SE.\smallskip
\item[g] ASE\textsubscript{ROB}: averaged robust SE.\smallskip
\item[h] ECP\textsubscript{MB}: empirical coverage probability based on the model-based SE.\smallskip
\item[i] ECP\textsubscript{ROB}: empirical coverage probability based on the robust SE.\smallskip
\item[j] ED: exponential decay correlation structure. SE: simple exchangeable correlation structure. ED/SE: ratio of Bias or ESE between the models under ED and SE.\smallskip
\item[k] ED: exponential decay correlation structure. NE: nested exchangeable correlation structure. ED/NE: ratio of Bias or ESE between the models under ED and NE.
\end{tablenotes}
\end{threeparttable}
}
\end{table}

\begin{table}[htbp]
\footnotesize
\caption{Simulation results for continuous outcomes under Simulation Scenario B2 with $I=30$. The working linear mixed model includes a discrete-time exponential decay correlation structure..}\label{tb:Tablew14}
\centering
\resizebox{\textwidth}{!}{%
\begin{threeparttable}
\begin{tabular}{lclcccccccccccc}
\toprule
\midrule
\multirow{2}[3]{*}{Scenario} & \multirow{2}[3]{*}{$I$} & \multirow{2}[3]{*}{Model\tnote{a}} & \multirow{2}[3]{*}{TE\tnote{b}} & \multirow{2}[3]{*}{Bias\tnote{c}} & \multirow{2}[3]{*}{ESE\tnote{d}} & \multirow{2}[3]{*}{RE\tnote{e}} & \multirow{2}[3]{*}{ASE\textsubscript{MB}\tnote{f}} & \multirow{2}[3]{*}{ASE\textsubscript{ROB}\tnote{g}} & \multirow{2}[3]{*}{ECP\textsubscript{MB}\tnote{h}} & \multirow{2}[3]{*}{ECP\textsubscript{ROB}\tnote{i}} & \multicolumn{2}{c}{ED/SE\tnote{j}} & \multicolumn{2}{c}{ED/NE\tnote{k}} \\
\cmidrule(lr){12-13} \cmidrule(lr){14-15}
& & & & & & & & & & & Bias & ESE & Bias & ESE \\
\midrule
B2 & 30 & B(i) & $\Delta$ & -1.043 & 0.233 & 1.000 & 0.272 & 0.228 & 0.023 & 0.010 & 0.979 & 0.974 & 1.034 & 0.987 \\
& & B(ii) & $\Delta^{D\text{-avg}}$ & -0.007 & 0.428 & 1.000 & 0.361 & 0.399 & 0.922 & 0.944 & 0.815 & 1.003 & 1.032 & 1.001 \\
& & & $\Delta(1)$ & 0.005 & 0.237 & 1.000 & 0.267 & 0.240 & 0.984 & 0.967 & 1.246 & 0.995 & 0.933 & 0.995 \\
& & & $\Delta(2)$ & -0.005 & 0.338 & 1.000 & 0.334 & 0.325 & 0.965 & 0.951 & 0.674 & 0.989 & 0.809 & 0.988 \\
& & & $\Delta(3)$ & -0.021 & 0.482 & 1.000 & 0.413 & 0.444 & 0.923 & 0.945 & 0.969 & 0.998 & 1.048 & 0.996 \\
& & & $\Delta(4)$ & -0.014 & 0.655 & 1.000 & 0.514 & 0.613 & 0.904 & 0.945 & 0.883 & 1.002 & 1.122 & 1.001 \\
& & & $\Delta(5)$ & -0.001 & 0.982 & 1.000 & 0.683 & 0.887 & 0.855 & 0.934 & 0.252 & 1.014 & 0.642 & 1.008 \\
& & B(iii) & $\Delta$ & -1.116 & 0.181 & 1.650 & 0.222 & 0.190 & 0.000 & 0.000 & 0.985 & 0.957 & 1.044 & 0.975 \\
& & B(iv) & $\Delta^{D\text{-avg}}$ & -0.006 & 0.374 & 1.311 & 0.317 & 0.345 & 0.929 & 0.939 & 0.787 & 1.004 & 0.925 & 0.999 \\
& & & $\Delta(1)$ & 0.003 & 0.189 & 1.582 & 0.216 & 0.198 & 0.986 & 0.965 & 1.417 & 0.989 & 1.123 & 0.992 \\
& & & $\Delta(2)$ & -0.005 & 0.276 & 1.501 & 0.279 & 0.265 & 0.972 & 0.952 & 0.659 & 0.979 & 0.758 & 0.976 \\
& & & $\Delta(3)$ & -0.015 & 0.399 & 1.458 & 0.353 & 0.365 & 0.949 & 0.941 & 0.911 & 0.992 & 0.985 & 0.990 \\
& & & $\Delta(4)$ & -0.009 & 0.553 & 1.403 & 0.445 & 0.505 & 0.914 & 0.942 & 0.867 & 1.000 & 1.064 & 0.998 \\
& & & $\Delta(5)$ & -0.005 & 0.803 & 1.495 & 0.589 & 0.725 & 0.879 & 0.927 & 0.723 & 1.021 & 0.859 & 1.010 \\
& & B(v) & $\Delta$ & -1.104 & 0.172 & 1.819 & 0.216 & 0.183 & 0.000 & 0.000 & 0.983 & 0.954 & 1.053 & 0.976 \\
& & B(vi) & $\Delta^{D\text{-avg}}$ & -0.011 & 0.355 & 1.460 & 0.299 & 0.334 & 0.927 & 0.946 & 0.864 & 1.001 & 0.946 & 0.997 \\
& & & $\Delta(1)$ & 0.000 & 0.176 & 1.814 & 0.206 & 0.190 & 0.987 & 0.970 & 0.409 & 0.987 & 2.759 & 0.993 \\
& & & $\Delta(2)$ & -0.007 & 0.258 & 1.707 & 0.267 & 0.255 & 0.972 & 0.959 & 0.714 & 0.974 & 0.813 & 0.973 \\
& & & $\Delta(3)$ & -0.018 & 0.378 & 1.630 & 0.336 & 0.353 & 0.951 & 0.951 & 0.918 & 0.986 & 0.981 & 0.988 \\
& & & $\Delta(4)$ & -0.015 & 0.530 & 1.525 & 0.421 & 0.491 & 0.915 & 0.951 & 0.879 & 0.996 & 0.999 & 0.996 \\
& & & $\Delta(5)$ & -0.013 & 0.778 & 1.594 & 0.557 & 0.710 & 0.870 & 0.937 & 0.908 & 1.021 & 0.937 & 1.010 \\
\bottomrule
\end{tabular}\smallskip
\begin{tablenotes}\linespread{1}\scriptsize
\item[a] TE: treatment effect. $\Delta = 2$; $\Delta^{D\text{-avg}}$ = $\sum_{d=1}^{5} \Delta(d)/5 = 2$; $\Delta(1) = 1$; $\Delta(2) = 1.5$; $\Delta(3) = 2$; $\Delta(4) = 2.5$; $\Delta(5) = 3$. \smallskip
\item[b] Model: six working linear mixed-models under the exponential decay correlation structure, which are combinations of covariate adjustment settings and treatment effect settings. B(i), no adjustment with constant TE; B(ii), no adjustment with duration-specific TE; B(iii), partial adjustment with constant TE; B(iv), partial adjustment with duration-specific TE; B(v), full adjustment with constant TE; B(vi), full adjustment with duration-specific TE. \smallskip
\item[c] Bias: bias as the average of (estimate - estimand) over simulations.\smallskip
\item[d] ESE: empirical standard error (SE).\smallskip
\item[e] RE: relative effeciency.\smallskip
\item[f] ASE\textsubscript{MB}: averaged model-based SE.\smallskip
\item[g] ASE\textsubscript{ROB}: averaged robust SE.\smallskip
\item[h] ECP\textsubscript{MB}: empirical coverage probability based on the model-based SE.\smallskip
\item[i] ECP\textsubscript{ROB}: empirical coverage probability based on the robust SE.\smallskip
\item[j] ED: exponential decay correlation structure. SE: simple exchangeable correlation structure. ED/SE: ratio of Bias or ESE between the models under ED and SE.\smallskip
\item[k] ED: exponential decay correlation structure. NE: nested exchangeable correlation structure. ED/NE: ratio of Bias or ESE between the models under ED and NE.
\end{tablenotes}
\end{threeparttable}
}
\end{table}

\begin{table}[htbp]
\footnotesize
\caption{Simulation results for continuous outcomes under Simulation Scenario B2 with $I=100$. The working linear mixed model includes a discrete-time exponential decay correlation structure.}\label{tb:Tablew15}
\centering
\resizebox{\textwidth}{!}{%
\begin{threeparttable}
\begin{tabular}{lclcccccccccccc}
\toprule
\midrule
\multirow{2}[3]{*}{Scenario} & \multirow{2}[3]{*}{$I$} & \multirow{2}[3]{*}{Model\tnote{a}} & \multirow{2}[3]{*}{TE\tnote{b}} & \multirow{2}[3]{*}{Bias\tnote{c}} & \multirow{2}[3]{*}{ESE\tnote{d}} & \multirow{2}[3]{*}{RE\tnote{e}} & \multirow{2}[3]{*}{ASE\textsubscript{MB}\tnote{f}} & \multirow{2}[3]{*}{ASE\textsubscript{ROB}\tnote{g}} & \multirow{2}[3]{*}{ECP\textsubscript{MB}\tnote{h}} & \multirow{2}[3]{*}{ECP\textsubscript{ROB}\tnote{i}} & \multicolumn{2}{c}{ED/SE\tnote{j}} & \multicolumn{2}{c}{ED/NE\tnote{k}} \\
\cmidrule(lr){12-13} \cmidrule(lr){14-15}
& & & & & & & & & & & Bias & ESE & Bias & ESE \\
\midrule
B2 & 100 & B(i) & $\Delta$ & -1.051 & 0.123 & 1.000 & 0.149 & 0.130 & 0.000 & 0.000 & 0.986 & 0.982 & 1.037 & 0.995 \\
& & B(ii) & $\Delta^{D\text{-avg}}$ & 0.003 & 0.231 & 1.000 & 0.201 & 0.229 & 0.919 & 0.950 & 0.776 & 1.011 & 0.887 & 1.004 \\
& & & $\Delta(1)$ & 0.000 & 0.128 & 1.000 & 0.146 & 0.137 & 0.975 & 0.962 & 0.676 & 1.000 & 0.830 & 0.997 \\
& & & $\Delta(2)$ & 0.005 & 0.178 & 1.000 & 0.184 & 0.185 & 0.962 & 0.959 & 0.755 & 0.994 & 0.777 & 0.994 \\
& & & $\Delta(3)$ & -0.003 & 0.246 & 1.000 & 0.228 & 0.254 & 0.928 & 0.959 & 1.584 & 1.001 & 1.235 & 0.997 \\
& & & $\Delta(4)$ & 0.000 & 0.359 & 1.000 & 0.285 & 0.354 & 0.888 & 0.954 & 0.043 & 1.008 & 0.274 & 1.002 \\
& & & $\Delta(5)$ & 0.013 & 0.545 & 1.000 & 0.378 & 0.520 & 0.830 & 0.932 & 0.949 & 1.012 & 1.016 & 1.007 \\
& & B(iii) & $\Delta$ & -1.119 & 0.097 & 1.594 & 0.122 & 0.108 & 0.000 & 0.000 & 0.989 & 0.965 & 1.045 & 0.978 \\
& & B(iv) & $\Delta^{D\text{-avg}}$ & 0.004 & 0.200 & 1.330 & 0.177 & 0.203 & 0.927 & 0.962 & 0.829 & 1.011 & 0.966 & 1.004 \\
& & & $\Delta(1)$ & 0.003 & 0.101 & 1.598 & 0.119 & 0.114 & 0.984 & 0.981 & 1.151 & 0.994 & 1.112 & 0.998 \\
& & & $\Delta(2)$ & 0.003 & 0.144 & 1.527 & 0.155 & 0.153 & 0.974 & 0.970 & 0.693 & 0.986 & 0.796 & 0.985 \\
& & & $\Delta(3)$ & 0.000 & 0.205 & 1.439 & 0.196 & 0.212 & 0.942 & 0.964 & 0.412 & 0.998 & 3.003 & 0.992 \\
& & & $\Delta(4)$ & 0.001 & 0.297 & 1.464 & 0.248 & 0.297 & 0.898 & 0.957 & 0.315 & 1.004 & 0.599 & 0.997 \\
& & & $\Delta(5)$ & 0.013 & 0.445 & 1.499 & 0.327 & 0.431 & 0.853 & 0.940 & 0.924 & 1.014 & 0.995 & 1.008 \\
& & B(v) & $\Delta$ & -1.107 & 0.095 & 1.674 & 0.118 & 0.104 & 0.000 & 0.000 & 0.987 & 0.973 & 1.054 & 0.987 \\
& & B(vi) & $\Delta^{D\text{-avg}}$ & 0.004 & 0.193 & 1.431 & 0.167 & 0.197 & 0.919 & 0.959 & 0.777 & 1.009 & 0.977 & 1.003 \\
& & & $\Delta(1)$ & 0.003 & 0.096 & 1.771 & 0.114 & 0.109 & 0.985 & 0.981 & 1.143 & 0.993 & 1.143 & 1.001 \\
& & & $\Delta(2)$ & 0.003 & 0.138 & 1.660 & 0.148 & 0.147 & 0.971 & 0.973 & 0.676 & 0.986 & 0.800 & 0.984 \\
& & & $\Delta(3)$ & 0.001 & 0.199 & 1.539 & 0.187 & 0.206 & 0.944 & 0.966 & 0.720 & 0.995 & 1.744 & 0.990 \\
& & & $\Delta(4)$ & 0.001 & 0.287 & 1.564 & 0.236 & 0.290 & 0.892 & 0.955 & 0.446 & 1.002 & 0.808 & 0.995 \\
& & & $\Delta(5)$ & 0.010 & 0.434 & 1.579 & 0.311 & 0.422 & 0.845 & 0.943 & 0.826 & 1.014 & 0.963 & 1.007 \\
\bottomrule
\end{tabular}\smallskip
\begin{tablenotes}\linespread{1}\scriptsize
\item[a] TE: treatment effect. $\Delta = 2$; $\Delta^{D\text{-avg}}$ = $\sum_{d=1}^{5} \Delta(d)/5 = 2$; $\Delta(1) = 1$; $\Delta(2) = 1.5$; $\Delta(3) = 2$; $\Delta(4) = 2.5$; $\Delta(5) = 3$. \smallskip
\item[b] Model: six working linear mixed-models under the exponential decay correlation structure, which are combinations of covariate adjustment settings and treatment effect settings. B(i), no adjustment with constant TE; B(ii), no adjustment with duration-specific TE; B(iii), partial adjustment with constant TE; B(iv), partial adjustment with duration-specific TE; B(v), full adjustment with constant TE; B(vi), full adjustment with duration-specific TE. \smallskip
\item[c] Bias: bias as the average of (estimate - estimand) over simulations.\smallskip
\item[d] ESE: empirical standard error (SE).\smallskip
\item[e] RE: relative effeciency.\smallskip
\item[f] ASE\textsubscript{MB}: averaged model-based SE.\smallskip
\item[g] ASE\textsubscript{ROB}: averaged robust SE.\smallskip
\item[h] ECP\textsubscript{MB}: empirical coverage probability based on the model-based SE.\smallskip
\item[i] ECP\textsubscript{ROB}: empirical coverage probability based on the robust SE.\smallskip
\item[j] ED: exponential decay correlation structure. SE: simple exchangeable correlation structure. ED/SE: ratio of Bias or ESE between the models under ED and SE.\smallskip
\item[k] ED: exponential decay correlation structure. NE: nested exchangeable correlation structure. ED/NE: ratio of Bias or ESE between the models under ED and NE.
\end{tablenotes}
\end{threeparttable}
}
\end{table}

\begin{table}[htbp]
\caption{Simulation scenarios and results for binary outcomes under Simulation Design C without covariate adjustment. The working linear mixed model includes a discrete-time exponential decay correlation structure and GLMM is fitted with an exponential decay random-effects structure.}\label{tb:Tablew16}
\centering
\resizebox{\textwidth}{!}{%
\begin{threeparttable}
\begin{tabular}{lccccccccccc}
\toprule
\midrule
\multirow{3}[4]{*}{Scenario} & \multirow{3}[4]{*}{$I$} & \multirow{3}[4]{*}{TE\tnote{a}} & \multirow{3}[4]{*}{Estimand} & \multicolumn{5}{c}{Linear Mixed Model} & \multicolumn{3}{c}{GLMM\tnote{b}} \\
\cmidrule(lr){5-9} \cmidrule(lr){10-12}
& & & & \multirow{2}[3]{*}{Bias\tnote{c}} & \multirow{2}[3]{*}{ASE/ESE\tnote{d}} & \multirow{2}[3]{*}{ECP\tnote{e}} & \multicolumn{2}{c}{ED/NE\tnote{f}} & \multirow{2}[3]{*}{Bias\tnote{c}} & \multirow{2}[3]{*}{ASE/ESE\tnote{d}} & \multirow{2}[3]{*}{ECP\tnote{e}} \\
\cmidrule(lr){8-9} 
& & & & & & & Bias & ESE\\
\midrule
C1 & 30 & $\Phi_1(1)$ & 2.299 & 0.061 & 0.875 & 0.905 & 0.998 & 1.006 & 0.139 & 0.839 & 0.915 \\
& & $\Phi_2(1)$ & 2.291 & 0.065 & 0.884 & 0.915 & 0.868 & 0.974 & 0.114 & 0.832 & 0.918 \\
& & $\Phi_2(2)$ & 2.716 & 0.089 & 0.866 & 0.899 & 1.005 & 1.011 & 0.196 & 0.895 & 0.941 \\
& 100 & $\Phi_1(1)$ & 2.299 & 0.024 & 0.985 & 0.951 & 0.999 & 1.000 & 0.109 & 0.903 & 0.930 \\
& & $\Phi_2(1)$ & 2.291 & 0.008 & 0.943 & 0.932 & 1.013 & 1.001 & 0.062 & 0.838 & 0.895 \\
& & $\Phi_2(2)$ & 2.716 & 0.010 & 0.957 & 0.932 & 1.009 & 1.001 & 0.124 & 0.933 & 0.925 \\
& 300 & $\Phi_1(1)$ & 2.299 & 0.008 & 0.976 & 0.939 & 1.000 & 1.000 & 0.092 & 0.893 & 0.886 \\
& & $\Phi_2(1)$ & 2.291 & 0.006 & 0.986 & 0.951 & 0.997 & 1.000 & 0.062 & 0.868 & 0.907 \\
& & $\Phi_2(2)$ & 2.716 & 0.004 & 0.991 & 0.939 & 0.996 & 1.000 & 0.119 & 0.959 & 0.902 \\
& 1000 & $\Phi_1(1)$ & 2.299 & 0.002 & 1.007 & 0.952 & 1.000 & 1.000 & 0.087 & 0.915 & 0.758 \\
& & $\Phi_2(1)$ & 2.291 & 0.002 & 0.926 & 0.932 & 1.000 & 1.000 & 0.057 & 0.808 & 0.814 \\
& & $\Phi_2(2)$ & 2.716 & 0.005 & 1.005 & 0.953 & 1.000 & 1.000 & 0.122 & 0.968 & 0.777 \\
C2 & 30 & $\Phi_1(1)$ & 2.340 & 0.031 & 0.922 & 0.908 & 0.959 & 0.988 & 0.070 & 0.961 & 0.934 \\
& & $\Phi_2(1)$ & 2.311 & 0.037 & 0.877 & 0.911 & 0.827 & 0.990 & 0.065 & 0.949 & 0.943 \\
& & $\Phi_2(2)$ & 2.732 & 0.070 & 0.843 & 0.903 & 0.932 & 0.998 & 0.122 & 0.901 & 0.938 \\
& 100 & $\Phi_1(1)$ & 2.340 & 0.029 & 0.975 & 0.939 & 0.994 & 1.000 & 0.071 & 0.969 & 0.946 \\
& & $\Phi_2(1)$ & 2.311 & 0.004 & 0.985 & 0.943 & 0.998 & 1.000 & 0.035 & 0.987 & 0.946 \\
& & $\Phi_2(2)$ & 2.732 & 0.009 & 0.986 & 0.948 & 1.095 & 1.001 & 0.061 & 0.991 & 0.948 \\
& 300 & $\Phi_1(1)$ & 2.340 & 0.003 & 0.986 & 0.948 & 1.000 & 1.000 & 0.044 & 0.980 & 0.941 \\
& & $\Phi_2(1)$ & 2.311 & 0.002 & 0.995 & 0.944 & 1.000 & 1.000 & 0.034 & 0.976 & 0.943 \\
& & $\Phi_2(2)$ & 2.732 & -0.002 & 0.975 & 0.945 & 1.000 & 1.000 & 0.052 & 0.964 & 0.940 \\
& 1000 & $\Phi_1(1)$ & 2.340 & 0.002 & 1.000 & 0.941 & 1.000 & 1.000 & 0.043 & 0.983 & 0.924 \\
& & $\Phi_2(1)$ & 2.311 & 0.001 & 0.923 & 0.929 & 1.000 & 1.000 & 0.033 & 0.902 & 0.910 \\
& & $\Phi_2(2)$ & 2.732 & 0.005 & 1.028 & 0.947 & 1.000 & 1.000 & 0.059 & 1.009 & 0.908 \\
\bottomrule
\end{tabular}\smallskip
\begin{tablenotes}\linespread{1}\scriptsize
\item[a] TE: treatment effect. \smallskip
\item[b] GLMM: Generalized linear mixed model. \smallskip
\item[c] Bias: bias as the average of (estimate - estimand) over simulations. \smallskip
\item[d] ASE/ESE: averaged standard error (SE) divided by empirical SE.\smallskip
\item[e] ECP: empirical coverage probability.\smallskip
\item[f] ED: exponential decay correlation structure. NE: nested exchangeable correlation structure. ED/NE: ratio of Bias or ESE between the models under ED and NE.
\end{tablenotes}
\end{threeparttable}
}
\end{table}

\begin{table}[htbp]
\caption{Simulation scenarios and results for binary outcomes under Simulation Design C with covariate adjustment. The working linear mixed model includes a discrete-time exponential decay correlation structure and GLMM is fitted with an exponential decay random-effects structure.}\label{tb:Tablew17}
\centering
\resizebox{\textwidth}{!}{%
\begin{threeparttable}
\begin{tabular}{lccccccccccc}
\toprule
\midrule
\multirow{3}[4]{*}{Scenario} & \multirow{3}[4]{*}{$I$} & \multirow{3}[4]{*}{TE\tnote{a}} & \multirow{3}[4]{*}{Estimand} & \multicolumn{5}{c}{Linear Mixed Model} & \multicolumn{3}{c}{GLMM\tnote{b}} \\
\cmidrule(lr){5-9} \cmidrule(lr){10-12}
& & & & \multirow{2}[3]{*}{Bias\tnote{c}} & \multirow{2}[3]{*}{ASE/ESE\tnote{d}} & \multirow{2}[3]{*}{ECP\tnote{e}} & \multicolumn{2}{c}{ED/NE\tnote{f}} & \multirow{2}[3]{*}{Bias\tnote{c}} & \multirow{2}[3]{*}{ASE/ESE\tnote{d}} & \multirow{2}[3]{*}{ECP\tnote{e}} \\
\cmidrule(lr){8-9} 
& & & & & & & Bias & ESE\\
\midrule
C1 & 30 & $\Phi_1(1)$ & 2.299 & 0.061 & 0.875 & 0.904 & 1.001 & 1.006 & 0.156 & 0.840 & 0.916 \\
& & $\Phi_2(1)$ & 2.291 & 0.065 & 0.888 & 0.920 & 0.872 & 0.973 & 0.129 & 0.836 & 0.919 \\
& & $\Phi_2(2)$ & 2.716 & 0.091 & 0.869 & 0.904 & 1.011 & 1.011 & 0.220 & 0.898 & 0.940 \\
& 100 & $\Phi_1(1)$ & 2.299 & 0.024 & 0.986 & 0.956 & 0.999 & 1.000 & 0.125 & 0.903 & 0.923 \\
& & $\Phi_2(1)$ & 2.291 & 0.008 & 0.944 & 0.930 & 0.996 & 1.000 & 0.077 & 0.839 & 0.894 \\
& & $\Phi_2(2)$ & 2.716 & 0.010 & 0.959 & 0.928 & 0.999 & 1.000 & 0.145 & 0.935 & 0.923 \\
& 300 & $\Phi_1(1)$ & 2.299 & 0.008 & 0.975 & 0.941 & 1.000 & 1.000 & 0.108 & 0.892 & 0.873 \\
& & $\Phi_2(1)$ & 2.291 & 0.007 & 0.986 & 0.955 & 0.997 & 1.000 & 0.077 & 0.867 & 0.892 \\
& & $\Phi_2(2)$ & 2.716 & 0.004 & 0.992 & 0.942 & 0.996 & 1.000 & 0.140 & 0.962 & 0.891 \\
& 1000 & $\Phi_1(1)$ & 2.299 & 0.002 & 1.006 & 0.954 & 1.000 & 1.000 & 0.102 & 0.915 & 0.700 \\
& & $\Phi_2(1)$ & 2.291 & 0.002 & 0.928 & 0.928 & 1.000 & 1.000 & 0.072 & 0.810 & 0.785 \\
& & $\Phi_2(2)$ & 2.716 & 0.005 & 1.006 & 0.952 & 1.000 & 1.000 & 0.142 & 0.970 & 0.721 \\
C2 & 30 & $\Phi_1(1)$ & 2.340 & 0.035 & 0.921 & 0.908 & 0.970 & 0.989 & 0.104 & 0.960 & 0.941 \\
& & $\Phi_2(1)$ & 2.311 & 0.037 & 0.877 & 0.909 & 0.836 & 0.990 & 0.092 & 0.947 & 0.941 \\
& & $\Phi_2(2)$ & 2.732 & 0.077 & 0.843 & 0.903 & 0.936 & 0.999 & 0.164 & 0.900 & 0.939 \\
& 100 & $\Phi_1(1)$ & 2.340 & 0.030 & 0.978 & 0.946 & 0.994 & 1.000 & 0.100 & 0.973 & 0.946 \\
& & $\Phi_2(1)$ & 2.311 & 0.004 & 0.988 & 0.943 & 1.000 & 1.000 & 0.061 & 0.990 & 0.949 \\
& & $\Phi_2(2)$ & 2.732 & 0.011 & 0.989 & 0.945 & 1.080 & 1.001 & 0.096 & 0.995 & 0.949 \\
& 300 & $\Phi_1(1)$ & 2.340 & 0.003 & 0.985 & 0.950 & 1.000 & 1.000 & 0.072 & 0.980 & 0.916 \\
& & $\Phi_2(1)$ & 2.311 & 0.002 & 0.995 & 0.946 & 1.000 & 1.000 & 0.059 & 0.976 & 0.941 \\
& & $\Phi_2(2)$ & 2.732 & -0.001 & 0.973 & 0.944 & 1.000 & 1.000 & 0.085 & 0.965 & 0.919 \\
& 1000 & $\Phi_1(1)$ & 2.340 & 0.001 & 1.000 & 0.943 & 1.000 & 1.000 & 0.071 & 0.984 & 0.850 \\
& & $\Phi_2(1)$ & 2.311 & 0.001 & 0.927 & 0.931 & 1.000 & 1.000 & 0.057 & 0.906 & 0.866 \\
& & $\Phi_2(2)$ & 2.732 & 0.005 & 1.030 & 0.946 & 1.000 & 1.000 & 0.092 & 1.011 & 0.862 \\
\bottomrule
\end{tabular}\smallskip
\begin{tablenotes}\linespread{1}\scriptsize
\item[a] TE: treatment effect. \smallskip
\item[b] GLMM: Generalized linear mixed model. \smallskip
\item[c] Bias: bias as the average of (estimate - estimand) over simulations. \smallskip
\item[d] ASE/ESE: averaged standard error (SE) divided by empirical SE.\smallskip
\item[e] ECP: empirical coverage probability.\smallskip
\item[f] ED: exponential decay correlation structure. NE: nested exchangeable correlation structure. ED/NE: ratio of Bias or ESE between the models under ED and NE.
\end{tablenotes}
\end{threeparttable}
}
\end{table}

\clearpage
\section{Additional results for data application}\label{sec: add-application}
\begin{figure}
    \centering
    \includegraphics[width=0.7\textwidth]{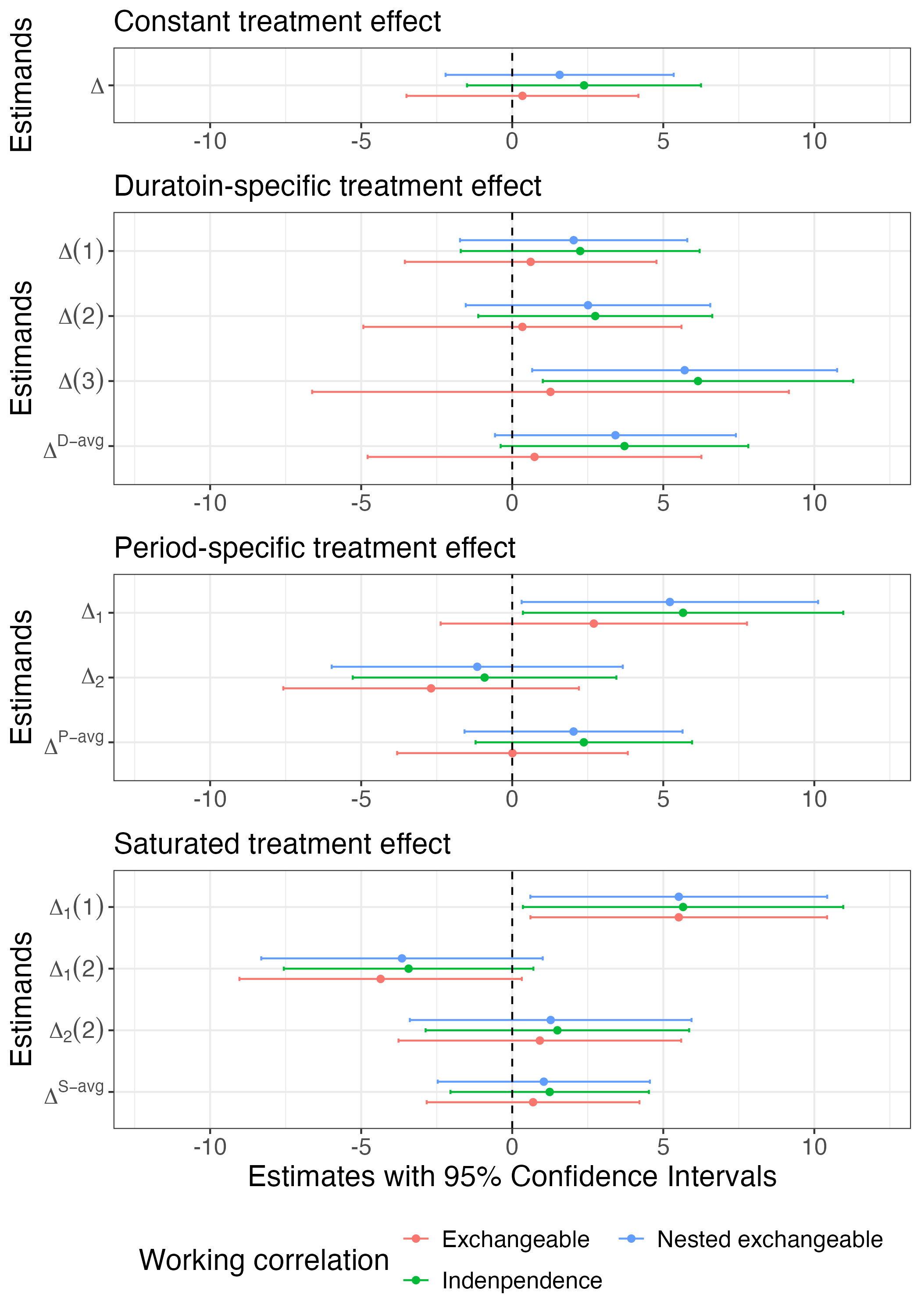}
    \caption{Point estimates and 95\% robust confidence intervals based on the analysis of SMARThealth India SW-CRT with covariate-adjusted linear mixed models. }
    \label{fig:data-analysis-adjusted}
\end{figure}

Figure \ref{fig:data-analysis-adjusted} presents the results of covariate-adjusted analyses for our data application. 

{
\bibliographystyle{apalike}
\bibliography{references}
}